\documentclass[10pt,aps,prd,notitlepage,superscriptaddress,nofootinbib,amsmath,amssymb,twocolumn,showpacs,pdftex]{revtex4-2}
\usepackage[T1]{fontenc} 
\usepackage{ulem}
\usepackage{lipsum}

\pdfoutput=1

\usepackage[outline]{contour}
\usepackage{xcolor,graphicx}
\usepackage[singlelinecheck=off, justification=centerlast]{caption}
\usepackage[singlelinecheck=off, justification=centerlast]{subcaption} 
\usepackage{bm,bbold}
\usepackage{cancel}

\usepackage{color,marvosym}
\usepackage{graphicx} 
\usepackage{comment}

\definecolor{darkred}{rgb}{0.5,0,0}
\definecolor{darkgreen}{rgb}{0,0.5,0}
\definecolor{darkblue}{rgb}{0,0,0.5}

\usepackage{hyperref}
\hypersetup{colorlinks,
linkcolor=darkblue,
filecolor=darkgreen,
urlcolor=darkblue,
citecolor=darkblue }
\newcommand{\stuff}{\mathcal{H}_0^2\Omega_{\rm m0}}

\newcommand{\UD}[2]{\ensuremath{^{#1}_{\phantom{#1} #2}}}

\newcommand{\UDD}[3]{\ensuremath{^{#1}_{\phantom{#1} #2 #3}}}
\newcommand{\UDDD}[4]{\ensuremath{^{#1}_{\phantom{#1} #2 #3 #4}}}

\newcommand{\calO}{\ensuremath{\mathcal{O}}}

\newcommand{\WXL}{\ensuremath{W_{XL}}}
\newcommand{\WXX}{\ensuremath{W_{XX}}}
\newcommand{\WLL}{\ensuremath{W_{LL}}}
\newcommand{\WLX}{\ensuremath{W_{LX}}}

\newcommand{\bwt}{\begin{widetext}}
\newcommand{\ewt}{\end{widetext}}

\newcommand{\beq}{\begin{equation}}
\newcommand{\eeq}{\end{equation}}
\newcommand{\bea}{\begin{eqnarray}}
\newcommand{\eea}{\end{eqnarray}}
\newcommand{\bit}{\begin{itemize}}
\newcommand{\eit}{\end{itemize}}
\newcommand{\bfi}{\begin{figure}}
\newcommand{\efi}{\end{figure}}
\newcommand{\bfic}{\begin{figure*}}
\newcommand{\efic}{\end{figure*}}
\newcommand{\bce}{\begin{center}}
\newcommand{\ece}{\end{center}}
\newcommand{\bt}{\begin{table}}
\newcommand{\et}{\end{table}}
\newcommand{\btb}{\begin{tabular}}
\newcommand{\etb}{\end{tabular}}

\newcommand{\calW}{\ensuremath{\mathcal{W}}}

\newcommand{\qed}{\nobreak \ifvmode \relax \else
      \ifdim\lastskip<1.5em \hskip-\lastskip
      \hskip1.5em plus0em minus0.5em \fi \nobreak
      \vrule height0.75em width0.5em depth0.25em\fi}

\begin{document}

\title{Isolating non-linearities of light propagation in inhomogeneous cosmologies}

\author{Michele Grasso}
\email{grasso@cft.edu.pl} 
\affiliation{Center for Theoretical Physics, Polish Academy of Science, \\ Al. Lotnik\' ow 32/46, 02-668 Warszawa, Poland.}

\author{Eleonora Villa}
\email{villa@cft.edu.pl}
\affiliation{Center for Theoretical Physics, Polish Academy of Science, \\ Al. Lotnik\' ow 32/46, 02-668 Warszawa, Poland.}

\author{Miko\l{}aj Korzy\' nski}
\email{korzynski@cft.edu.pl}
\affiliation{Center for Theoretical Physics, Polish Academy of Science, \\ Al. Lotnik\' ow 32/46, 02-668 Warszawa, Poland.}

\author{Sabino Matarrese} 
\email{sabino.matarrese@pd.infn.it}
\affiliation{Dipartimento di Fisica e Astronomia G. Galilei, Universit\`a degli Studi di Padova, I-35131 Padova, Italy.}
\affiliation{INFN, Sezione di Padova, via F. Marzolo 8, I-35131 Padova, Italy.}
\affiliation{INAF - Osservatorio Astronomico di Padova, vicolo dell'Osservatorio 5, I-35122 Padova, Italy.}
\affiliation{Gran Sasso Science Institute, viale F. Crispi 7, I-67100 L'Aquila, Italy.}

\date{\today}

\begin{abstract} 
A new formulation for light propagation in geometric optics by means of the Bi-local Geodesic Operators is considered. We develop the {\tt BiGONLight Mathematica} package, uniquely designed to apply this framework to compute optical observables in Numerical Relativity. Our package can be used for light propagation on a wide range of scales and redshifts and accepts numerical as well as analytical input for the spacetime metric.
In this paper we focus on two cosmological observables, the redshift and the angular diameter distance, specializing our analysis to a wall universe modeled within the post-Newtonian approximation. With this choice and the input metric in analytical form, we are able to estimate non-linearities of light propagation by comparing and isolating the contributions coming from Newtonian and post-Newtonian approximations as opposed to linear perturbation theory. We also clarify the role of the dominant post-Newtonian contribution represented by the linear initial seed which, strictly speaking, is absent in the Newtonian treatment. We found that post-Newtonian non-linear corrections are below $1\%$, in agreement with previous results in the literature.
\end{abstract}


\maketitle
%\flushbottom

%%%%%%%%%%%%%%%%%%%%%%%%%%%%%%%%%%%%%%%%%%%%%%%%%%%%%%%%%%%%%%%%%%%%%%%
%%%%%%%%%%%%%%%%%%%%%%%%%%%%%%%%%%%%%%%%%%%%%%%%%%%%%%%%%%%%%%%%%%%%%%%
\section{Introduction}
\label{sec:intro}
%%%%%%%%%%%%%%%%%%%%%%%%%%%%%%%%%%%%%%%%%%%%%%%%%%%%%%%%%%%%%%%%%%%%%%%
%%%%%%%%%%%%%%%%%%%%%%%%%%%%%%%%%%%%%%%%%%%%%%%%%%%%%%%%%%%%%%%%%%%%%%%
Upcoming galaxy surveys like Euclid, LSST, SKA and others\footnote{{\color{blue}{\tt {http://sci.esa.int/euclid/}}}, {\color{blue}{\tt {https://www.lsst.org}}}, {\color{blue}{\tt {http://skatelescope.org/}}}} mark the beginning of a new exciting era, dubbed \textit{precision cosmology}. The reason behind this name is twofold: on one side these future observations will map almost all the visible universe with the unprecedented precision of $1\%$ and on the other side cosmological modelling aim at the same precision target. 

In this view treating non-linearities, i.e. going beyond (linear) cosmological perturbation theory is of crucial importance and new approximation schemes were developed specifically or approximations used in other contexts were applied to cosmology. They include: the post-Newtonian (PN) approximation (see \cite{mater, carbone2005unified} for formulations of PN cosmology in two different gauges), the post-Friedmann approximation (see \cite{Milillo:2015cva, Rampf:2016wom} for a different approach, which adapts to cosmology the weak-field post-Minkowskian approximation and reproduces linear-order cosmological perturbation theory at their zeroth-order), the weak-field approximation\footnote{The leading order of the last two approximation schemes were shown to be equivalent for a dust universe in the Poisson gauge in \cite{kopp2014newton}, whereas \cite{mater, carbone2005unified} were constructed on purpose to include second-order perturbation theory at their PN order.} (see \cite{green2011new} for the development of the framework and \cite{Adamek:2013wja} for estimations with the use of Newtonian simulations for a plane-symmetric universe), and, more recently, a two-parameters gauge-invariant approximation (see \cite{Goldberg:2016lcq}). In addition, over the past few decades, numerical simulations have increasingly become a powerful tool in cosmology to model the growth of non-linear structures. Since Newtonian dynamics seems to be a good approximation to describe late-time structure formation, the first generation of cosmological simulations adopted Newtonian gravity to simulate cosmological dynamics. Then, Newtonian simulations were used to feed approximate field equations coming from General Relativity (GR) as e.g. in \cite{Bruni:2013mua, Adamek:2014xba, Fidler:2017pnb}. Only recently we assist to a revolution in cosmological simulations with the birth of codes aiming at simulating fully general relativistic dynamics, \cite{loffler2012einstein, mertens2016integration, adamek2016general, macpherson2017, barrera2020gramses}: for the state of the art and the comparison among different codes, see \cite{Adamek:2020jmr}.

However, a sophisticated general relativistic (exact or approximated) description of cosmological dynamics is not the end of the story. The key point is how (much) it affects light propagation, the final aim being to characterize and (hopefully) measure non-linear GR effects in the observables on cosmological scales or, at least, quantify their bias in observations. These studies are still in their infancy but they are addressed with several approaches most of which we briefly sketched above. A non-comprehensive list includes \cite{Thomas:2014aga, Barreira:2016wqo, Borzyszkowski:2017ayl, Sanghai:2017yyn, Giblin:2017ezj, Adamek:2018rru, Gressel:2019jxw, Lepori:2020ifz}.
Despite being too early to draw definitive conclusions, it seems that the codes that approximate GR dynamics are in agreement with Newtonian simulations for what concerns weak-lensing observables \cite{Thomas:2014aga, Lepori:2020ifz} but a modification in the statistics of the luminosity distance \cite{Adamek:2018rru} was found. In addition, the PN approximation for some models gives predictions different from $\Lambda$CDM \cite{Sanghai:2017yyn}. A bit of work is still needed to adapt to (observational) cosmology the truly GR numerical codes.

In this paper we examine the differences between linear and non-linear light propagation. An accurate treatment of the problem would require to analyse light propagation in a realistic model of the universe. However, our aim is not to make general predictions, but rather to deeply investigate the various factors and effects on observables coming from non-linearities. For this purpose, we decided to employ a toy-model of the universe in which light rays pass through a series of plane-symmetric perturbations around a Friedmann-Lema\^{i}tre-Robertson-Walker (FLRW) background. This model is known as plane-parallel or wall universe, and it was used in the past to study the back-reaction from the small-scale inhomogeneities \cite{Villa:2011vt, diDio2012back, adamek2014distance, Clifton:2019cep}.
We start by extending the results of \cite{Villa:2011vt} by providing the so-called Zel'dovich solution with a $\Lambda$CDM background. In this model, we compute the redshift and the angular diameter distance within three different approximation schemes: linear, Newtonian and post-Newtonian. In order to quantify and isolate non-linear contributions, we present our results in terms of the relative differences between observables computed with these three different approximations (see Sec. \ref{sec:method} for details). We also analyse different aspects of non-linearities, e.g. scale-dependence, non-Gaussianity, etc. Even if our modelling is very simple, we believe that this kind of analysis is representative of more general configurations. 

Besides, an important novelty of this work is that we make use of the new {\tt BiGONLight Mathematica} package to study light propagation in GR and compute observables numerically, \cite{Grasso:BGO}. Contrary to other software, {\tt BiGONLight} implements light propagation within the new Bi-local Geodesic Operator (BGO) framework, which is applicable to more general situations than the standard formalism and it is also suitable to construct new observables, \cite{Grasso:2018mei, Korzynski:2019oal}. This unique design makes the package adaptable to study various light propagation problems in numerical simulations.

We begin by presenting in Sec. \ref{sec:plane//} the plane-parallel toy-model as introduced in \cite{Villa:2011vt}. Then, in Sec. \ref{sec:lightprop} we briefly describe the BGO framework, pointing out to \cite{Grasso:2018mei} and \cite{Korzynski:2019oal} for further details. In Sec. \ref{sec:method}, we introduce the goals of our analysis and the method which led to the results presented in Sec. \ref{sec:results}. Finally, we address our conclusions in Sec. \ref{sec:concl}.

Notation: Greek indices ($\alpha, \beta, ...$) run from 0 to 3, while Latin indices ($i,j, ...$) run from 1 to 3 and refer to spatial coordinates only. Latin indices ($A,B, ...$) run from 1 to 2. Tensors and bitensors expressed in a semi-null frame are denoted using boldface indices: Greek boldface indices ($\boldsymbol{\alpha}, \boldsymbol{\beta}, ...$) run from 0 to 3, Latin boldface indices ($\mathbf{a}, \mathbf{b}, ...$) run from 1 to 3 and capital Latin boldface indices ($\mathbf{A}, \mathbf{B}, ...$) run from 1 to 2. A dot denotes derivative with respect to conformal time.
Quantities with a subscript 0 are meant to be evaluated at present, whereas the subscript $``{\rm in}"$ indicates the initial time. 
Similarly, we indicate with a subscript $\mathcal{O}$ ($\mathcal{S}$) quantities defined at the observer (source). An overbar indicates quantities evaluated in the $\Lambda$CDM model.
In this paper we use three different approximations and consequently three different notations: ``N'' for Newtonian, ``PN'' for post-Newtonian, ``Lin'' for first-order perturbation theory. We place these abbreviations up or down depending on convenience.

%%%%%%%%%%%%%%%%%%%%%%%%%%%%%%%%%%%%%%%%%%%%%%%%%%%%%%%%%%%%%%%%%%%%%%%
%%%%%%%%%%%%%%%%%%%%%%%%%%%%%%%%%%%%%%%%%%%%%%%%%%%%%%%%%%%%%%%%%%%%%%%
\section{The plane-parallel dynamics in three approximations}
\label{sec:plane//}
%%%%%%%%%%%%%%%%%%%%%%%%%%%%%%%%%%%%%%%%%%%%%%%%%%%%%%%%%%%%%%%%%%%%%%%
%%%%%%%%%%%%%%%%%%%%%%%%%%%%%%%%%%%%%%%%%%%%%%%%%%%%%%%%%%%%%%%%%%%%%%%
We consider a toy-model characterized by the choice of globally plane-parallel configurations, i.e. the case where the initial perturbation field depends on a single coordinate. The dynamics of this very simple universe consists of a collection of parallel planes that collapse along the direction of their normal to form a pancake. For the purposes of our work, we are not interested in a more realistic modelling of the Universe; rather our main aim is to estimate, isolate and compare purely non-linear and non-Newtonian contributions in light propagation, e.g. in fundamental observables such as  redshift and angular diameter distance. 

We work in the synchronous-comoving gauge and leave to future work the gauge issue of every perturbation scheme that affects the observables as well as the estimate of the related contributions in other gauges. Despite gauge effects in the observables are known in standard cosmological perturbation theory (see Ref.~\cite{Yoo:2017svj} for a recent discussion of gauge invariance of cosmological observables up to second order), the issue is more delicate for non-standard approximations, such as those considered in this paper.

The starting point of our analysis is the results of Ref.~\cite{Villa:2011vt}: the authors started from the Newtonian background given by the well-known Zel'dovich approximation, \cite{zeldovich70a}, which, for plane-parallel perturbations in the Newtonian limit, represents an exact solution. They then obtained the exact analytical form for the PN metric, thereby providing the exact PN extension of the Zel'dovich solution. Let us remark how the Zel'dovich approximation is constructed: in its conformal version, it is an expansion around the three-dimensional spatial displacement vector of the CDM particles between the position comoving with the Hubble flow and the true position governed by perturbations. The peculiarity is the following: the solution for the displacement vector is strictly linear, as it is found from the linear Newtonian equations of motion. But all other dynamical quantities, such as the mass density, are written in terms of such a displacement vector, as if it was exact, i.e. from their non-perturbative definition. The same construction was first extended to the PN approximation of General Relativity, where the metric tensor also is a dynamical variable, in Ref.~\cite{mater} and specialized in the plane-parallel case in Ref.~\cite{Villa:2011vt}.
The Zel'dovich specific feature is evident in the form of the metric tensor \eqref{eq:metricNWT}, which is quadratic in the perturbations, and the density contrast in Eq.~\eqref{eq:density_N} for the Newtonian background and in Eq.~\eqref{eq:metricPN} and Eq.~\eqref{eq:density_PN} for the PN solution found in Ref.~\cite{Villa:2011vt}.

We provide here the $\Lambda$CDM extension of the PN metric found in Ref.~\cite{Villa:2011vt}, which was obtained for the Einstein-de Sitter background model, i.e. the dust-only universe. 
Starting from the line element
\begin{widetext}
\beq
ds^2=a^2(\eta)\left\{ -c^2 d\eta^2 + \gamma^{\rm{PN}}_{11}(\eta, q_{\rm 1}) dq_{\rm 1}^2 +\gamma^{\rm{PN}}_{22}(\eta, q_{\rm 1}) dq_{\rm 2}^2+\gamma^{\rm{PN}}_{33}(\eta, q_{\rm 1}) dq_{\rm 3}^2\right\}
\eeq
\end{widetext}
we then obtain the conformal metric given by\footnote{We take this chance to point out a typo in Eq.~(4.37) of Ref.~\cite{Villa:2011vt}: in the first term of the second line of the expression for $\gamma_{11}$ the correct coefficient is $5/756$ instead of $5/576$.}
\bwt
\begin{equation} \label{eq:metricPN}
\begin{split}
\gamma^{\rm{PN}}_{11} =& \left(1-\frac{2}{3}\frac{\partial_{q_1}^2 \phi_0}{ \stuff} \mathcal{D} \right)^2 +\\
 & +\frac{1}{c^2}\left[-\frac{10}{3} \phi_0 +(4 a_{\rm nl}-5)\frac{ 10}{9}\frac{(\partial_{\rm q_{\rm 1}} \phi_0)^2}{ \stuff}\mathcal{D}  +(a_{\rm nl}-1)\frac{40}{9}\frac{ \phi_0  \partial_{\rm q_{\rm 1}}^2 \phi_0}{ \stuff} \mathcal{D}  + \right.\\
& \left.  \left(\frac{41}{7}-4 a_{\rm nl}\right)\frac{20}{27} \frac{(\partial_{\rm q_{\rm 1}} \phi_0)^2  \partial_{\rm q_{\rm 1}}^2 \phi_0}{(\stuff)^2}\mathcal{D}^2 + \left(3-2 a_{\rm nl}\right) \frac{40}{27}\frac{\phi_0  (\partial_{\rm q_{\rm 1}}^2 \phi_0)^2}{ (\stuff)^2} \mathcal{D}^2 -\frac{80}{189}\frac{(\partial_{\rm q_{\rm 1}} \phi_0)^2 ( \partial_{\rm q_{\rm 1}}^2\phi_0)^2}{(\stuff)^3}\mathcal{D}^3 \right]\\
\gamma^{\rm{PN}}_{22}= & 1+\frac{1}{c^2}\left[ \frac{10}{9}\left( \frac{\mathcal{D} (\partial_{\rm q_{\rm 1}} \phi_0)^2 }{\stuff}-3\phi_0\right)\right]\\
\gamma^{\rm{PN}}_{33}=& 1+\frac{1}{c^2}\left[ \frac{10}{9}\left( \frac{\mathcal{D} (\partial_{\rm q_{\rm 1}} \phi_0)^2 }{\stuff}-3\phi_0\right)\right].
 \end{split}
\end{equation}
\ewt
In the above expression $\eta$ is the conformal time, $a$ is the scale-factor encoding the evolution of the $\Lambda$CDM  background, $\mathcal{H}_0$, $\Omega_{\rm m_0}$, and $\phi_0$ are the (conformal) Hubble parameter $\mathcal{H} \equiv \dot{a}/a$, the matter (ordinary plus dark) densaity parameter, and the peculiar gravitational potential, respectively, all evaluated at present. The dot denotes differentiation with respect to conformal time. $\mathcal{D}$ is the growing mode solution of the first-order equation for the density contrast which is defined as 
\begin{equation}
\delta(\eta, q_{\rm 1})\equiv \frac{\rho(\eta, q_{\rm 1})}{\bar{\rho}(\eta)}-1,
\end{equation}
where $\bar{\rho}$ the $\Lambda$CDM  background matter density.
At first order in standard perturbation theory and without loss of generality, the space and time dependence of the expression of the growing density contrast can be factored out. In our 
one-dimensional case we have $\delta^{\rm Lin}(\eta, q_{\rm 1})=\mathcal{D}(\eta)\delta^{\rm Lin}_0(q_{\rm 1})$, where we fix the constant $\delta_0$ at the present time, and the growing mode $\mathcal{D}$ obeys the well-known equation
\begin{equation} \label{eqforD+}
\ddot{\cal D} +\mathcal{H}\dot{\cal D}-\frac{3}{2}\mathcal{H}_0^2\Omega_{m_0}\frac{\cal D}{a}=0\,.
\end{equation}
It is worth noticing that these quantities are all connected via the cosmological Poisson equation
\begin{equation}
\mathcal{D} \nabla^2 \phi_{\rm 0} -\frac{3}{2} \stuff  \delta_{\rm Lin} =0 \,.
\label{eq:Poisson_eq}
\end{equation}
Finally, we follow here the parametrization for primordial non-Gaussianity defined in Ref.~\cite{Bartolo:2005kv}: the number $a_{\rm nl}$ parametrizes local primordial non-Gaussianity of the gauge-invariant curvature perturbation of uniform density hypersurfaces. This is linked to the parametrization of the primordial gravitational potential by a simple relation between the respective parameters: $f_{\rm nl}= (5/3) (a_{\rm nl} -1)$.

The metric in~\eqref{eq:metricPN} corresponds to the most sophisticated approximation that we will use in this paper: although being referred to the 1D toy-model, it is fully non-linear in the standard perturbative sense, i.e. it is not assumed that density perturbations are small. On the contrary, taking advantage of the Zel'dovich prescription, we calculate the density contrast non-perturbatively, see Eq.~\eqref{eq:density_PN} below.
The PN approximation extends standard perturbation theory including the leading-order corrections to the Newtonian treatment, which are the terms proportional to $1/c^2$.   
We will compare light propagation in the spacetime described by~\eqref{eq:metricPN} with other two cases, that are both extended in ~\eqref{eq:metricPN}: the linear order of standard cosmological perturbation theory and the Newtonian approximation. The linear spacetime metric in the synchronous-comoving gauge is very well known and in 1D it reads
\begin{equation} \label{eq:metricIPT}
\begin{split}
\gamma^{\rm{Lin}}_{11} =& 1-\frac{4}{3}\frac{ \mathcal{D} \partial_{\rm q_{\rm 1}}^2 \phi_0}{ \stuff} -\frac{10}{3\,c^2}\phi_0\\
\gamma^{\rm{Lin}}_{22}= & 1-\frac{10}{3\,c^2}\phi_0\\
\gamma^{\rm{Lin}}_{33}=& 1-\frac{10}{3\,c^2}\phi_0.
 \end{split}
\end{equation}
This metric is the solution of the Einstein's equations expanded at first order around the FLRW background. Note however that the planar symmetry reduces the degrees of freedom to be only scalar (there are no vector or tensor mode in 1D, by construction) and confines the dynamical part in $\gamma^{\rm{Lin}}_{11}$ only, i.e. only in the direction of the perturbations, while in the other two directions we have just the (PN) initial conditions.
On the other hand, in the Newtonian approximation we have
\begin{equation} \label{eq:metricNWT} 
\begin{split}
\gamma^{\rm{N}}_{11} =& \left(1-\frac{2}{3}\frac{ \mathcal{D} \partial_{\rm q_{\rm 1}}^2 \phi_0}{ \stuff} \right)^2 \\
\gamma^{\rm{N}}_{22}= & 1\\
\gamma^{\rm{N}}_{33}=& 1.
 \end{split}
\end{equation}
This metric can be read off~\eqref{eq:metricPN} by discarding the PN corrections proportional to $1/c^2$. 

\begin{figure*}[ht]
    \centering
    \begin{subfigure}{0.49\linewidth}%{0.8\columnwidth}%0.40\textwidth
        \includegraphics[width=\linewidth]{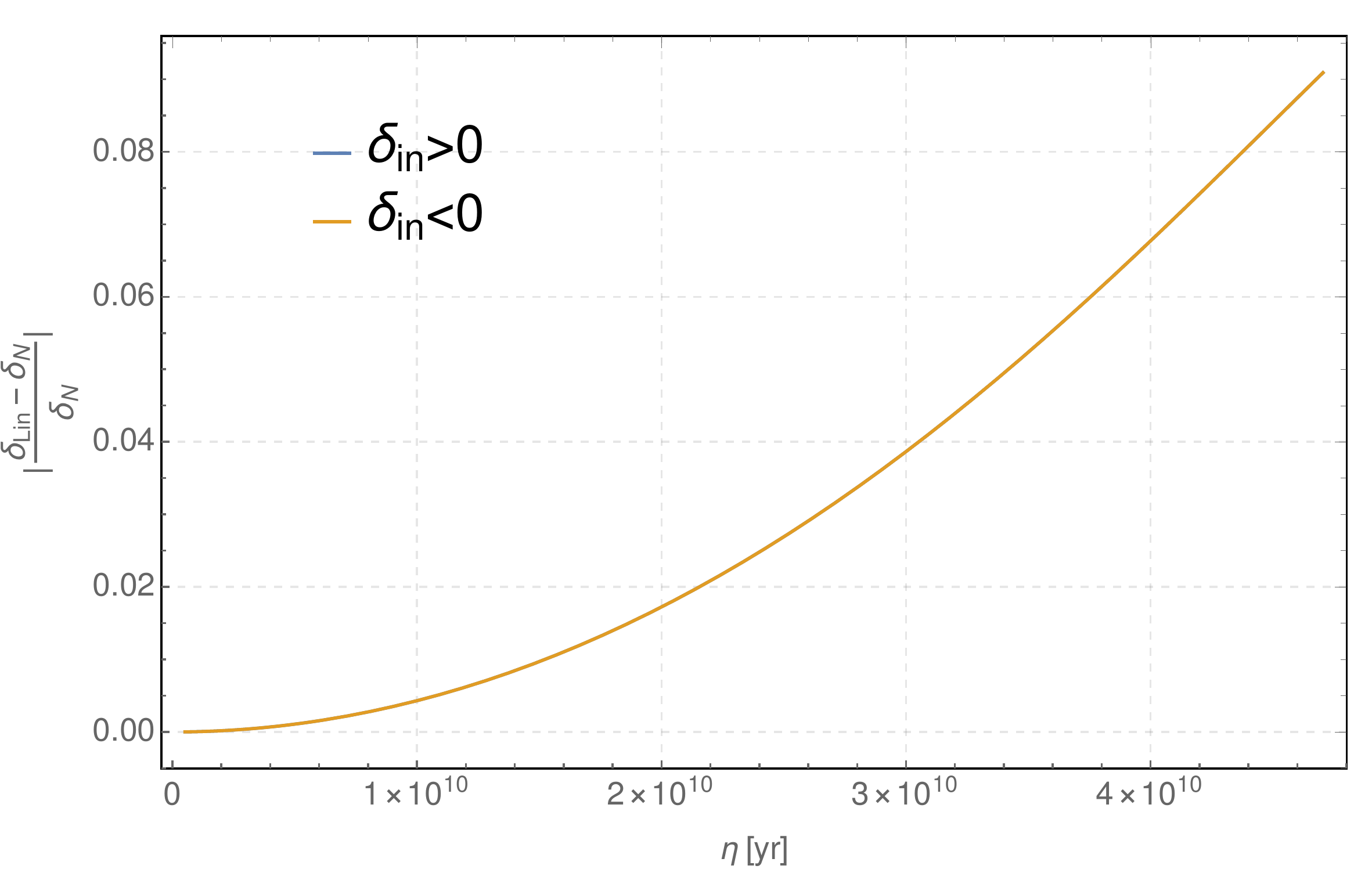}
       \caption{$\left|\frac{\delta_{\rm Lin}-\delta_{\rm N}}{\delta_{\rm N}}\right|$}
        \label{fig:delta_Lin_N}
    \end{subfigure}
    \begin{subfigure}{0.49\linewidth}%{0.8\columnwidth}
        \includegraphics[width=\linewidth]{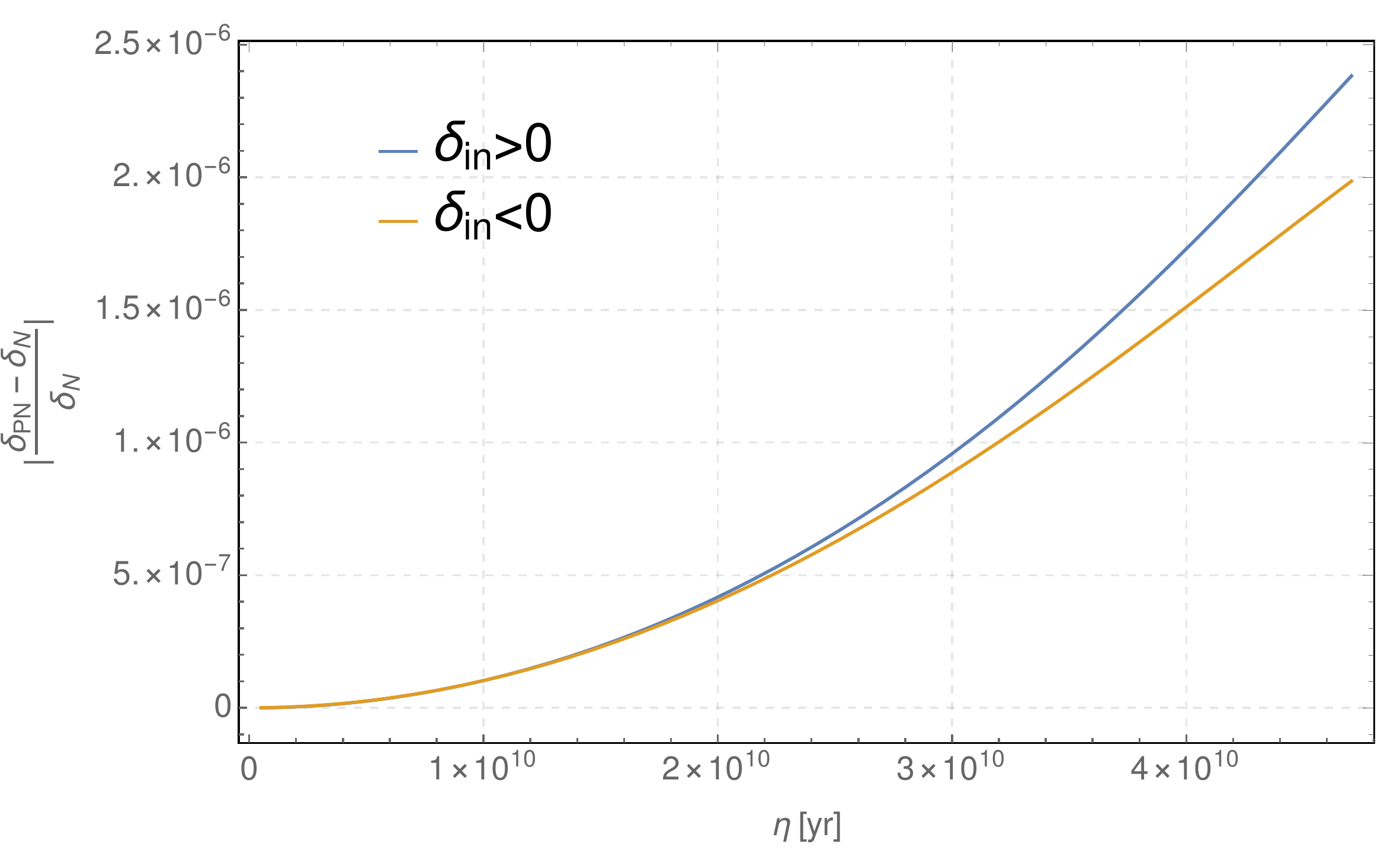}
        \caption{$\left|\frac{\delta_{\rm PN}-\delta_{\rm N}}{\delta_{\rm N}}\right|$ }
        \label{fig:delta_PN_N}
    \end{subfigure}
    \\
    \begin{subfigure}{0.5\linewidth}%{0.8\columnwidth}
        \includegraphics[width=\linewidth]{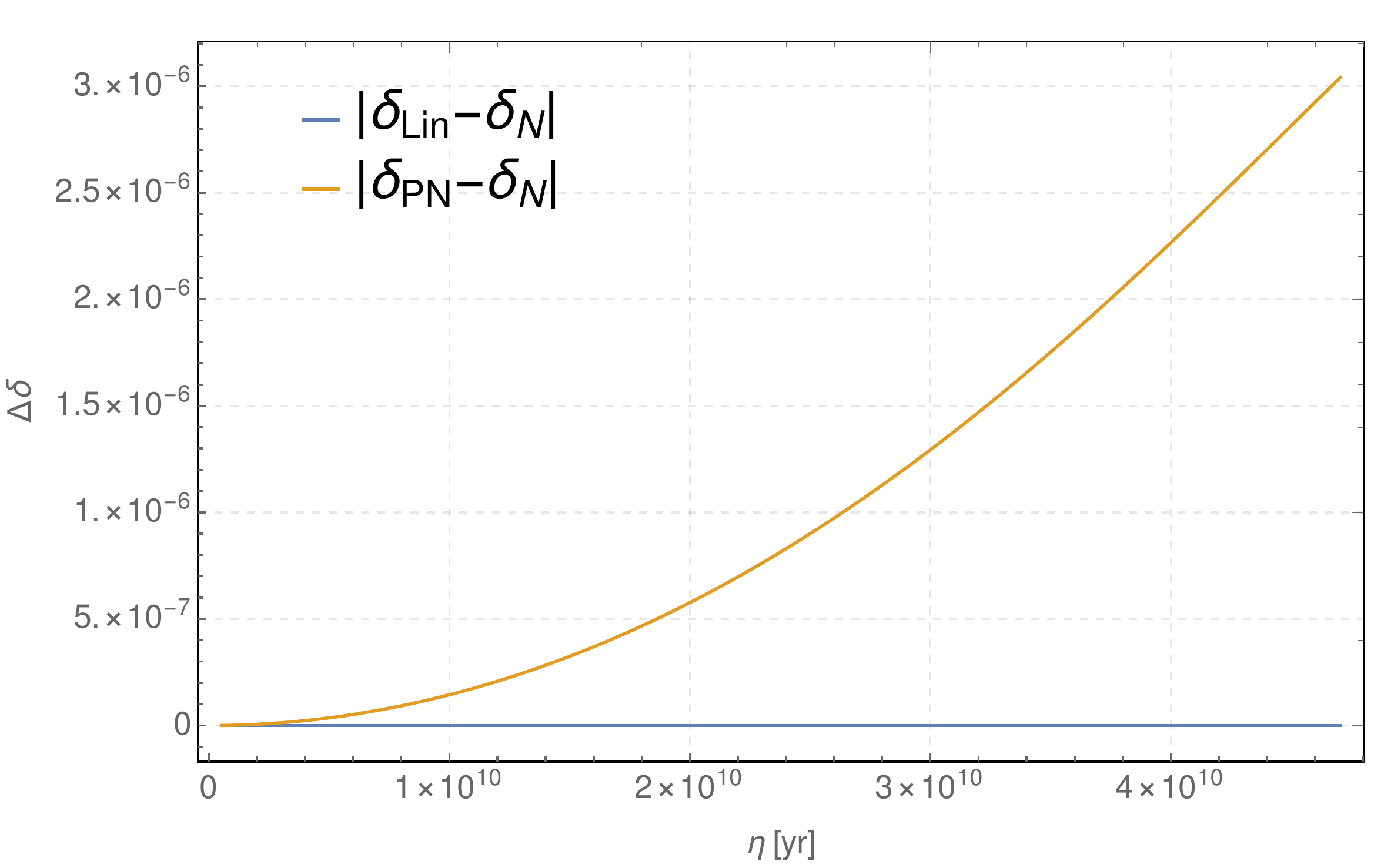}
        \caption{ Difference $\left|\delta_{\rm PN}-\delta_{\rm N}\right|$ and $\left|\delta_{\rm PN}-\delta_{\rm N}\right|$ for regions with $\delta_{\rm in} = 0$ }
        \label{fig:delta_homo}
    \end{subfigure}
    \caption{Comparison between $\delta_{\rm Lin}$, $\delta_{\rm N}$ and $\delta_{\rm PN}$ as functions of conformal time $\eta$. The comparisons  \ref{fig:delta_Lin_N} and \ref{fig:delta_PN_N} are evaluated for an initial over-density $\delta_{\rm in}>0$ and under-density $\delta_{\rm in}<0$, while \ref{fig:delta_homo} is plotted for $\delta_{\rm in}=0$. All the three approximations are obtained for $\phi_{\rm 0} =\mathcal{I} \sin(\omega q_{\rm 1})$ with $\omega=\frac{2 \pi}{500 \, \rm Mpc} $,  amplitude $\mathcal{I}$ such that ${\rm max} \Big(\delta_{\rm PN}(\eta_{\rm 0}, q_{\rm 1})\Big)=0.1$. and cosmological parameters taken from~\cite{planck2018param, planck2019anl}.}\label{fig:density_variations}
\end{figure*}

For completeness we report here the expressions of the density contrast in the three cases:
\begin{equation}
\delta_{\rm Lin}=\frac{2}{3}\frac{\mathcal{D} \partial^2_{\rm q_{\rm 1}} \phi_0}{\stuff}
\label{eq:density_lin}
\end{equation}
\begin{equation}
\delta_{\rm N}=\frac{\frac{2}{3}\frac{\mathcal{D} \partial^2_{\rm q_{\rm 1}} \phi_0}{\stuff}}{1-\frac{2}{3}\frac{\mathcal{D} \partial^2_{\rm q_{\rm 1}} \phi_0}{\stuff}} 
\label{eq:density_N}
\end{equation}
\bwt
\begin{equation}
\begin{array}{l}
\delta_{\rm PN}= \frac{\frac{2}{3}\frac{\mathcal{D} \partial_{\rm q_{\rm 1}}^2 \phi_0}{\mathcal{H}_0^2 \Omega_{\rm m0}}}{\left(1-\frac{2}{3}\frac{\mathcal{D} \partial_{\rm q_{\rm 1}}^2 \phi_0}{\mathcal{H}_0^2 \Omega_{\rm m0}} \right)}+\frac{1}{c^2}\frac{1}{\left(1-\frac{2}{3}\frac{\mathcal{D} \partial_{\rm q_{\rm 1}}^2 \phi_0}{\mathcal{H}_0^2 \Omega_{\rm m0}} \right)^2}\left[\frac{5}{9}(3-4 a_{\rm nl})\frac{\mathcal{D}}{\mathcal{H}_0^2 \Omega_{\rm m0}}(\partial_{\rm {q_{\rm 1}}} \phi_{\rm 0})^2+\frac{20}{9}(2-a_{\rm nl})\frac{\mathcal{D}}{\mathcal{H}_0^2 \Omega_{\rm m0}}(\phi_{\rm 0} \partial^2_{\rm {q_{\rm 1}}} \phi_{\rm 0})\right. \\
 \left. +\frac{20}{21}\left(\frac{2}{3}\frac{\mathcal{D}}{\mathcal{H}_0^2 \Omega_{\rm m0}}\right)^2 \partial_{\rm q_{\rm 1}}^2 \phi_0 (\partial_{\rm q_{\rm 1}} \phi_0)^2 \right]
\end{array}
\label{eq:density_PN}
\end{equation}
\ewt
Note that the Newtonian density contrast, according to the Zel'dovich approximation, is calculated exactly from the continuity equation in the synchronous-comoving gauge (see Eq.~\eqref{eq:delta_ex}) as the PN one, which is just expanded in powers of $1/c^2$.
\begin{figure}[h]
\includegraphics[width=\columnwidth]{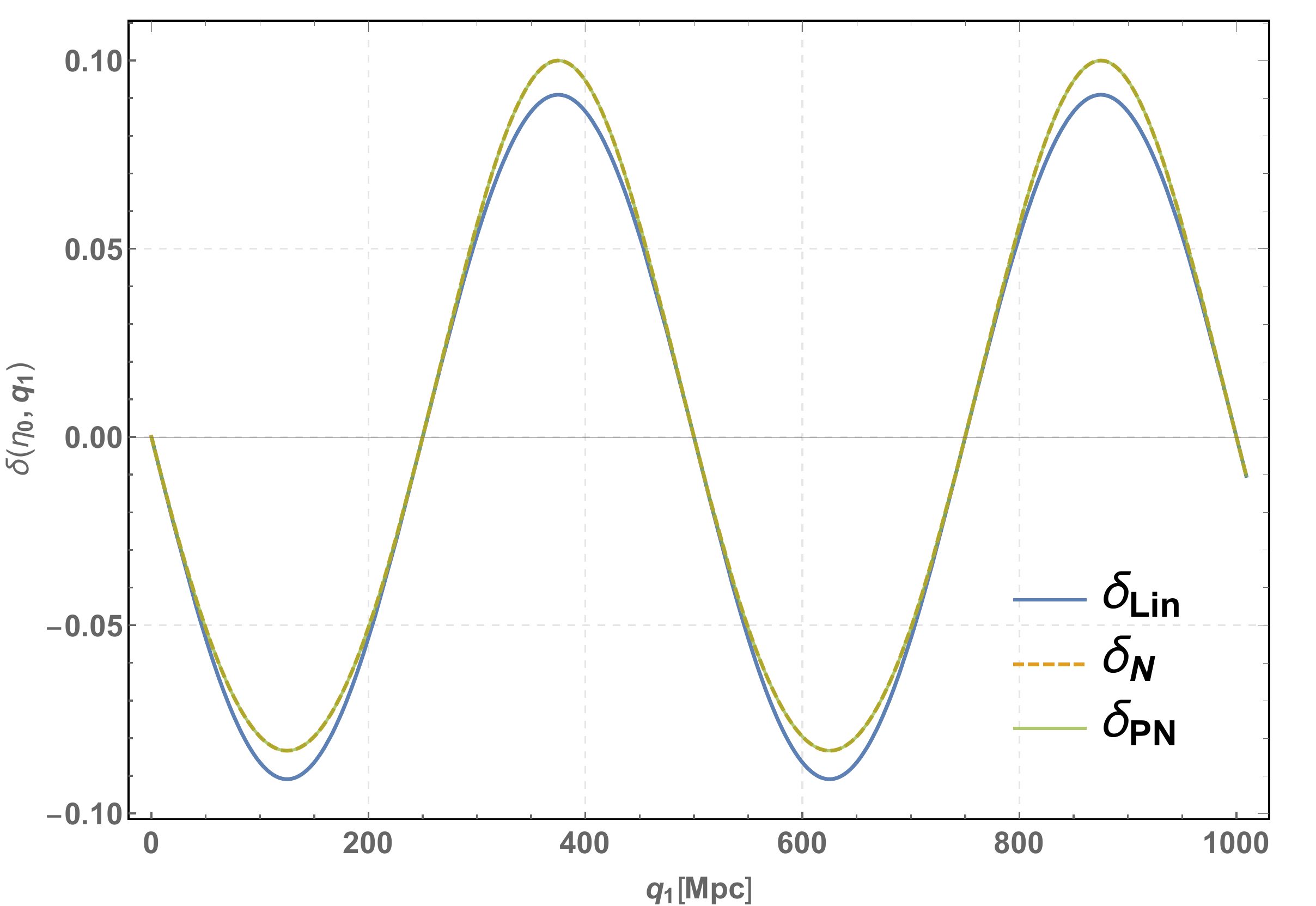}
\caption{ 
Density contrast at present (conformal) time $\eta_0$ in the three approximations $\delta_{Lin}$, $\delta_{N}$ and $\delta_{PN}$, as in Eqs.~\eqref{eq:density_lin},~\eqref{eq:density_N} and~\eqref{eq:density_PN} respectively. The plots are obtained setting up the potential as $\phi_{\rm 0} =\mathcal{I} \sin(\omega q_{\rm 1})$ with $\omega=\frac{2 \pi}{500 \, \rm Mpc} $ and amplitude $\mathcal{I}$ such that ${\rm max} \Big(\delta_{\rm PN}(\eta_{\rm 0}, q_{\rm 1})\Big)=0.1$. The values for $\Omega_{\rm m0}$, $\Omega_{\rm \Lambda}$, $f_{\rm nl}$ and $\mathcal{H}_{\rm 0}$ are taken from~\cite{planck2018param, planck2019anl}.}
        \label{fig:delta_profile}
\end{figure}
We take our initial conditions at $\eta_{\rm in}$, after the end of inflation and in the matter-dominated era, when linear theory around the Einstein-de Sitter model is still a good approximation. The explicit expression for the initial density contrast is thus
\begin{equation}
\delta_{\rm in}=\frac{2}{3}\frac{\mathcal{D}_{\rm in} \partial^2_{\rm q_{\rm 1}} \phi_0}{\stuff}\,
\label{eq:density_in}
\end{equation}
where $\mathcal{D}_{\rm in}\propto \eta^2_{\rm in}$ is the linear growing mode of the Einstein-de Sitter model. 
We model the profile of the gravitational potential at present as $\phi_{\rm 0} = \mathcal{I} \sin(\omega q_{\rm 1})$, with frequency $\omega=\frac{2 \pi}{500 \, \rm{Mpc}}$ and amplitude $\mathcal{I}$ such that ${\rm max} \Big(\delta_{\rm PN}(\eta_{\rm 0}, q_{\rm 1})\Big)=0.1$. We set the fiducial values of the cosmological parameters from~\cite{planck2018param,planck2019anl} with $\Omega_{\rm m0} =0.3153$, $\Omega_{\rm \Lambda}=0.6847$, $\mathcal{H}_{\rm 0}=67.36$ and $a_{\rm nl}=\frac{3}{5} f_{\rm nl}+1 = 0.46$. 
The three profiles~\eqref{eq:density_lin}, ~\eqref{eq:density_N} and ~\eqref{eq:density_PN} of the density contrast at the present time are shown in Fig.~\ref{fig:delta_profile}. The plot shows the amplitudes of the curves of the N and PN density contrast are shifted to higher values compared to the linear one, namely the N and PN corrections have the effect of increasing both the under- and the over-density peaks by the same amount $\approx4.15\times 10^{-3}$.
The differences in the evolution of the density contrast in the three approximations are more evident from Fig.~\ref{fig:density_variations}, in which we analyse the growth of an initial under-density, $\delta_{\rm in}<0$, over-density, $\delta_{\rm in}>0$, and the case of vanishing $\delta_{\rm in}$.
In Figs.~\ref{fig:delta_Lin_N} and  \ref{fig:delta_PN_N} we show the deviations $\left|\frac{\delta_{\rm Lin}-\delta_{\rm N}}{\delta_{\rm N}}\right|$ and $\left|\frac{\delta_{\rm PN}-\delta_{\rm N}}{\delta_{\rm N}}\right|$ respectively, at fixed position as a function of time for the two cases of initial over- and under-densities. 
In Fig.~\ref{fig:delta_Lin_N} we clearly see that the variation Lin vs N grows with time,
reaching $\approx 9\%$ at present, which is exactly the shift of $4.15\times 10^{-3}$ that we see in Fig.~\ref{fig:delta_profile}. The variation PN vs N grows with time as well, Fig.~\ref{fig:delta_PN_N}, but the value at present is 4 orders of magnitude less. Furthermore, in this figure we can also appreciate how the over-densities accrete faster than the under-densities, as one should expect. This is not visible in Fig.~\ref{fig:delta_Lin_N}, due to the fact that the difference between the linear and the Newtonian approximation is dominant.
%%%%%%%%%%%%%%%%%%%%%%%%%%%%%%%%%%%%%%%%%%%%%%%%%%%%%%%%%%%%%%%%%%%%%%%%%%%%%%%%%%%%%%%%%%%%%%%%%%%%%%%
%%%%%%%%%%%%%%%%%%%%%%%%%%%%%%%%%%%%%%%%%%%%%%%%%%%%%%%%%%%%%%%%%%%%%%%%%%%%%%%%%%%%%%%%%%%%%%%%%%%%%%%
The case of $\delta_{\rm in}=0$ is presented in Fig.~\ref{fig:delta_homo}, where we plot the differences $\left|\delta_{\rm Lin}-\delta_{\rm N}\right|$ and $\left|\delta_{\rm PN}-\delta_{\rm N}\right|$. The reason why we took the difference instead of the variation (as was done in the over- and under-density cases) is to avoid the operation of dividing by zero, since we are considering regions with $\delta=0$.
We begin by noticing that for $\delta_{\rm in}=0$ both $\delta_{\rm Lin}$ and $\delta_{\rm N}$ vanish at any time, as evident from~\eqref{eq:density_lin} and~\eqref{eq:density_N}: both $\delta_{\rm Lin}$ and $\delta_{\rm N}$ are proportional to $\partial^2_{\rm q_{\rm 1}}\phi_{\rm 0}$ and they are exactly zero at the same value of $q_{\rm 1}$, in our toy model $\phi_{\rm 0}\propto \sin(\omega q_{\rm 1})$. Conversely, the term $\propto\partial_{\rm q_{\rm 1}}\phi_{\rm 0}$ in $\delta_{\rm PN}$~\eqref{eq:density_PN} implies that $\delta_{\rm PN}(\eta_{\rm in})\neq 0$ at the position where $\delta_{\rm in}=0$. Therefore, what is actually shown in Fig.~\ref{fig:delta_homo} is the evolution of the density contrast in the post-Newtonian approximation, whose absolute value increases up to $\approx 3\ \times 10^{-6}$.

%%%%%%%%%%%%%%%%%%%%%%%%%%%%%%%%%%%%%%%%%%%%%%%%%%%%%%%%%%%%%%%%%%%%%%%
%%%%%%%%%%%%%%%%%%%%%%%%%%%%%%%%%%%%%%%%%%%%%%%%%%%%%%%%%%%%%%%%%%%%%%%
\section{Light propagation in the BGO framework}
\label{sec:lightprop}
%%%%%%%%%%%%%%%%%%%%%%%%%%%%%%%%%%%%%%%%%%%%%%%%%%%%%%%%%%%%%%%%%%%%%%%
%%%%%%%%%%%%%%%%%%%%%%%%%%%%%%%%%%%%%%%%%%%%%%%%%%%%%%%%%%%%%%%%%%%%%%%

\begin{figure}
\includegraphics[width=\columnwidth]{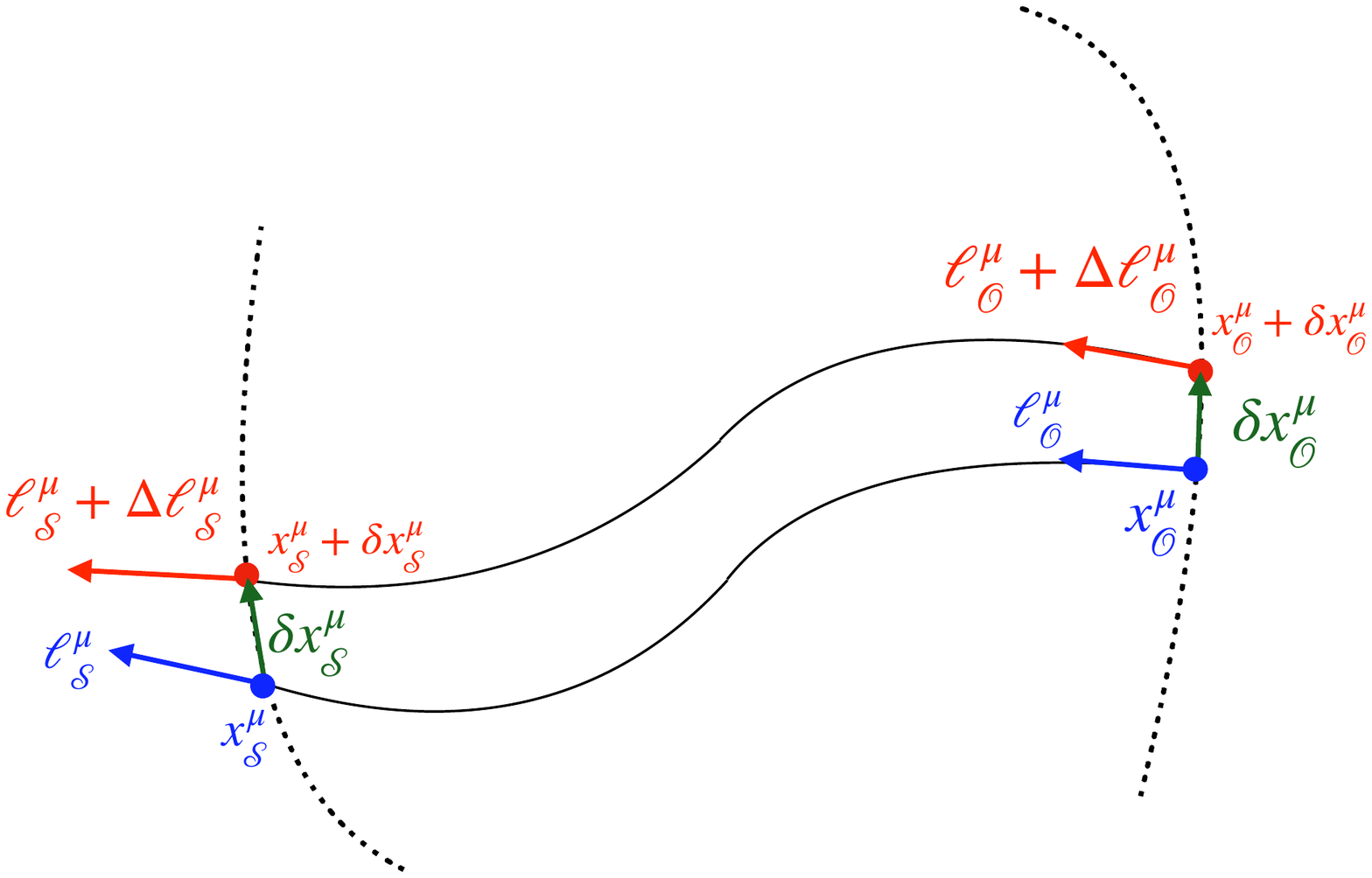}
\caption{Sketchy representation of the geometric set-up. Both the source $\mathcal{S}$ and the observer $\mathcal{O}$ are free to move along their worldlines, with the condition that at every proper time $\tau_\mathcal{O}$ there exists a null geodesic connecting $x^{\mu}_\mathcal{O}$ and $x^{\mu}_\mathcal{S}$.}
\label{fig:phys_syst}
\end{figure}

In this section we present the key elements of the formulation of light propagation which we are going to use in this paper: the bi-local geodesic operator (BGO) framework (for a more extended discussion of this formalism, see \cite{Grasso:2018mei}).
The physical situation we want to study is depicted in Fig. \ref{fig:phys_syst}: an observer $\mathcal{O}$ placed at $x^{\mu}_{\mathcal{O}}$ is connected to the source $\mathcal{S}$ placed at $x^{\mu}_{\mathcal{S}}$ through a null geodesic $\gamma$. Both $\mathcal{S}$ and $\mathcal{O}$ are free to move along their timelike worldlines, but we assume that the typical length scale of the regions in which their motion takes place is small compared to the {distance} between them, so that their local geometry can be treated as flat. Therefore, we can safely assume that all gravitational effects on light propagation are due to the curvature of the spacetime between $\mathcal{S}$ and $\mathcal{O}$.  

In general, geodesics are uniquely specified by giving the initial position and the initial tangent vector, that can be assigned at the observation point. In other words, a fiducial null geodesic $\gamma$ can be identified through its initial conditions ($x^{\mu}_{\mathcal{O}}$, $\ell^{\mu}_{\mathcal{O}}$). Now, if the observer is displaced by $\delta x^{\mu}_{\mathcal{O}}$, a new geodesic connects $\mathcal{S}$ and $\mathcal{O}$ and it is characterized by the new initial conditions ($x^{\mu}_{\mathcal{O}}+\delta x^{\mu}_{\mathcal{O}}$, $\ell^{\mu}_{\mathcal{O}} + \Delta \ell^{\mu}_{\mathcal{O}}$), where we define $\Delta \ell^{\mu}_{\mathcal{O}}$ as the covariant deviation of the tangent vector $\ell^{\mu}_{\mathcal{O}}$ at the observer position, namely
\begin{equation}
\Delta \ell^{\mu}_{\mathcal{O}}=\delta \ell^{\mu}_{\mathcal{O}}+\Gamma\UDD{\mu}{\alpha}{\beta}(x_\mathcal{O})\ell^{\alpha}_{\mathcal{O}} \delta x^{\beta}_{\mathcal{O}}\, .
\label{eq:cov_variation_ell}
\end{equation} 
The deviations ($\delta x^{\mu}_{\mathcal{O}}$, $\Delta \ell^{\mu}_{\mathcal{O}}$) can be used to parametrize a family of null geodesics around the fiducial geodesic $\gamma$, provided that the geodesics of the family stay close enough to $\gamma$, such that it can be studied by keeping all the equations linear in the displacements.

The deviation vector $\delta x^{\mu}$, which is the displacement between $\gamma$ and infinitesimally separated geodesics, propagates accordingly to the geodesic deviation equation (GDE) 
\begin{equation}
\nabla_{\ell} \nabla_{\ell} \delta x^{\mu} - R\UDDD{\mu}{\alpha}{\beta}{\nu}\ell^{\alpha}\ell^{\beta} \delta x^{\nu}=0
\end{equation}
with initial conditions
\begin{equation}
\begin{array}{l}
\delta x^{\mu}(x_{\mathcal{O}})=\delta x^{\mu}_{\mathcal{O}}\\
\nabla_{\ell} \delta x^{\mu}(x_{\mathcal{O}})= \Delta \ell^{\mu}_{\mathcal{O}}
\end{array}.
\end{equation}
Using the linearity of the GDE and considering its projection into the parallel-propagated frame\footnote{The frame $\left( u^{\mu}, \phi\UD{\mu}{\mathbf{A}},\ell^{\mu}\right)$ is called semi-null frame and it is composed by two parallel-propagated Sachs screen vectors $\phi\UD{\mu}{\mathbf{A}}$, both orthogonal to $u^{\mu}$ and $\ell^{\mu}$. See \citep{Grasso:2018mei} for a detailed discussion on semi-null frames properties.} $\phi\UD{\mu}{\boldsymbol{\alpha}}=\left( u^{\mu}, \phi\UD{\mu}{\mathbf{A}},\ell^{\mu}\right)$ (with $\boldsymbol{\alpha}=0,1,2,3$ and $\mathbf{A}=1,2$ frame indices), the deviations at the source ($\delta x^{\boldsymbol{\mu}}_{\mathcal{S}}$, $\Delta \ell^{\boldsymbol{\mu}}_{\mathcal{S}}$) can be given as a linear combination of the initial deviations ($\delta x^{\boldsymbol{\mu}}_{\mathcal{O}}$, $\Delta \ell^{\boldsymbol{\mu}}_{\mathcal{O}}$)
\begin{equation}
\begin{array}{l}
\delta x^{\boldsymbol{\mu}}_{\mathcal{S}}= \WXX{}\UD{\boldsymbol{\mu}}{\boldsymbol{\nu}} \delta x^{\boldsymbol{\nu}}_{\mathcal{O}} + \WXL{}\UD{\boldsymbol{\mu}}{\boldsymbol{\nu}} \Delta \ell^{\boldsymbol{\nu}}_{\mathcal{O}}\\
\Delta \ell^{\boldsymbol{\mu}}_{\mathcal{S}}= \WLX{}\UD{\boldsymbol{\mu}}{\boldsymbol{\nu}} \delta x^{\boldsymbol{\nu}}_{\mathcal{O}} + \WLL{}\UD{\boldsymbol{\mu}}{\boldsymbol{\nu}} \Delta \ell^{\boldsymbol{\nu}}_{\mathcal{O}}\,,
\end{array}
\label{eq:deltax_deltal_BGO}
\end{equation}
where the BGO $\WXX$, $\WXL$, $\WLX$, $\WLL$ are bi-tensors acting from $\mathcal{O}$ to $\mathcal{S}$. 
Equation~\eqref{eq:deltax_deltal_BGO} can then be written in the more compact form
\begin{equation}
\begin{split}
\left(\begin{array}{l}
\delta x_{\mathcal{S}}\\
\Delta \ell_{\mathcal{S}}
\end{array}\right)&=\begin{pmatrix} 
\WXX & \WXL\\
\WLX & \WLL\\
\end{pmatrix}\left(\begin{array}{l}
\delta x_{\mathcal{O}}\\
\Delta \ell_{\mathcal{O}}
\end{array}\right)\\
&=\mathcal{W}(\mathcal{S}, \mathcal{O})\left(\begin{array}{l}
\delta x_{\mathcal{O}}\\
\Delta \ell_{\mathcal{O}}
\end{array}\right),
\label{eq:compact_W(S,O)}
\end{split}
\end{equation}
where $\mathcal{W}(\mathcal{S}, \mathcal{O})$ is the resolvent of the GDE acting from $\mathcal{O}$ to $\mathcal{S}$ and satisfying the properties:
\begin{equation}
\begin{array}{l}
\mathcal{W}(\mathcal{O}, \mathcal{S})=\left(\mathcal{W}(\mathcal{S}, \mathcal{O})\right)^{-1}\\
\mathcal{W}(\mathcal{S}, \mathcal{O})=\mathcal{W}(\mathcal{S}, p_{\lambda}) \, \mathcal{W}(p_{\lambda}, \mathcal{O}),\\
\end{array}
\label{eq:BGO_properties}
\end{equation}
with $p_{\lambda}$ being an arbitrary point on $\gamma$. 
A third key property of the BGO is that $\mathcal{W}$ is symplectic. To be precise, this property is written as
\begin{equation}
\calW^T{}\UD{\tilde{m}}{\tilde{a}} \Omega_{\tilde{m} \tilde{s}} \calW\UD{\tilde{s}}{\tilde{b}} =\Omega_{\tilde{a} \tilde{b}}
\label{eq:propW_symplectic}
\end{equation}
where $\Omega$ is the $8 \times 8$ non-singular, skew-symmetric matrix 
\begin{equation}
\Omega_{\tilde{a} \tilde{b}}=\begin{pmatrix}
0 & h_{\boldsymbol{\alpha} \boldsymbol{\beta}}\\
-h_{\boldsymbol{\gamma} \boldsymbol{\delta}} & 0
\end{pmatrix}\, ,
\end{equation}
with $h_{\boldsymbol{\alpha} \boldsymbol{\beta}}$ the metric associated to the parallel-transported frame $\phi\UD{\mu}{\boldsymbol{\alpha}}$, tilded indices run from $0$ to $7$ and bold indices $\boldsymbol{\alpha}=0,1,2,3$ are those associated with the frame. 
Inserting~\eqref{eq:compact_W(S,O)} in the GDE equation projected in the parallel transported frame $\phi\UD{\mu}{\boldsymbol{\alpha}}$, we obtain the propagation equation for the BGO
\begin{equation}
\frac{d}{d \lambda} \mathcal{W}=\begin{pmatrix}
0 & \mathbb{1}_{4 \times 4}\\
R_{\ell \ell} & 0
\end{pmatrix} \mathcal{W}
\label{eq:GDE_for_BGO}
\end{equation}
with initial conditions
\begin{equation}
\mathcal{W} \left|_{\mathcal{O}} \right. = \begin{pmatrix}
\mathbb{1}_{4 \times 4} & 0\\
0 & \mathbb{1}_{4 \times 4}
\end{pmatrix},
\label{eq:IC_GDE_for_BGO}
\end{equation}
where $\lambda$ is the affine parameter spanning the geodesic $\gamma$ and $R_{\ell \ell}$ is a short-hand notation to express the optical tidal matrix in the frame $R\UDDD{\boldsymbol{\mu}}{\alpha}{\beta}{\boldsymbol{\nu}}\ell^{\alpha}\ell^{\beta}$.

The usual procedure for studying light propagation in numerical simulations is that the spacetime dynamics is integrated forward in time, while the study of light propagation is done 
in post-processing, tracing the light beam backwards from the observer $\cal O$ to the source $\cal S$. By solving \eqref{eq:GDE_for_BGO} with initial conditions \eqref{eq:IC_GDE_for_BGO} 
at $\cal O$, one obtains the BGO $\mathcal{W}(p_\lambda,\mathcal{O})$ connecting the observer with the point $p_{\lambda}$ up to the source $\mathcal{S}=p_{\lambda_{\mathcal{S}}}$ and this is the procedure to compute observables, since real observations are made from the observer position. 
Nevertheless, in the framework we present here, one can choose to give initial conditions at $\mathcal{S}$ (or anywhere else) and integrate forward in time to $\mathcal{O}$. The key advantage is that in this way one is able to integrate \eqref{eq:GDE_for_BGO} for light propagation on-the-fly with Einstein's equations for spacetime dynamics. In this case one obtains the BGO $\mathcal{W}(p_{\lambda},\mathcal{S})$ relating the point $p_{\lambda}$ with the source. The two procedures for light propagation are fully equivalent and the relation between them, namely between $\mathcal{W}(p_{\lambda},\mathcal{O})$ and $\mathcal{W}(p_{\lambda},\mathcal{S})$, simply follows from the BGO properties \eqref{eq:BGO_properties} and reads\footnote{The symplectic property of $\mathcal{W}$, Eq. \eqref{eq:propW_symplectic} simplifies a lot the computation of $\mathcal{W}^{-1}$.}
\begin{equation}
\mathcal{W}(p_{\lambda}, \mathcal{O})=\mathcal{W}(p_{\lambda}, \mathcal{S})\left(\mathcal{W}(\mathcal{O}, \mathcal{S})\right)^{-1}
\label{eq:W_in_two_procedure}
\end{equation}
where $\mathcal{W}(\mathcal{O}, \mathcal{S})=\mathcal{W}(p_{\lambda_{\mathcal{O}}}, \mathcal{S})$.

In this work the input is an analytic form of the spacetime metric and we integrate the equation for the BGO \eqref{eq:GDE_for_BGO} backwards in time and obtain directly the L.H.S. of Eq. \eqref{eq:W_in_two_procedure}.
The procedure to compute observables can be summarized in the following steps:
\begin{enumerate}
\item compute the null geodesic connecting $\mathcal{O}$ and $\mathcal{S}$;
\item perform the parallel transport of a reference frame;
\item solve the evolution equation for the BGO, Eq.~\eqref{eq:GDE_for_BGO}, with initial conditions, Eq. \eqref{eq:IC_GDE_for_BGO} from the observer to the source;
\item combine the BGO with the four-velocity of source and observer to obtain the observables we are interested in, which are redshift and angular diameter distance $D_{\rm ang}$, written 
in terms of the BGO as, see \cite{Korzynski:2019oal}
\begin{align}
1+z &=\frac{\ell_{\sigma } u^{\sigma}|_{\cal S}}{\ell_{\sigma} u^{\sigma}|_{\cal O}} \label{eq:z_in_BGO}\\
D_{ \rm ang} &= \ell_{\sigma} u^{\sigma}|_{\cal O} \left| \det \left(\WXL {}\UD{\bm A}{\bm B}\right) \right|^{\frac{1}{2}} \label{eq:Dang_in_BGO}\,.
\end{align}
\end{enumerate}
The advantage of the BGO formalism is that it provides a unified approach to geometric optics. Furthermore, it extends the standard Sachs formalism, allowing also to describe what happens when the observation occurs for a prolonged period of time and the slow temporal variations of the optical observables, called the drift effects, could become measurable. 

All the steps $1$ - $4$ require the ability of solving systems of coupled ODEs, that can be done either using analytical methods (exact or perturbative approach) or numerical methods. In our work we will use both methods, as we are going to explain in the next section.

%%%%%%%%%%%%%%%%%%%%%%%%%%%%%%%%%%%%%%%%%%%%%%%%%%%%%%%%%%%%%%%%%%%%%%%
%%%%%%%%%%%%%%%%%%%%%%%%%%%%%%%%%%%%%%%%%%%%%%%%%%%%%%%%%%%%%%%%%%%%%%%
\section{Method}
\label{sec:method}
%%%%%%%%%%%%%%%%%%%%%%%%%%%%%%%%%%%%%%%%%%%%%%%%%%%%%%%%%%%%%%%%%%%%%%%
%%%%%%%%%%%%%%%%%%%%%%%%%%%%%%%%%%%%%%%%%%%%%%%%%%%%%%%%%%%%%%%%%%%%%%%
The core of our analysis is to estimate the magnitude of the non-linear effects on light propagation, through the comparison of some cosmological observables calculated within different approximation schemes. In particular, we will compare the redshift $z$ and the angular diameter distance $D_{\rm ang}$ computed in the following three cases:
\begin{enumerate}
\item using the first-order expansion in standard cosmological perturbation theory of the plane-parallel metric \eqref{eq:metricIPT} and performing light propagation perturbatively, up to first order. We will denote as $O^{\rm Lin}$, the generic observable $O$ obtained in this way, which only includes effects linear in the perturbations; ($z^{\rm Lin}$ and $D_{\rm ang}^{\rm Lin}$ are derived in App. \ref{apx:linear_obs});
\item using the Newtonian part of the plane-parallel metric, namely the metric in Eq.~\eqref{eq:metricNWT}, and performing exact light propagation\footnote{The term ``exact'' refers to the fact that no perturbative approach is used when we derive and solve the equations describing the propagation of light and observables. In other words, even if the spacetime metric was obtained using some perturbation scheme, we use it as if it were exact for the entire procedure to calculate the observables, starting from the very beginning, i.e. the geodesic equation. We will discuss this approach further on in this section.} using numerical integration. The observables calculated in this way will be indicated as $O^{\rm N}$;
\item using the full PN plane-parallel metric \eqref{eq:metricPN} and performing exact light propagation via numerical integration. We denote the observables calculated with this method as $O^{\rm PN}$.
\end{enumerate}
The observables calculated with the last two methods, $O^{\rm N}$ and $O^{\rm PN}$, are obtained using \emph{\texttt{BiGONLight.m}} (\textbf{Bi}-local \textbf{G}eodesic \textbf{O}perators framework for \textbf{N}umerical \textbf{Light} propagation), a publicly available {\tt Mathematica} package ({\color{blue}{\tt {https://github.com/MicGrasso/bigonlight1.0}}}) developed to study light propagation in numerical simulations using the BGO framework. The package contains a collection of function definitions, including those to compute geodesics, parallel transported frames and solve the BGO's equation~\eqref{eq:GDE_for_BGO}. \texttt{BiGONLight.m} works as an independent package that, once is called by a {\tt Mathematica} notebook, can be used to compute numerically the BGO along the line of sight, given the spacetime metric, the four-velocities and accelerations of source and observer as inputs: a sample of the notebook we used for our analysis can be found in the repository folder \emph{Plane-parallel}. An exhaustive description of \texttt{BiGONLight} and several tests of the package are presented in \cite{Grasso:BGO}. Here we just report in App. \ref{apx:bigonlight} two case studies of code testing, the $\Lambda$CDM and the Szekeres model. 

Let us now comment about the fact that we use exact light propagation for the Newtonian and post-Newtonian observables, despite the fact that the respective spacetime metric is obtained with perturbative techniques. Firstly, we notice that this method used to compute $O^{\rm PN}$ does not produce observables strictly of PN order: the observables $O^{\rm PN}$ will contain also some of higher than PN contributions, coming from the fact that we start from the PN metric \eqref{eq:metricPN} but we do not expand further the equations for light propagation or the expressions for the observables in powers of $1/c^2$ (we set $c=1$ everywhere). One would naively expect that the higher than PN terms are always sub-leading with respect to the PN ones, as in any well-defined perturbation scheme. The key point here is if this hierarchy, which starts at the level of metric perturbations, is preserved throughout the full calculation to the final results, especially in our case where the equations to compute the observables are fully non-linear. We find that this is indeed the case, as indicated in similar investigations in the literature. In order to show this explicitly and to give an estimate of the higher than PN corrections, we have compared the density contrast calculated strictly up to PN order $\delta_{\rm PN}$ \eqref{eq:density_PN} and the density contrast $\delta_{\rm ex}$ obtained from its exact expression from the continuity equation in synchronous-comoving gauge, i.e.
\begin{equation}
\delta_{\rm ex}(\eta,q_{\rm 1})= (\delta(\eta_{\rm in},q_{\rm 1})+1)\sqrt{\frac{|\gamma(\eta_{\rm in},q_{\rm 1})|}{|\gamma(\eta
,q_{\rm 1})|}}-1 \, ,
\label{eq:delta_ex}
\end{equation} 
where $|\gamma|$ is the short-hand notation for the determinant of the metric \eqref{eq:metricPN}, calculated here without expanding in powers of $1/c^2$.
%%%%%%%%%%%%%%%%%%%%%%%%%%%%%%%%%%%%%%%%%%%%%%%%%%%%%%%%%%%%%
\begin{figure}[h]
    \centering
    \begin{subfigure}[h]{0.49\textwidth}
        \includegraphics[width=1.01\textwidth]{delta_PN_N.pdf}
       \caption{$\left|\frac{\delta_{\rm PN}-\delta_{\rm N}}{\delta_{\rm N}}\right|$}
    \end{subfigure}
    \!\!\! 
    \begin{subfigure}[h]{0.49\textwidth}
        \includegraphics[width=1.01\textwidth]{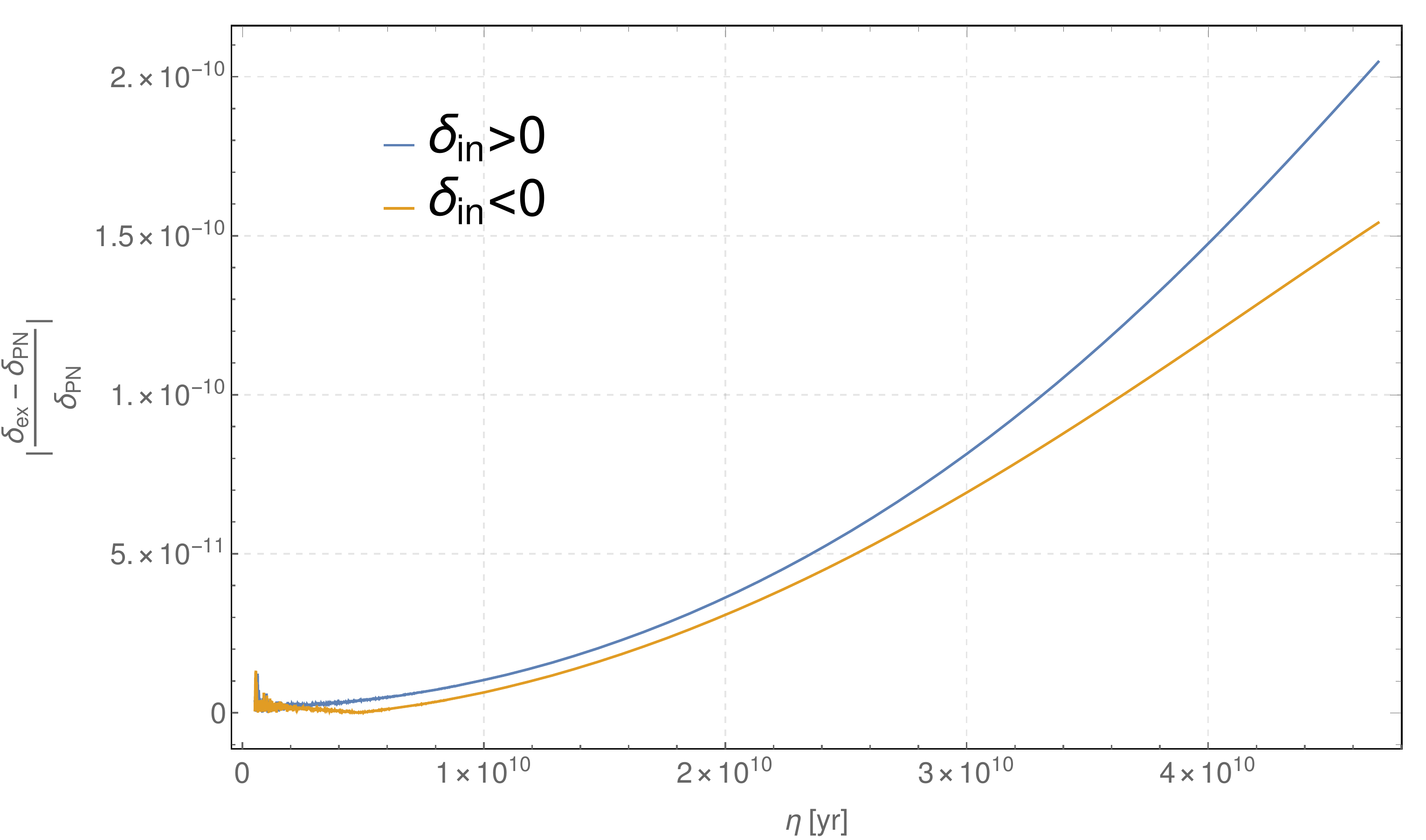}
        \caption{$\left|\frac{\delta_{\rm ex}-\delta_{\rm PN}}{\delta_{\rm PN}}\right|$ }
    \end{subfigure}
    \caption{Evolution of the variation $\delta_{\rm PN}$ vs $\delta_{\rm N}$ (a) and $\delta_{\rm ex}$ vs $\delta_{\rm PN}$ (b) for initial over-density $\delta_{\rm in}>0$ and under-density $\delta_{\rm in}<0$ regions for $k = 500\, \rm Mpc$. The variation $\delta_{\rm ex}$ vs $\delta_{\rm PN}$ (b) is 4 orders of magnitude smaller than the variation $\delta_{\rm PN}$ vs $\delta_{\rm N}$ (a).}\label{fig:over-under_delta_ex_VS_PN}
\end{figure}

\begin{figure}[h]
    \centering
    \begin{subfigure}[h]{0.49\textwidth}
        \includegraphics[width=1.01\textwidth]{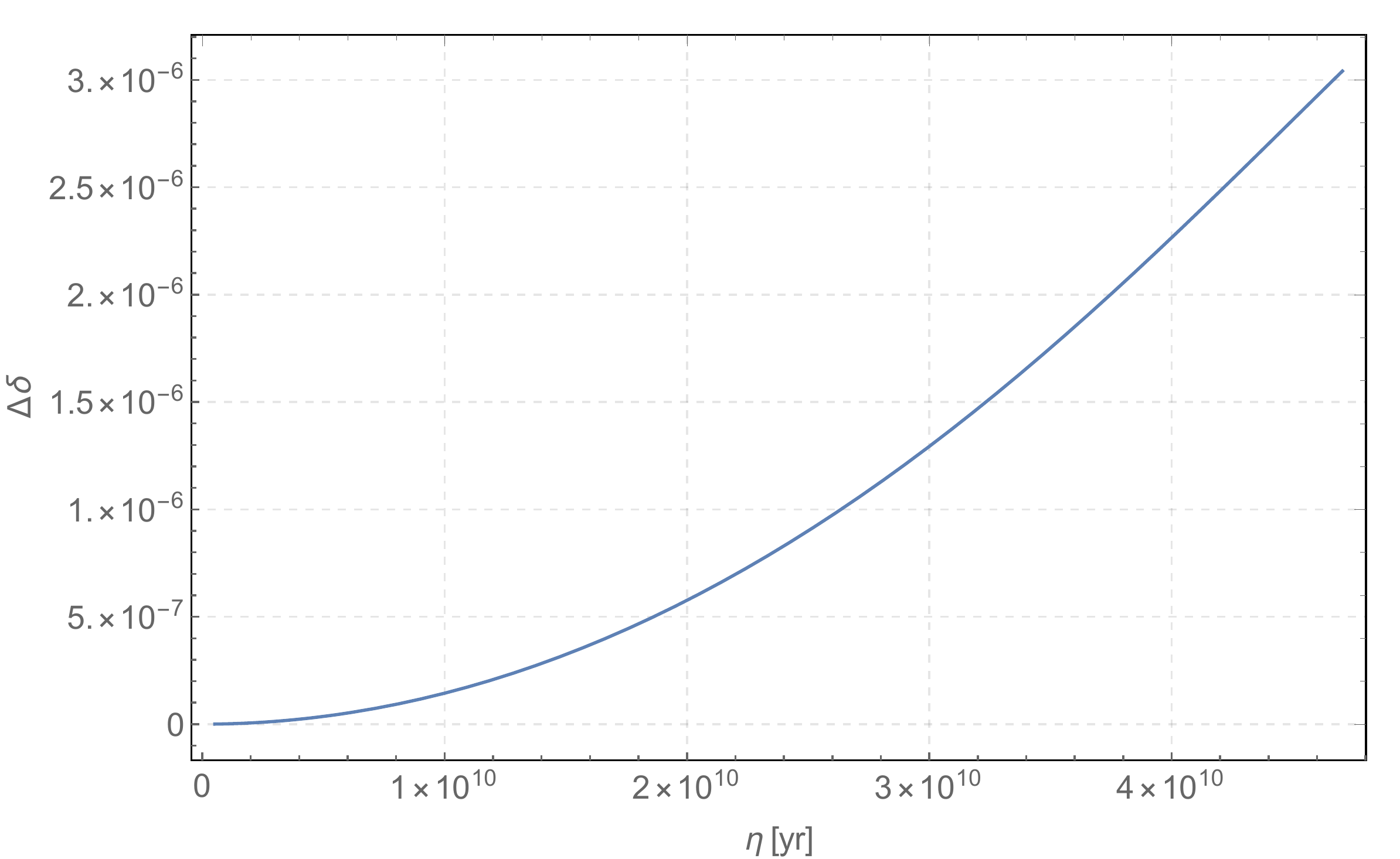}
        \caption{$\left|\delta_{\rm PN}-\delta_{\rm N}\right|$ }
       
    \end{subfigure}
    \!\!\! 
    \begin{subfigure}[h]{0.49\textwidth}
        \includegraphics[width=1.01\textwidth]{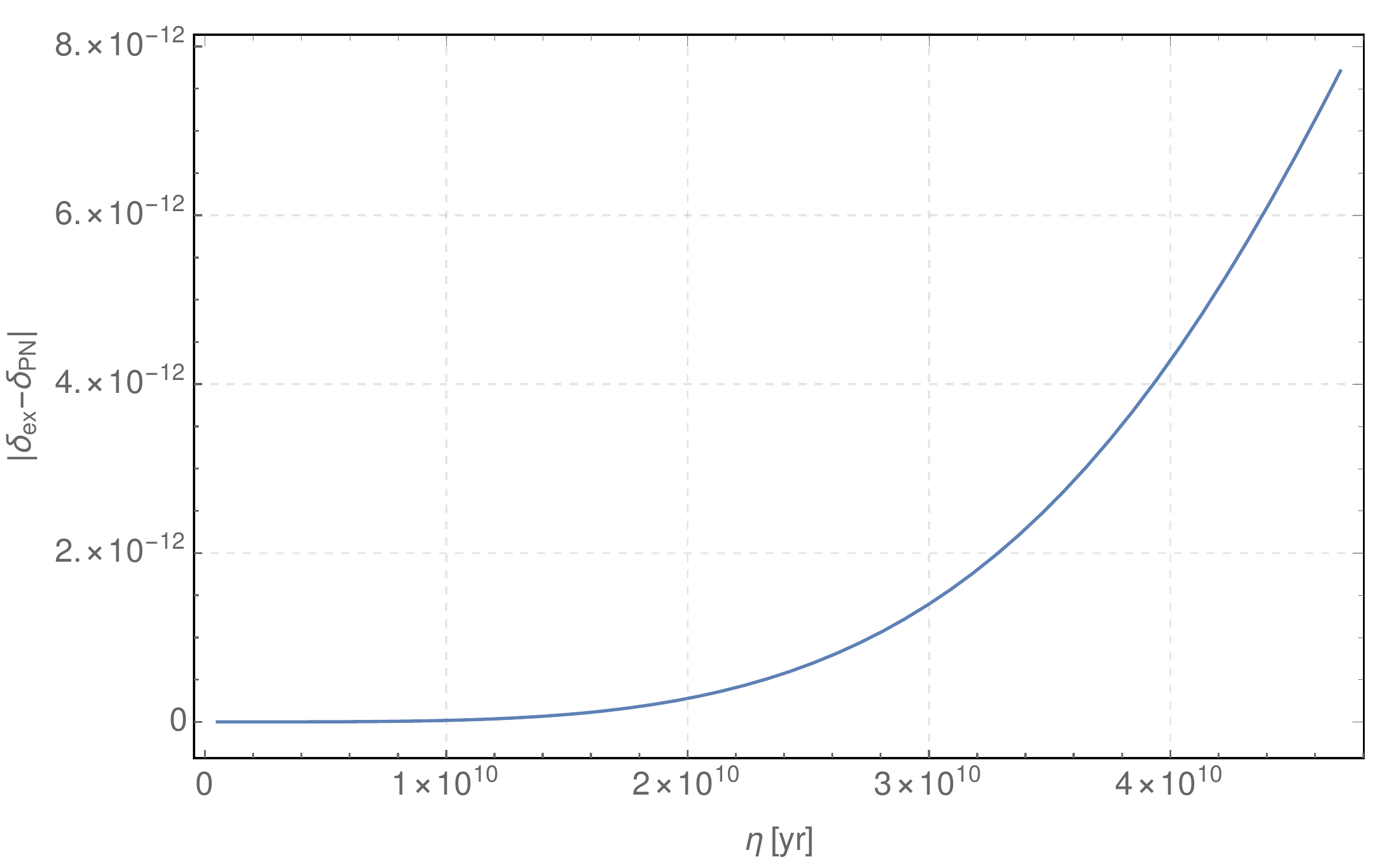}
        \caption{$\left|\delta_{\rm ex}-\delta_{\rm PN}\right|$ }
    \end{subfigure}
    \caption{Evolution of the variation $\delta_{\rm PN}$ vs $\delta_{\rm N}$ (a) and $\delta_{\rm ex}$ vs $\delta_{\rm PN}$ (b) for regions with $\delta_{\rm in}=0$ and $k = 500\, \rm Mpc$. The variation $\delta_{\rm ex}$ vs $\delta_{\rm PN}$ (b) is 6 orders of magnitude smaller than the variation $\delta_{\rm PN}$ vs $\delta_{\rm N}$ (a).}\label{fig:d=0_delta_ex_VS_PN}
\end{figure}
%%%%%%%%%%%%%%%%%%%%%%%%%%%%%%%%%%%%%%%%%%%%%%%%%%%%%%%%%%%%%
The plots in Fig.~\ref{fig:over-under_delta_ex_VS_PN} show that the variation between $\delta_{\rm PN}$ and $\delta_{\rm ex}$ is 4 orders smaller than the variation between $\delta_{\rm N}$ and $\delta_{\rm PN}$ for initial over- and under-dense regions and it is 6 orders smaller when we consider regions with vanishing initial density contrast, Fig.\ref{fig:d=0_delta_ex_VS_PN}. This is something we expected, since the impact of the corrections gets smaller and smaller with the increase of the order in the expansion and, more importantly, we were able to isolate and quantify the corrections coming from the higher than PN terms. This specific result holds for the density contrast but it is perfectly reasonable that this estimation is roughly valid for the observables too, even if the calculation to get them is different. We believe that the argument just presented validates our method of performing exact light propagation.

In order to compare the observables calculated within different approximations, we introduce the dimensionless variation $\Delta O$ for the generic observable $O$ calculated in the two approximations $\mathit{a}$ and $\mathit{b}$ defined as:
\begin{equation}
\Delta O (b, a)= \frac{O^{\mathit{b}}-O^{\mathit{a}}}{O^{\mathit{a}}}
\label{eq:variation}
\end{equation}
where $\mathit{a}$ and $\mathit{b}$ stand for $\rm \Lambda CDM$, $\rm Lin$, $\rm N$ or $\rm PN$, namely the $\rm \Lambda CDM$ background, the linear order in standard PT, Newtonian or post-Newtonian approximations, respectively.

Having introduced the general method we use for our analysis and we defined the key quantity for our comparisons, we have to specify the free functions and the parameters of the plane-parallel universe we are considering, of the $\rm \Lambda CDM$ background model and its perturbations. We recall that the evolution of the inhomogeneities in our model is governed by the growing mode solution $\mathcal{D}$ \eqref{eqforD+}, while the spatial part of the matter distribution is determined by the gravitational potential $\phi_{\rm 0}$, which is the only free function. We use a sinusoidal profile for the gravitational potential $\phi_{\rm 0}$ defined as:
\begin{equation}
\phi_{\rm 0}= \mathcal{I} \sin(\omega q_{\rm 1})
\label{eq:phi}
\end{equation}
where the frequency $\omega=2 \pi/k$ is determined from the scale of the inhomogeneities $k$, while the amplitude $\mathcal{I}$ is obtained from \eqref{eq:density_PN} for a certain value of the maximum of post-Newtonian density contrast today $\delta^{\rm max}_{0}$. The scale $k$ and the maximum of the density contrast $\delta^{\rm max}_{0}$ are linked by the matter power spectrum and we will repeat our analysis for different values of $(k, \delta^{\rm max}_{0})$ (this will be discussed in the section \ref{sec:results}). In Tab. \ref{tab:k-delta} we report the chosen  values for the scales and the corresponding maximum of the density contrast today.
\begin{table}\caption{\label{tab:k-delta}Values $(k, \, \delta^{\rm max}_{0})$ used in our analysis.}
\begin{ruledtabular}
\begin{tabular}{lccccc}
$k\, (\rm Mpc)\, $ & $\, 500\, $ & $\, 300 \,$ & $\, 100 \,$ & $\, 50 \,$ & $\, 30 \,$\\
$\delta_{\rm 0}^{\rm max}$ & $0.1$ & $0.35$ & $1$ & $1.5$ & $1.8$\\
\end{tabular}
\end{ruledtabular}
\end{table}
The cosmological parameters are set using the fiducial values from~\cite{planck2018param}, i.e. $\Omega_{\rm m0} =0.3153$, $\Omega_{\rm \Lambda}=0.6847$ and $\mathcal{H}_{\rm 0}=67.36$. For primordial non-Gaussianity we use the parameter $a_{\rm nl}$ introduced in \cite{Bartolo:2005kv}. It is linked to the parameter $f_{nl}$ by:
\begin{equation}
a_{\rm nl}=\frac{3}{5}f_{\rm nl}+1
\end{equation}
where $a_{\rm nl}=1$, i.e. $f_{\rm nl}=0$, correspond to the case of exact Gaussian fluctuations. The latest measurement of $f_{\rm nl}$ from the Planck collaboration \cite{planck2019anl} gives $a_{\rm nl}=0.46 \pm 3.06$ that will fix $a_{\rm nl}=0.46$ as the fiducial value for our analysis. However, since in our case we take deterministic initial conditions, $a_{\rm nl}$ merely represents an extra free parameter of our approach which tunes the post-Newtonian corrections\footnote{This is evident, since $a_{\rm nl}$ appears only in the post-Newtonian terms and some of them can be cancelled or dimmed with an appropriate choice of the $a_{\rm nl}$'s value.}. Given that $a_{\rm nl}$ has a lot of room to vary inside its confidence interval of $\pm 3.06$, we have also investigated how the comparison Newtonian vs post-Newtonian gets modified if we take different values of $a_{\rm nl}$ to calculate post-Newtonian observables $O^{\rm PN}$ (see section \ref{sec:results}).

\begin{figure}[h]
    \centering
        \includegraphics[width=0.35\textwidth]{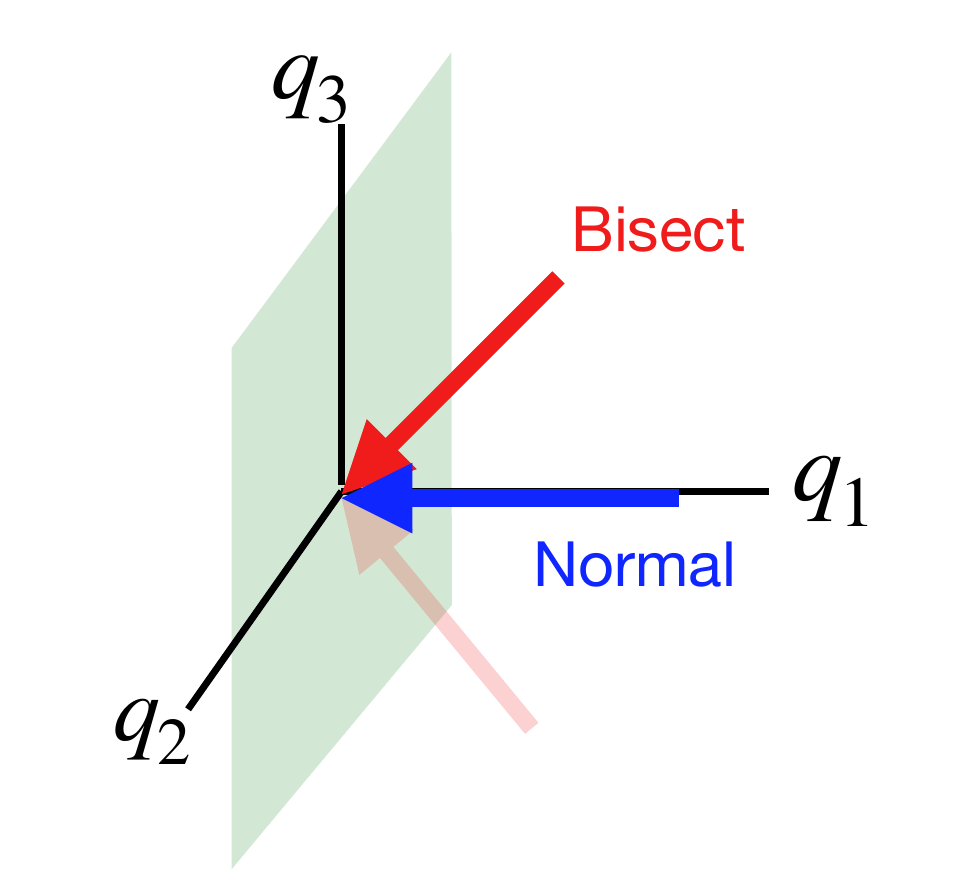}
       \caption{Graphic representation of the direction normal to the planes (blue) and the direction along the bisect (red). Two geodesics with these directions will intersect the uniform density planes with different angles. Therefore the matter distribution profiles along the geodesics are also different. }
        \label{fig:directions}
\end{figure}

\begin{figure}[!hb]
    \centering
    \begin{subfigure}[h]{0.49\textwidth}
        \includegraphics[width=1.01\textwidth]{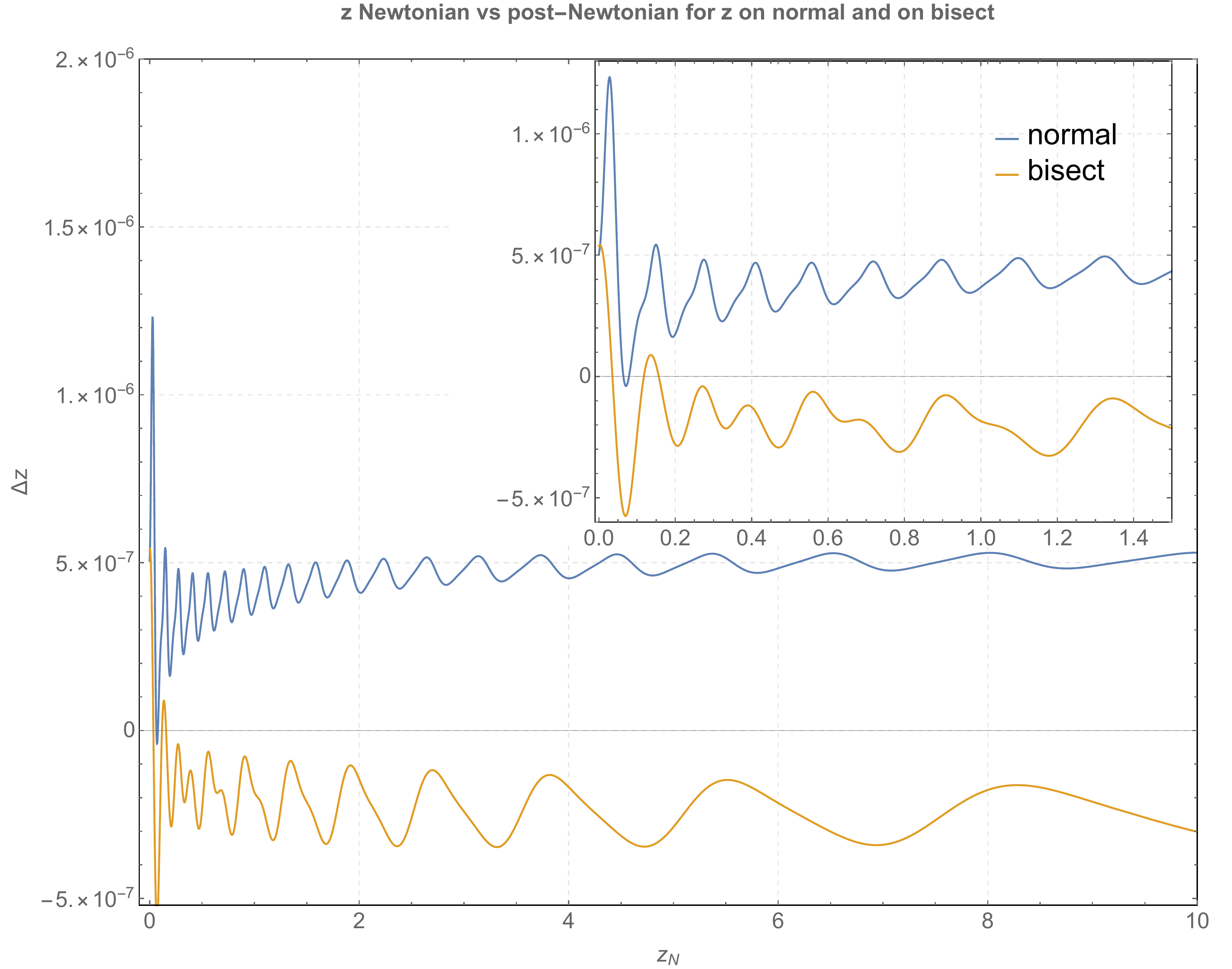}
       \caption{$\Delta z$}
        \label{fig:delta_z_direz}
    \end{subfigure}
    \!\!\! 
    \begin{subfigure}[h]{0.49\textwidth}
        \includegraphics[width=1.01\textwidth]{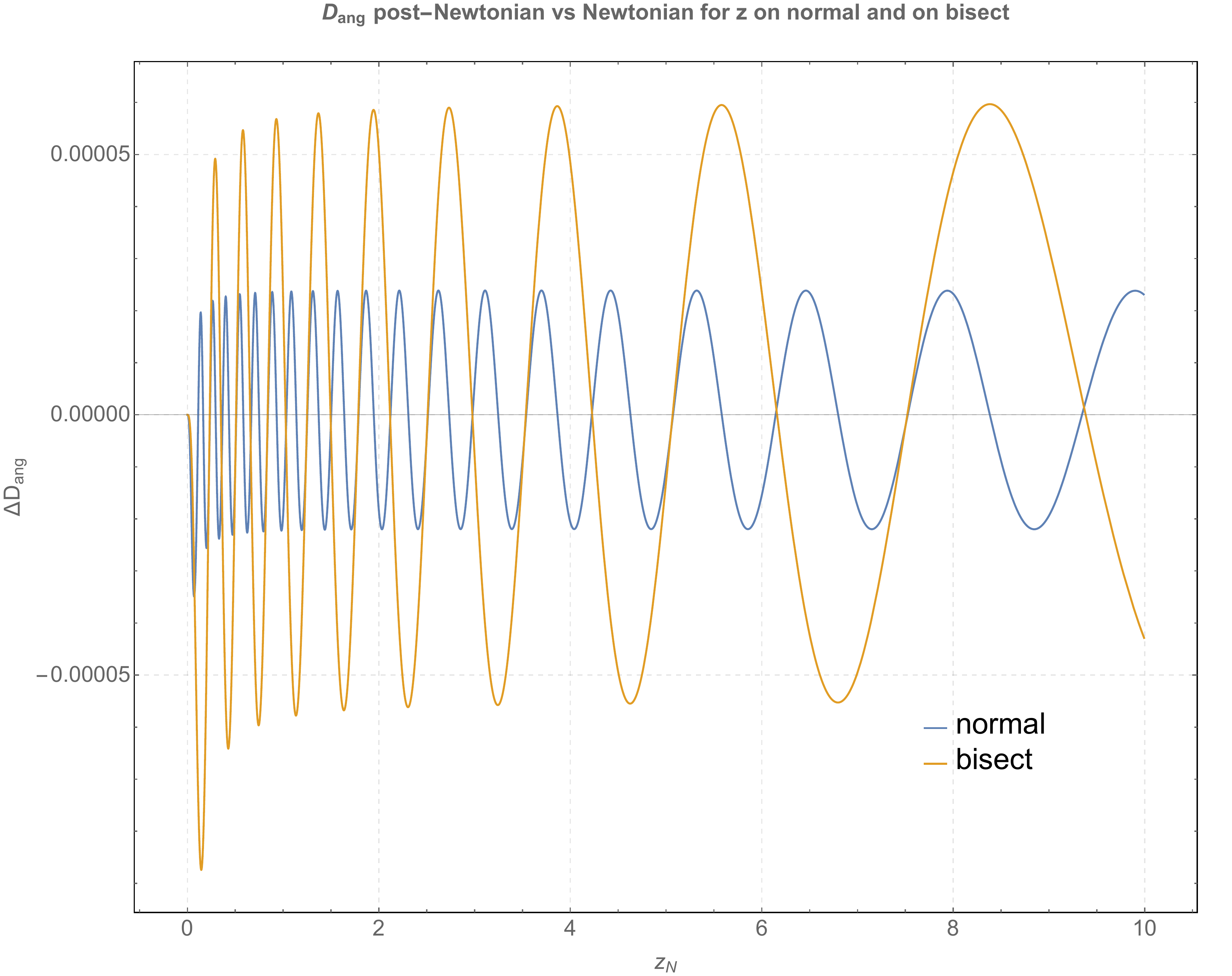}
        \caption{$\Delta D_{\rm ang}$ }
        \label{fig:delta_D_direz}
    \end{subfigure}
    \caption{$\Delta z(PN, N)$ \eqref{fig:delta_z_direz} and $\Delta D_{\rm ang}(PN, N)$ \eqref{fig:delta_D_direz} according to our definition \eqref{eq:variation} for the two geodesics with directions normal to the planes (blue lines) and parallel to the bisect (orange lines).}\label{fig:diff_directions}
\end{figure}

The last things we need to specify  are  the observer and emitter positions and their kinematics. In our study we place the observer in a position with vanishing initial density contrast $\delta_{\rm in}=0$ and we will leave the analysis on how the comparison change when the observer is located in an initial overdensity or underdensity for future investigations. 
The geodesic equations and the BGO equations \eqref{eq:GDE_for_BGO} are solved giving the initial conditions at the observer position and they are integrated backwards in time up to redshift $z=10$.
The choice of analysing only sources at $z=10$ still leaves us the freedom in selecting the direction from which the light is coming. The difference between geodesics with different directions is mainly due to the way in which the geodesics cross the parallel planes with uniform density.  
To investigate this effect, we have considered two geodesics, one with direction normal to the planes and one with direction parallel to the bisect as represented in Fig. \ref{fig:directions}, considering in both cases the observer in a position with $\delta_{\rm in}=0$ and the gravitational potential \eqref{eq:phi} set such that $k=500 {\rm Mpc}$ and $\delta^{\rm max}_{\rm 0}=0.1$. For both geodesics we have analysed what are the effects of the direction on the variations Newtonian vs post-Newtonian for $\Delta z$ (Fig. \ref{fig:delta_z_direz}) and $\Delta D_{\rm ang}$ (Fig. \ref{fig:delta_D_direz}).
%%%%%%%%%%%%%%%%%%%%%%%%%%%%%%%%%%%%%%%%%%%%%%%%%%%%%%%%%%%%%

%%%%%%%%%%%%%%%%%%%%%%%%%%%%%%%%%%%%%%%%%%%%%%%%%%%%%%%%%%%%%
From the plots we can conclude that there are small differences in the comparison post-Newtonian vs Newtonian for geodesics with different directions. However, the change in the matter distribution on the geodesic induced by the different directions does not modify the magnitude of the variations too much, but only their shapes. In conclusion, when we consider geodesics along the normal, the effects of the non-linearities are somewhat smaller than for the geodesics along the bisect direction. Nevertheless, from now on, we will consider only geodesics directed along the bisect: this will not affect our conclusions because we will make all the comparisons using geodesics along the bisect in all the cases under study. 

For clarity, the following list summarizes the conditions we set for our analysis:
\begin{itemize}
\item If not specified, the observer $\mathcal{O}$ is placed in a position with initial vanishing density contrast $\delta_{\rm in}=0$.
\item The sources are at redshift $z=10$ and such that the observer receives the light with direction parallel to the bisect.
\item Our analysis is performed in synchronous comoving gauge implying that both emitter and observer are comoving with the cosmic flow. 
\item The primordial non-Gaussianity parameter $a_{\rm nl}$ is set using the fiducial value from Planck \cite{planck2019anl}, i.e. $a_{\rm nl}=0.46$. However, in section \ref{sec:results} we will also consider the case when $a_{\rm nl}$ is set equal to the extreme of its confidence interval.
\end{itemize}

%%%%%%%%%%%%%%%%%%%%%%%%%%%%%%%%%%%%%%%%%%%%%%%%%%%%%%%%%%%%%%%%%%%%%%%
%%%%%%%%%%%%%%%%%%%%%%%%%%%%%%%%%%%%%%%%%%%%%%%%%%%%%%%%%%%%%%%%%%%%%%%
\section{Results}
\label{sec:results}
\begin{figure*}[ht]
    \centering
    \begin{subfigure}{0.49\linewidth}%{0.8\columnwidth}%0.40\textwidth
        \includegraphics[width=\linewidth]{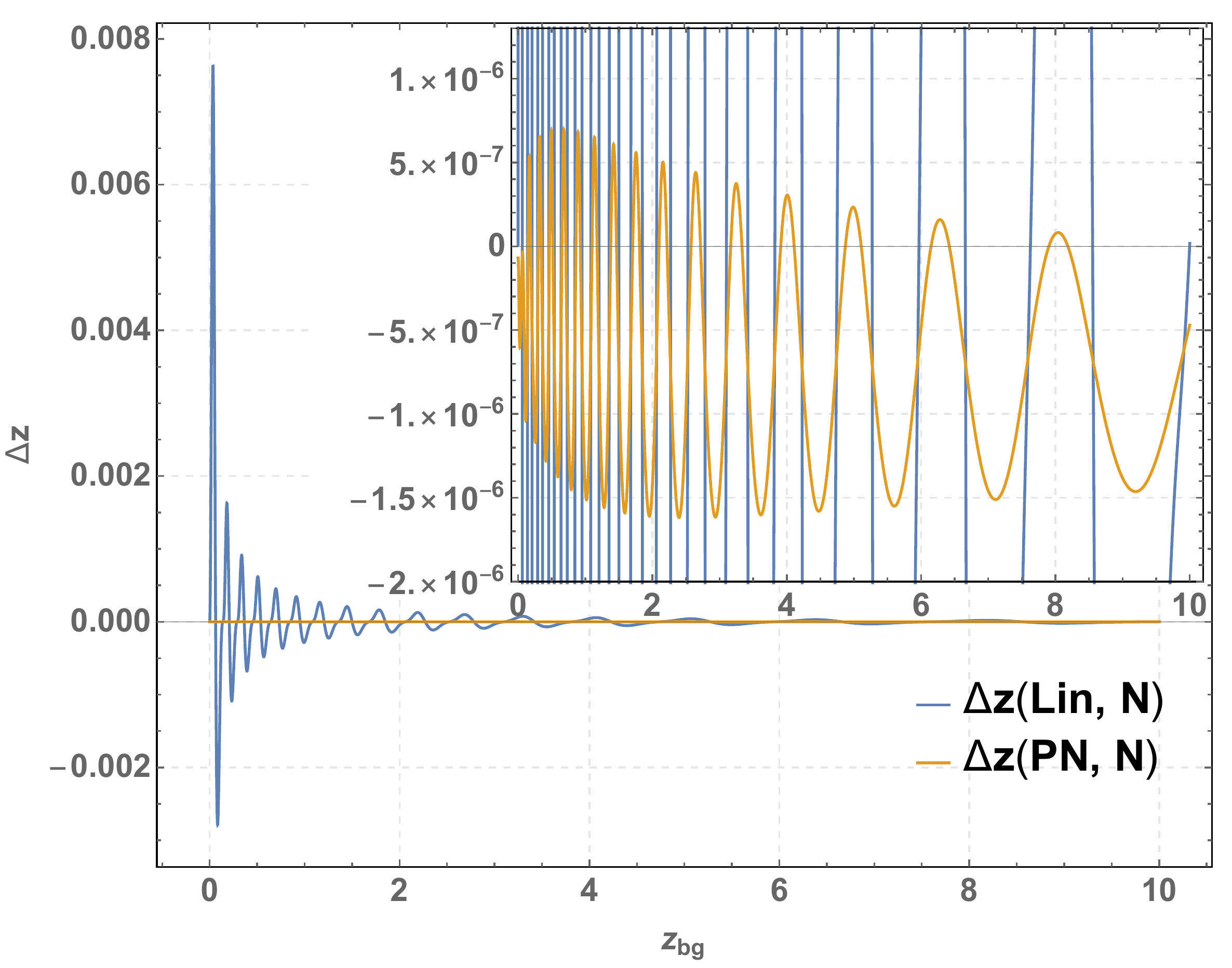}
       \caption{ k=300}
        \label{fig:dz_point_1_k300}
    \end{subfigure}
    \begin{subfigure}{0.49\linewidth}%{0.8\columnwidth}
        \includegraphics[width=\linewidth]{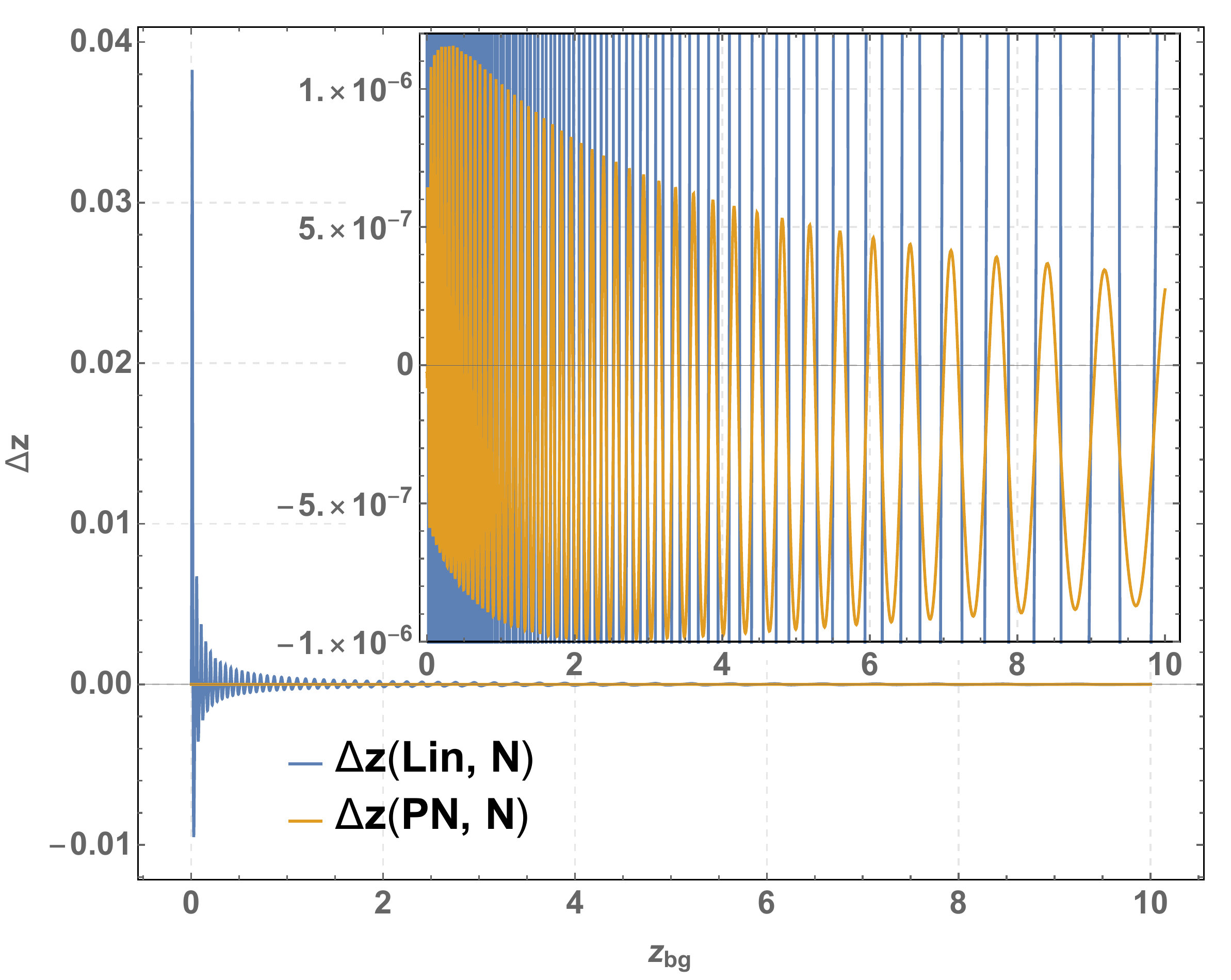}
        \caption{ k=100 }
        \label{fig:dz_point_1_k100}
    \end{subfigure}
    \\
    \begin{subfigure}{0.5\linewidth}%{0.8\columnwidth}
        \includegraphics[width=\linewidth]{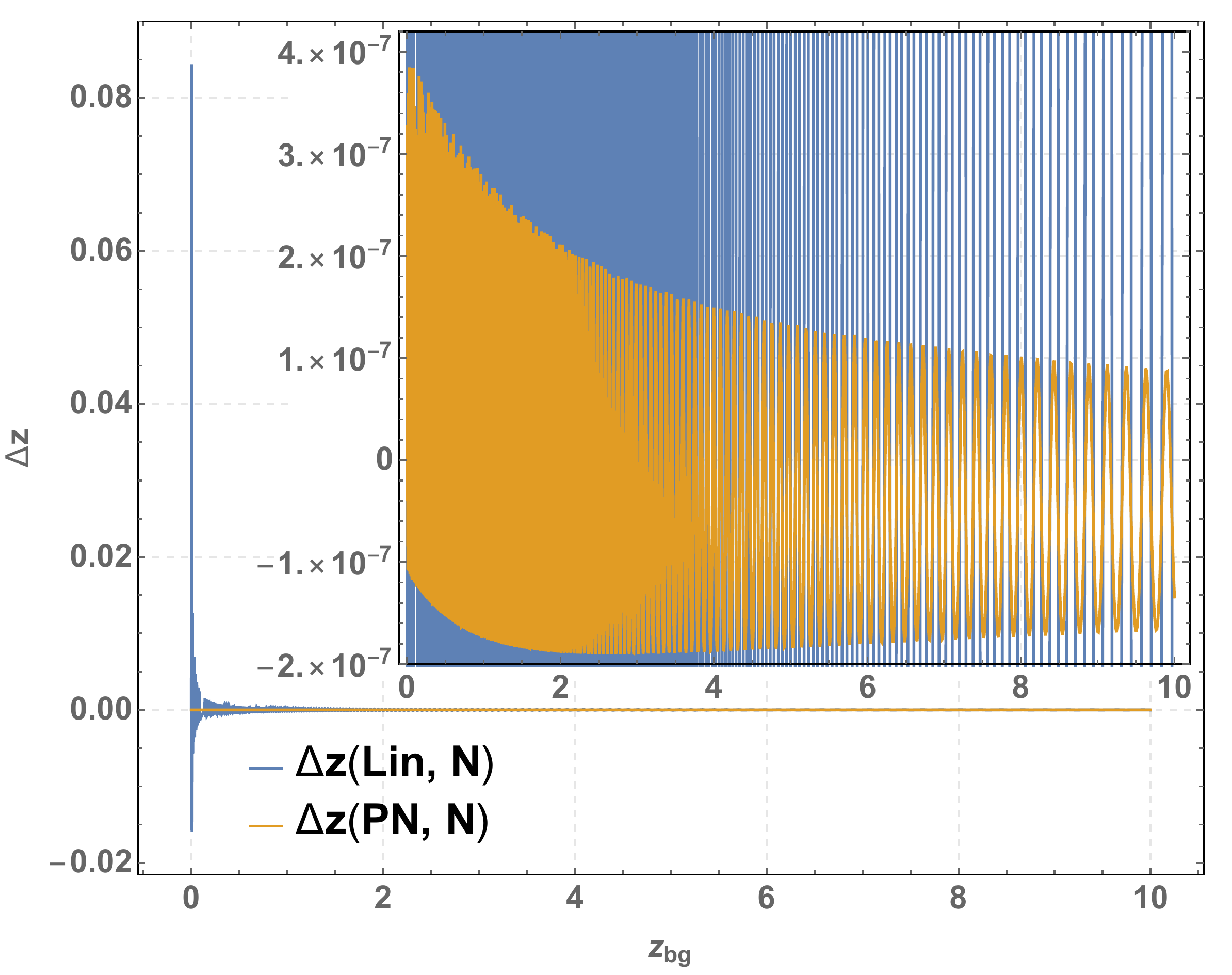}
        \caption{ k=30 }
        \label{fig:dz_point_1_k30}
    \end{subfigure}
    \caption{Redshift variations, as defined in Eq.~\eqref{eq:variation_res}, Linear vs Newtonian (blue) and post-Newtonian vs Newtonian (orange) on three different scales $k=30\, , 100\, , 300\, \rm Mpc$. We see that $\Delta z (Lin, N) \sim 10^2 \, \Delta z (PN, N)$ on every scale $k$. The variable on the horizontal axis is the $\Lambda CDM$ redshift.}\label{fig:dz_point_1}
\end{figure*}

\begin{figure*}[ht]
    \centering
    \begin{subfigure}[h]{0.49\linewidth}%{0.8\columnwidth}
        \includegraphics[width=\linewidth]{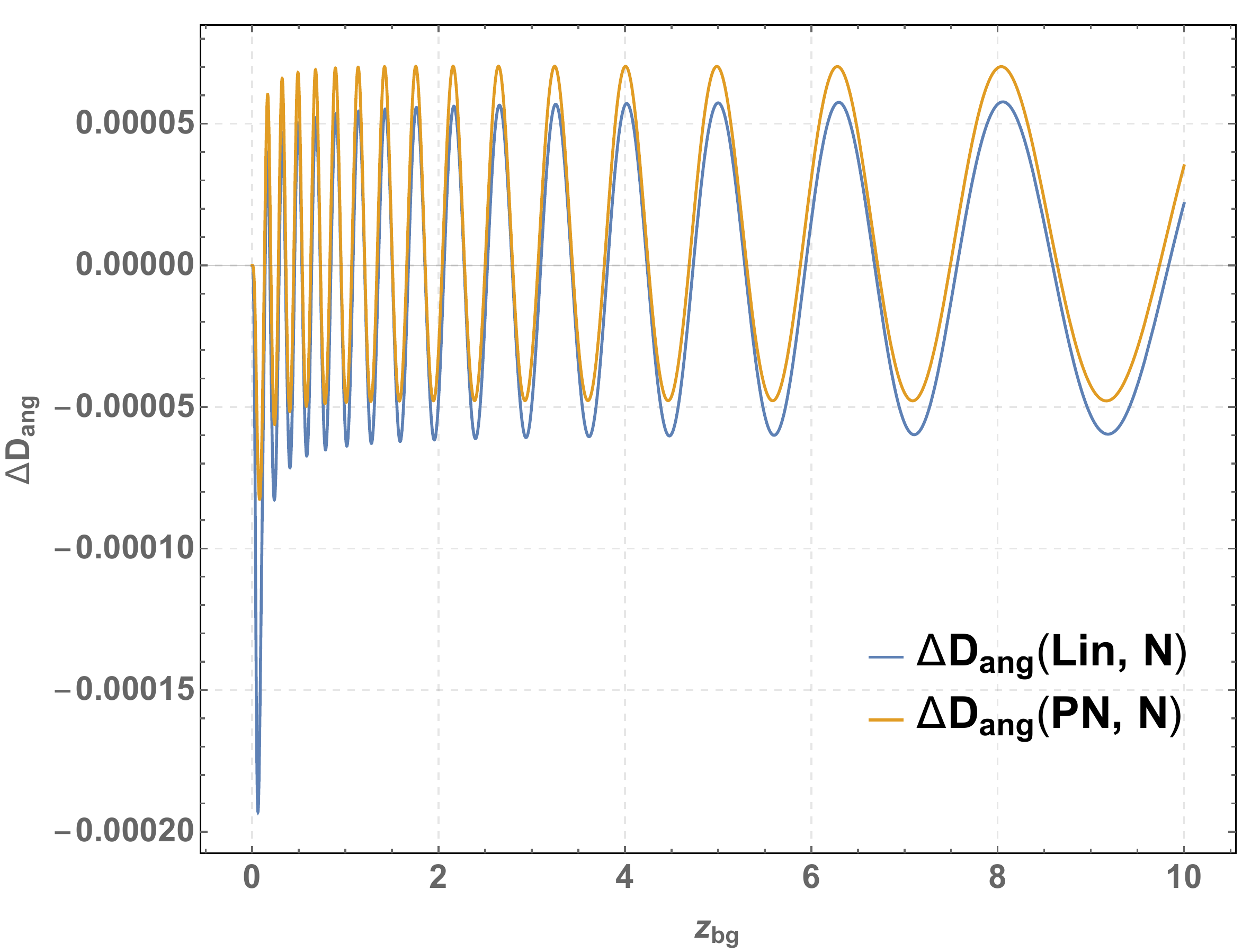}
       \caption{ k=300}
        \label{fig:dD_point_1_k300}
    \end{subfigure}
    \begin{subfigure}[h]{0.49\linewidth}%{0.8\columnwidth}
        \includegraphics[width=\linewidth]{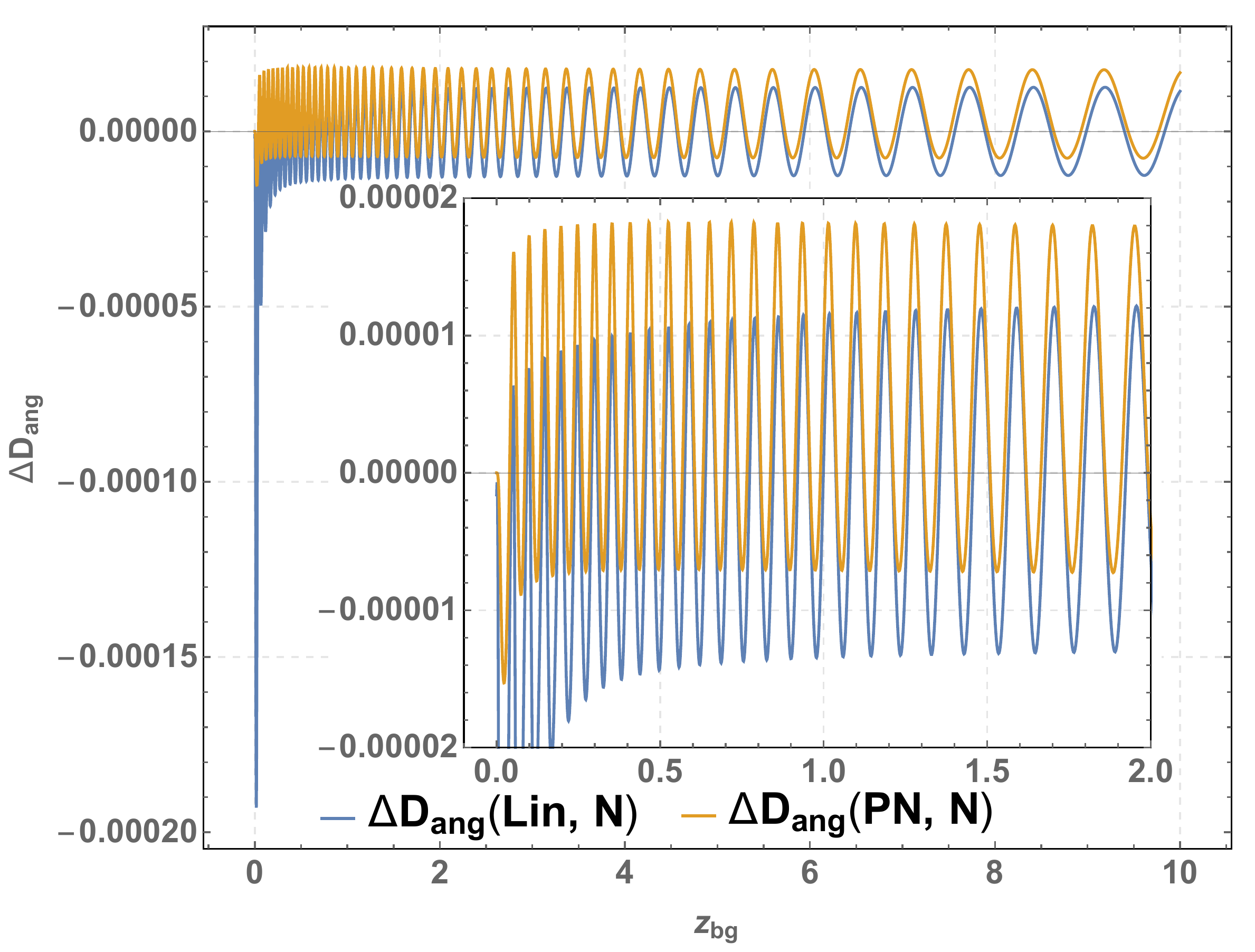}
        \caption{ k=100 }
        \label{fig:dD_point_1_k100}
    \end{subfigure}
    \begin{subfigure}[h]{0.49\linewidth}%{0.8\columnwidth}
        \includegraphics[width=\linewidth]{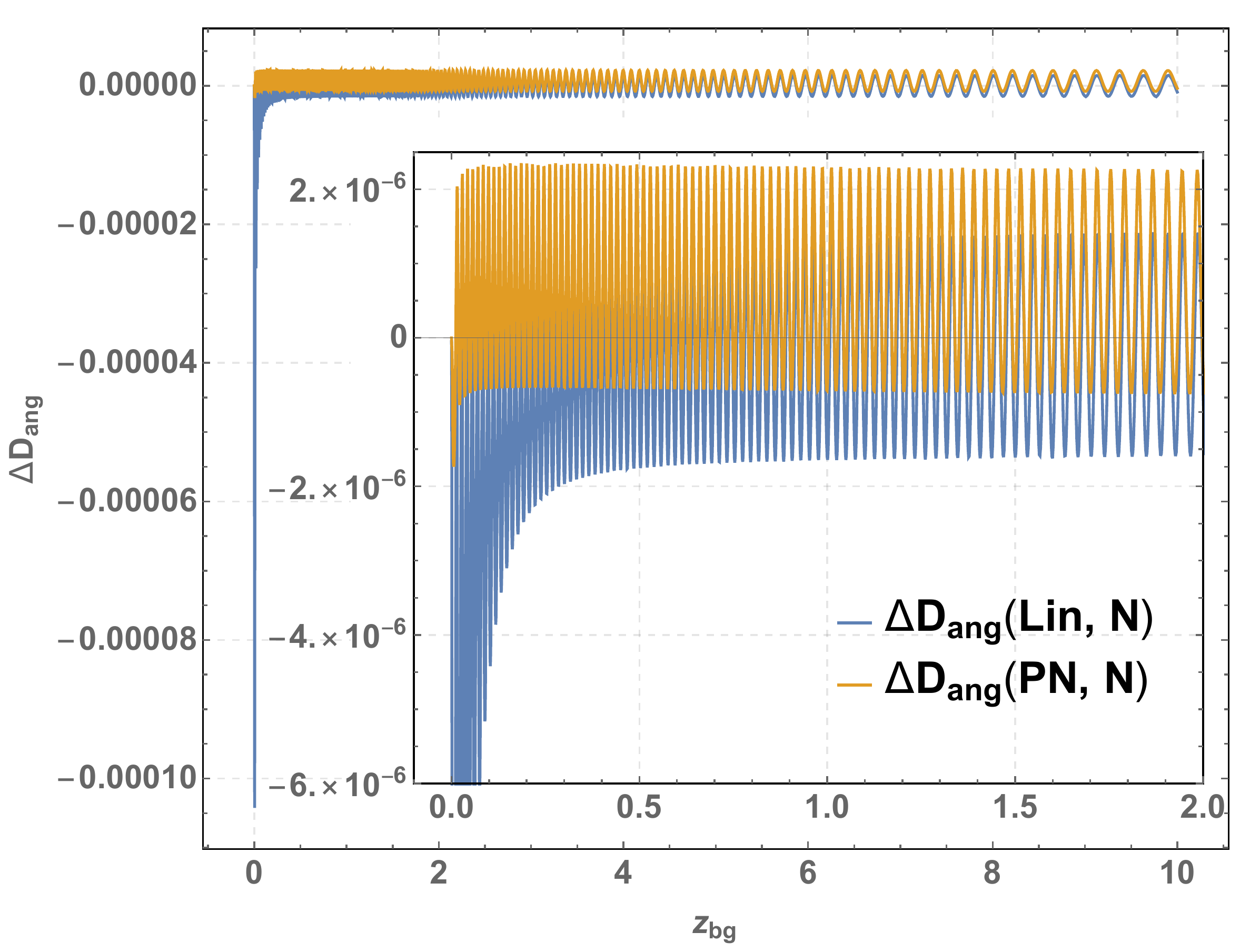}
        \caption{ k=30 }
        \label{fig:dD_point_1_k30}
    \end{subfigure}
    \caption{Angular diameter distance variations, as defined in Eq.~\eqref{eq:variation_res}, Linear vs Newtonian (blue) and post-Newtonian vs Newtonian (orange) on three different scales $k=30\, , 100\, , 300\, \rm Mpc$. We see that $\Delta D_{ \rm ang} (Lin, N) \sim \Delta D_{\rm ang} (PN, N)$ on every scale $k$. The variable on the horizontal axis is the $\Lambda CDM$ redshift.}\label{fig:dD_point_1}
\end{figure*}

%%%%%%%%%%%%%%%%%%%%%%%%%%%%%%%%%%%%%%%%%%%%%%%%%%%%%%%%%%%%%%%%%%%%%%%
%%%%%%%%%%%%%%%%%%%%%%%%%%%%%%%%%%%%%%%%%%%%%%%%%%%%%%%%%%%%%%%%%%%%%%%
In this section we present the results of our study that we plot in terms of the quantity
\begin{equation}
\Delta O (b, a)= \frac{O^{\mathit{b}}-O^{\mathit{a}}}{O^{\mathit{a}}}\, ,
\label{eq:variation_res}
\end{equation} 
where our observables $O$ are the redshift $z$ and the angular diameter distance $D_{\rm ang}$ and $a, b$ stand for the approximations used in turn.
Let us start with Figs. \ref{fig:dz_point_1} and \ref{fig:dD_point_1} in which we plot the variation between linear and Newtonian approximations, $\Delta z (Lin, N)$ and $\Delta D_{\rm ang} (Lin, N)$, and the PN corrections to the Newtonian approximation, $\Delta z (PN, N)$ and $\Delta D_{\rm ang} (PN, N)$ for three different scales, $k=30\, , 100\, , 300\, \rm Mpc$.
%%%%%%%%%%%%%%%%%%%%%%%%%%%%%%%%%%%%%%%%%%%%%%%%%%%%%%%%%%%%%%%%%%%%%%%%%%%%%%%%%%%%%%%%%%%%%%%%%%%%%%%
%%%%%%%%%%%%%%%%%%%%%%%%%%%%%%%%%%%%%%%%%%%%%%%%%%%%%%%%%%%%%%%%%%%%%%%%%%%%%%%%%%%%%%%%%%%%%%%%%%%%%%%
%%%%%%%%%%%%%%%%%%%%%%%%%%%%%%%%%%%%%%%%%%%%%%%%%%%%%%%%%%%%%%%%%%%%%%%%%%%%%%%%%%%%%%%%%%%%%%%%%%%%%%%
The main result here is that the variations behave differently for the redshift and for the angular diameter distance. Indeed, while for $z$ the post-Newtonian corrections are two orders of magnitudes smaller than the non-linear Newtonian contributions with respect to linear theory, for $D_{\rm ang}$ the two corrections are of the same order. This can be clearly seen on $k=300 \, \rm Mpc$, Figs. \ref{fig:dz_point_1_k300} and \ref{fig:dD_point_1_k300}, and the same behavior also holds on smaller scales, Figs. \ref{fig:dz_point_1_k100} - \ref{fig:dz_point_1_k30} and \ref{fig:dD_point_1_k100} - \ref{fig:dD_point_1_k30}.
For $z\lesssim 2$ we have that $\Delta z(\rm PN, N)\sim 10^{-6}$ on $k=300\, \rm Mpc$ with oscillation dumped as the redshift increases. On the other hand, for the angular diameter distance $\Delta D_{ \rm ang} (Lin, N) \sim \Delta D_{ \rm ang} (PN, N) \sim 10^{-4}$ on $k=300\, \rm Mpc$ in the full redshift range $[0, \, 10]$. 

\begin{figure}[!ht]
    \centering
    \begin{subfigure}[h]{0.49\textwidth}
        \includegraphics[width=\textwidth]{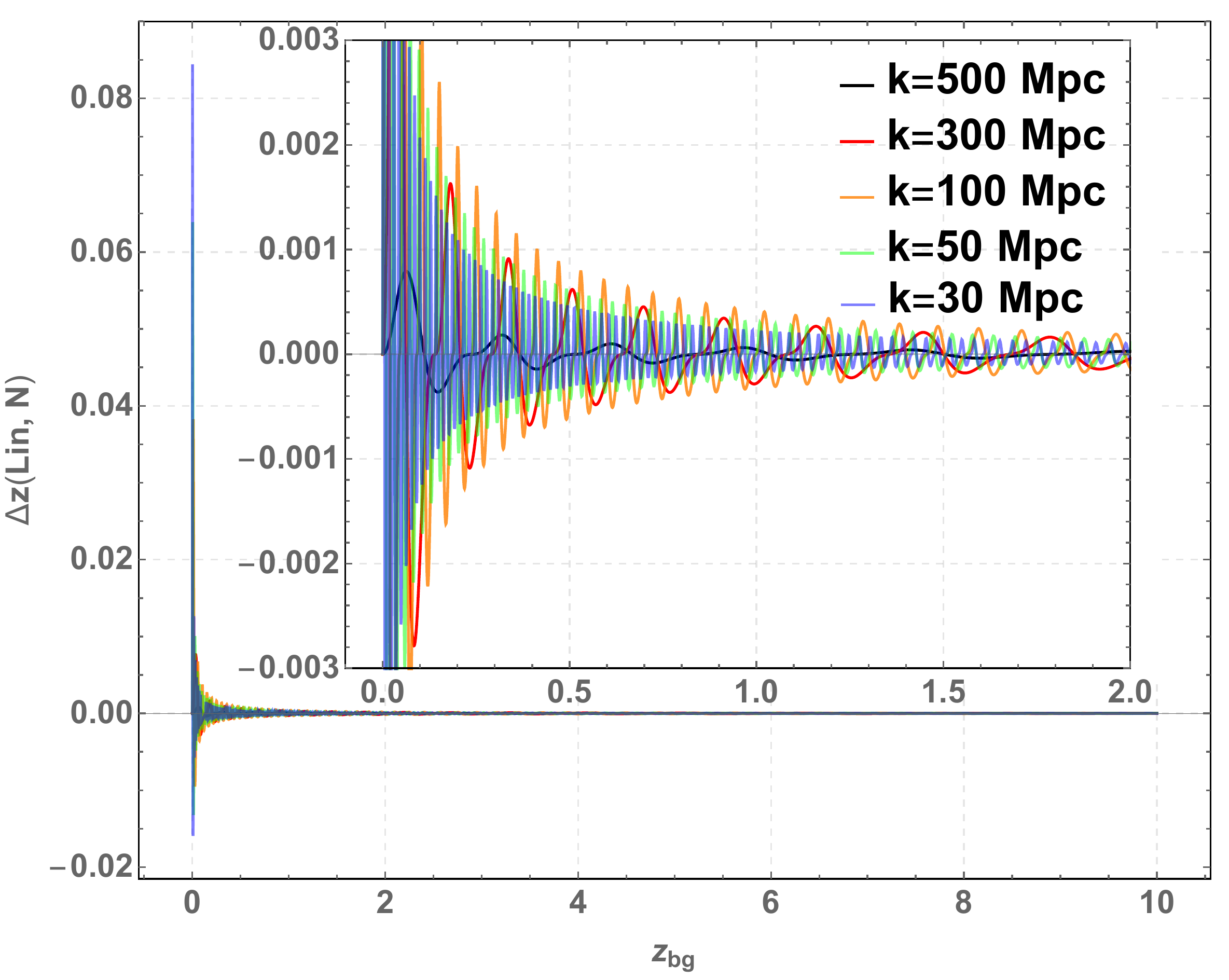}
       \caption{$\Delta z (\rm Lin, N)$ }
        \label{fig:dz_LvsN_point2}
    \end{subfigure}
    \begin{subfigure}[h]{0.49\textwidth}
        \includegraphics[width=\textwidth]{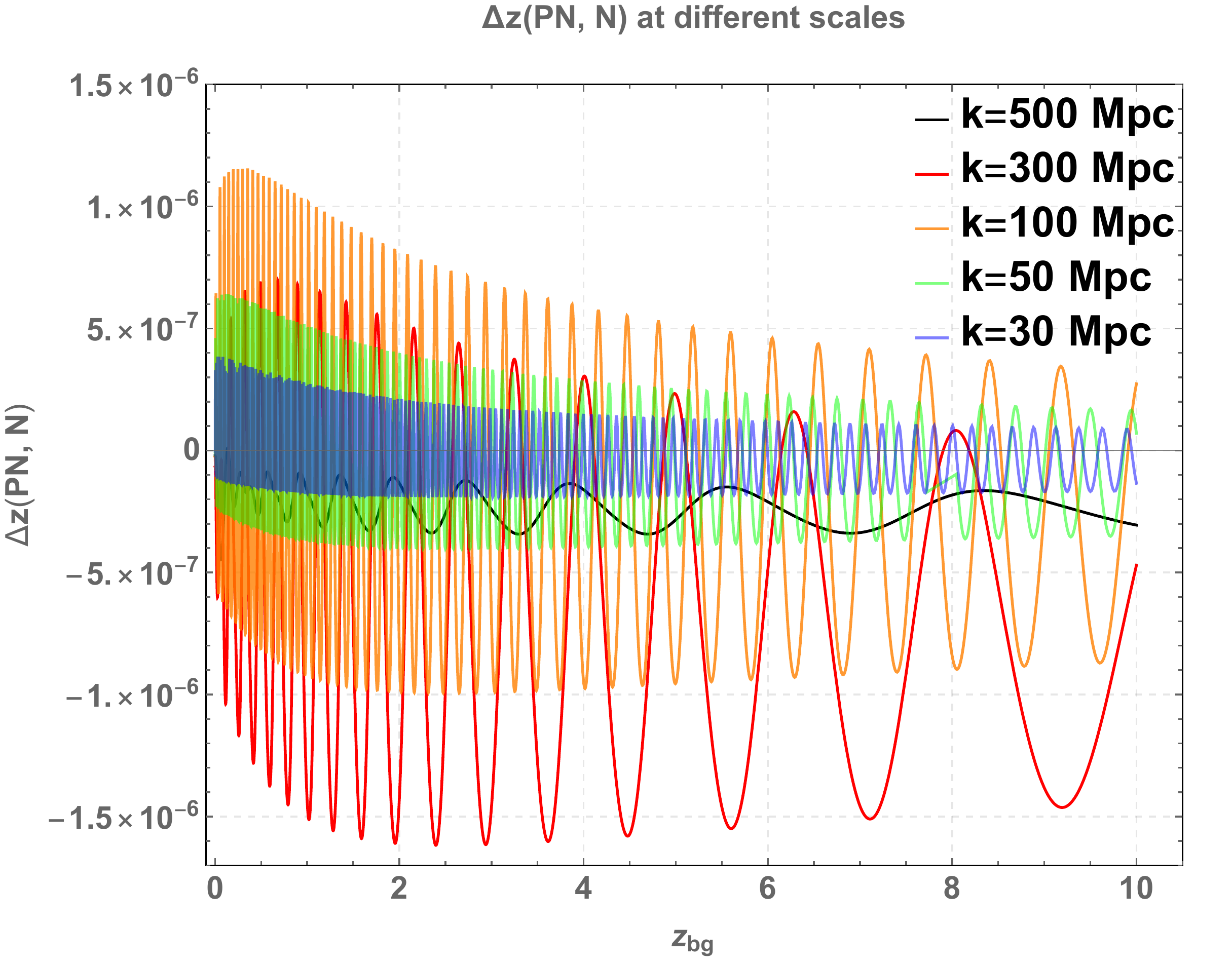}
        \caption{$\Delta z (\rm PN, N)$}
        \label{fig:dz_PNvsN_point2}
    \end{subfigure}
    \caption{Variations $\Delta z (\rm Lin, N)$ and $\Delta z (\rm PN, N)$, as defined in Eq.~\eqref{eq:variation_res}, on different scales in the range $[30, \, 500] \, \rm Mpc$. Both the variations show a maximum around $k=100 \, \rm Mpc$. The variable on the horizontal axis is the $\Lambda CDM$ redshift.}\label{fig:dz_point_2}
\end{figure}

\begin{figure}[!ht]
    \centering
    \begin{subfigure}[h]{0.49\textwidth}
        \includegraphics[width=\textwidth]{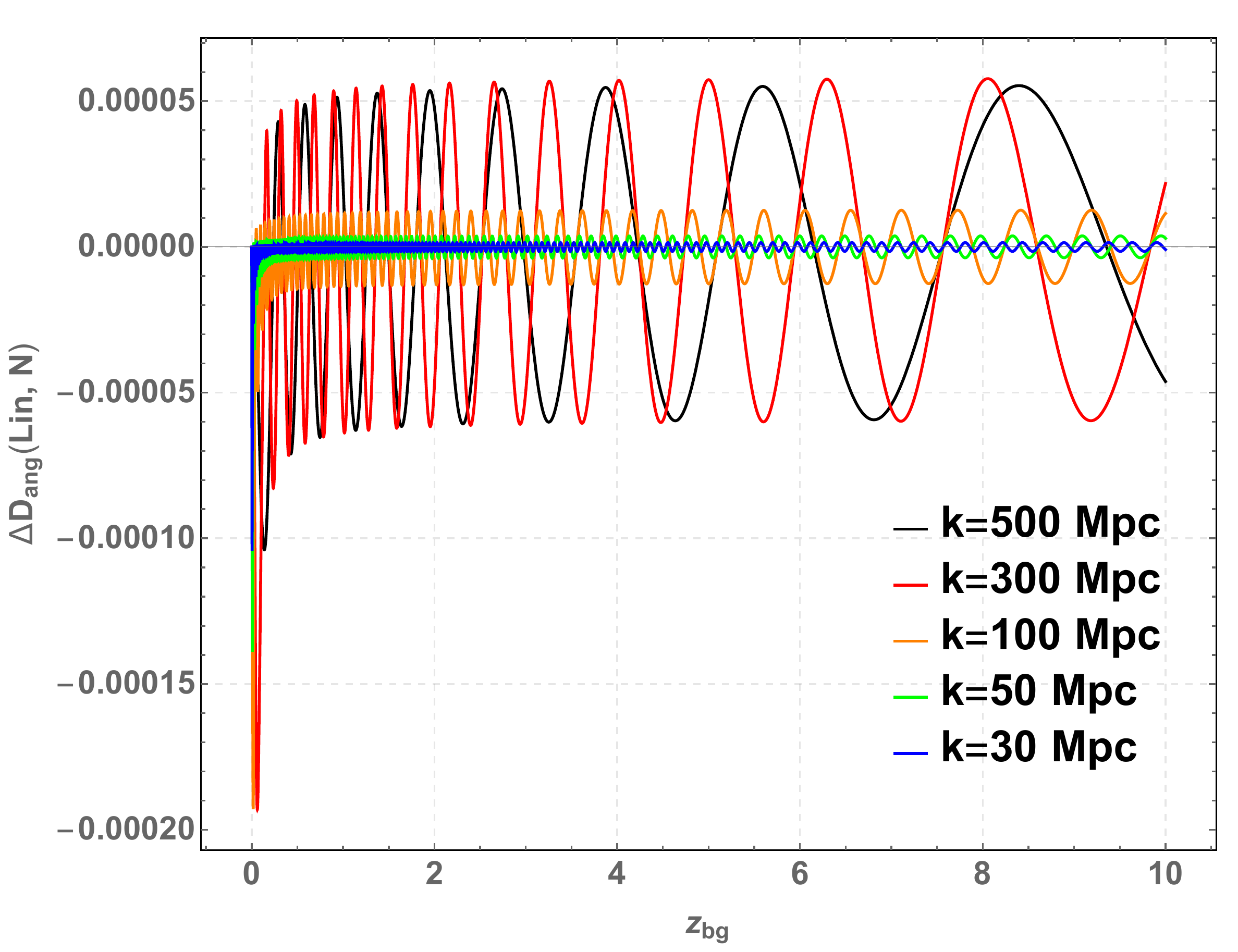}
       \caption{$\Delta D_{\rm ang} (\rm Lin, N)$}
        \label{fig:dD_LvsN_point2}
    \end{subfigure}
    \begin{subfigure}[h]{0.49\textwidth}
        \includegraphics[width=\textwidth]{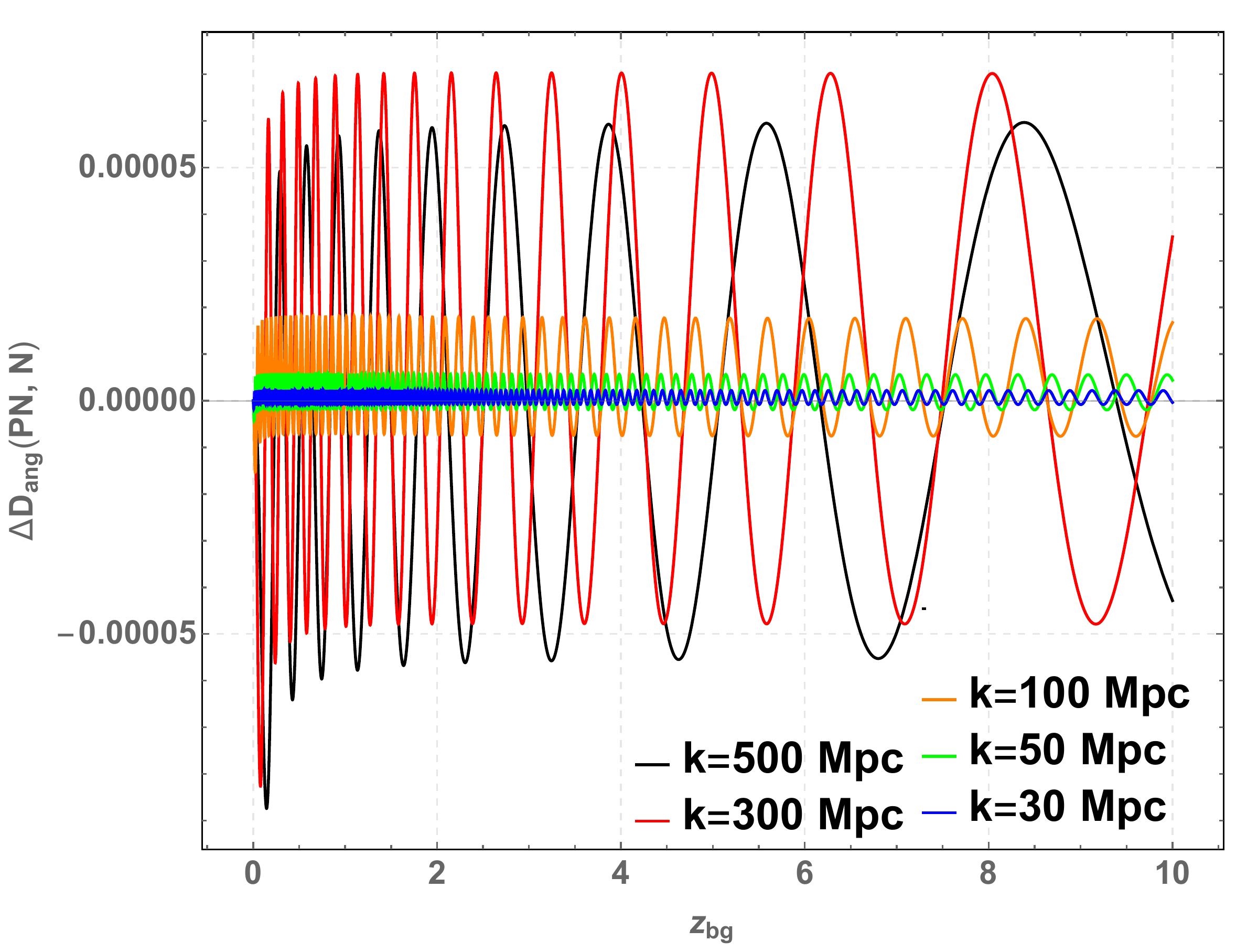}
        \caption{$\Delta D_{\rm ang} (\rm PN, N)$}
        \label{fig:dD_PNvsN_point2}
    \end{subfigure}
    \caption{Variations $\Delta D_{\rm ang} (\rm Lin, N)$ and $\Delta D_{\rm ang} (\rm PN, N)$, as defined in Eq.~\eqref{eq:variation_res}, on different scales in the range $[30, \, 500] \, \rm Mpc$. The amplitudes monotonically decrease as the scale $k$ becomes smaller. The variable on the horizontal axis is the $\Lambda CDM$ redshift.}\label{fig:dD_point_2}
\end{figure}

We dedicate a separate study, reported in Figs. \ref{fig:dz_point_2} and \ref{fig:dD_point_2}, to the change of the amplitude of all the variations with the scale. We consider inhomogeneities scales of $k= \, 500, \, 300, \, 100, \, 50,\, 30\, \rm Mpc$.
%%%%%%%%%%%%%%%%%%%%%%%%%%%%%%%%%%%%%%%%%%%%%%%%%%%%%%%%%%%%%%%%%%%%%%%%%%%%%%%%%%%%%%%%%%%%%%%%%%%%%%%
%%%%%%%%%%%%%%%%%%%%%%%%%%%%%%%%%%%%%%%%%%%%%%%%%%%%%%%%%%%%%%%%%%%%%%%%%%%%%%%%%%%%%%%%%%%%%%%%%%%%%%%
%%%%%%%%%%%%%%%%%%%%%%%%%%%%%%%%%%%%%%%%%%%%%%%%%%%%%%%%%%%%%%%%%%%%%%%%%%%%%%%%%%%%%%%%%%%%%%%%%%%%%%%
Fig. \ref{fig:dz_point_2} shows that both the variations $\Delta z(\rm Lin, \, N)$ and $\Delta z(\rm PN, \, N)$ increase from $k=500\, \rm Mpc$ to reach the maximum amplitude on $k=100 \, \rm Mpc$ and then decreases down to $k=30\, \rm Mpc$. In terms of amplitudes we have: $\Delta z(\rm Lin, \, N) \sim 10^{-4}$, with a maximum $\sim 10^{-3}$ around $k=100\, \rm Mpc$ and $\Delta z(\rm PN, \, N) \sim 10^{-7}$, with a maximum $\sim 10^{-6}$ around $k=100\, \rm Mpc$. We again note that the variations for the redshift are damped as $z$ increases. This is most evident for $\Delta z(\rm Lin, \, N)$.
The angular diameter distance shows in Fig. \ref{fig:dD_point_2} a different behavior: the amplitude of both variations $\Delta D_{\rm ang} (\rm Lin, N)$ and $\Delta D_{\rm ang} (\rm PN, N)$ decreases monotonically as the scale $k$ become smaller. Both the amplitudes start from $\Delta D_{\rm ang} \sim 10^{-4}$ on $k=500 \, \rm Mpc$ and decrease to $10^{-6}$ on $k=30 \, \rm Mpc$.

%%%%%%%%%%%%%%%%%%%%%%%%%%%%%%%%%%%%%%%%%%%%%%%%%%%%%%%%%%%%%%%%%%%%%%%%%%%%%%%%%%%%%%%%%%%%%%%%%%%%%%%
\begin{figure}[!ht]
    \centering
     \begin{subfigure}[h]{0.49\textwidth}
        \includegraphics[width=\textwidth]{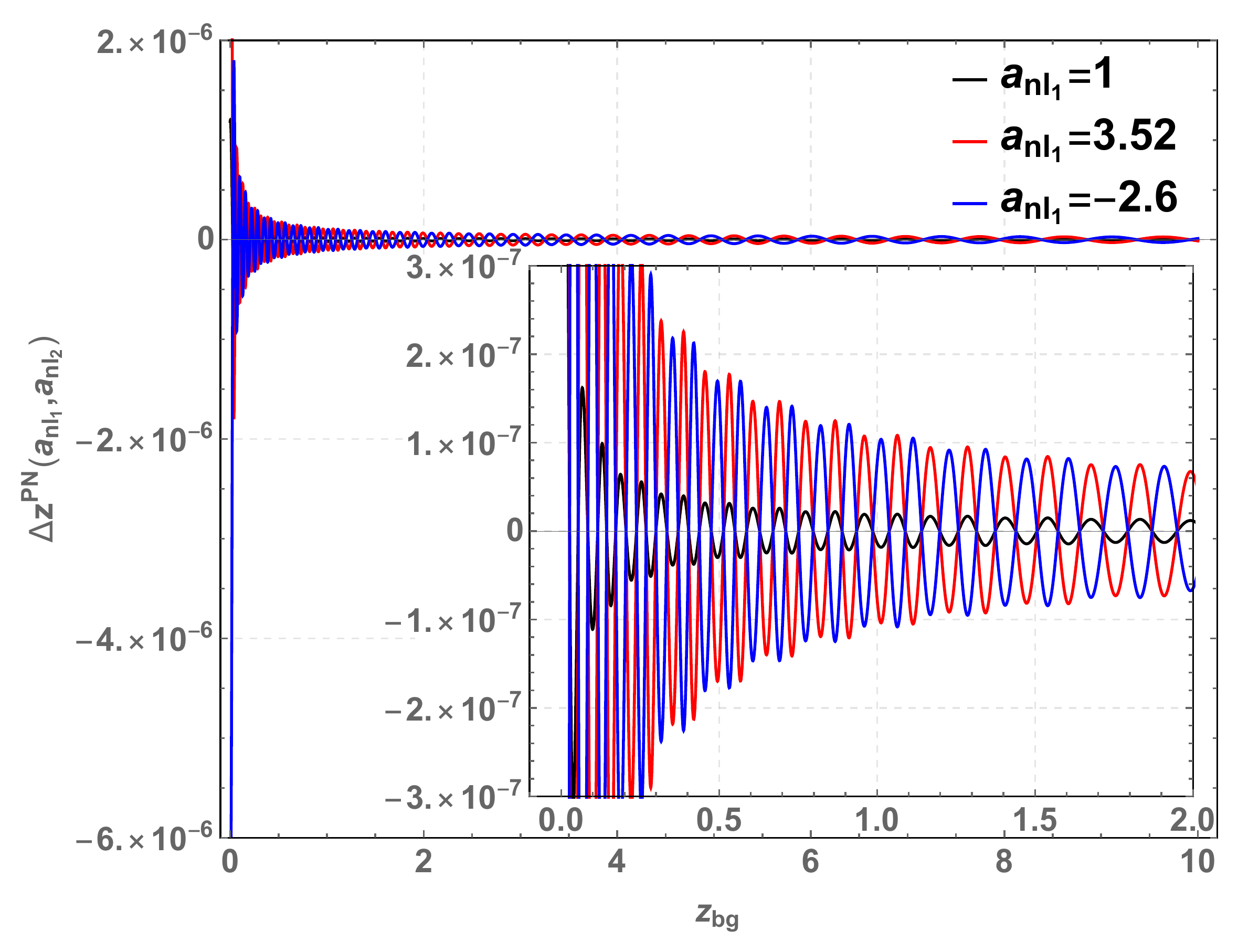}
        \caption{$\Delta z^{\rm PN} (a_{\rm nl_{\rm 1}},\, a_{\rm nl_{\rm 2}})$}
        \label{fig:zPN_a1_vs_a2}
    \end{subfigure}
   \begin{subfigure}[h]{0.49\textwidth}
        \includegraphics[width=\textwidth]{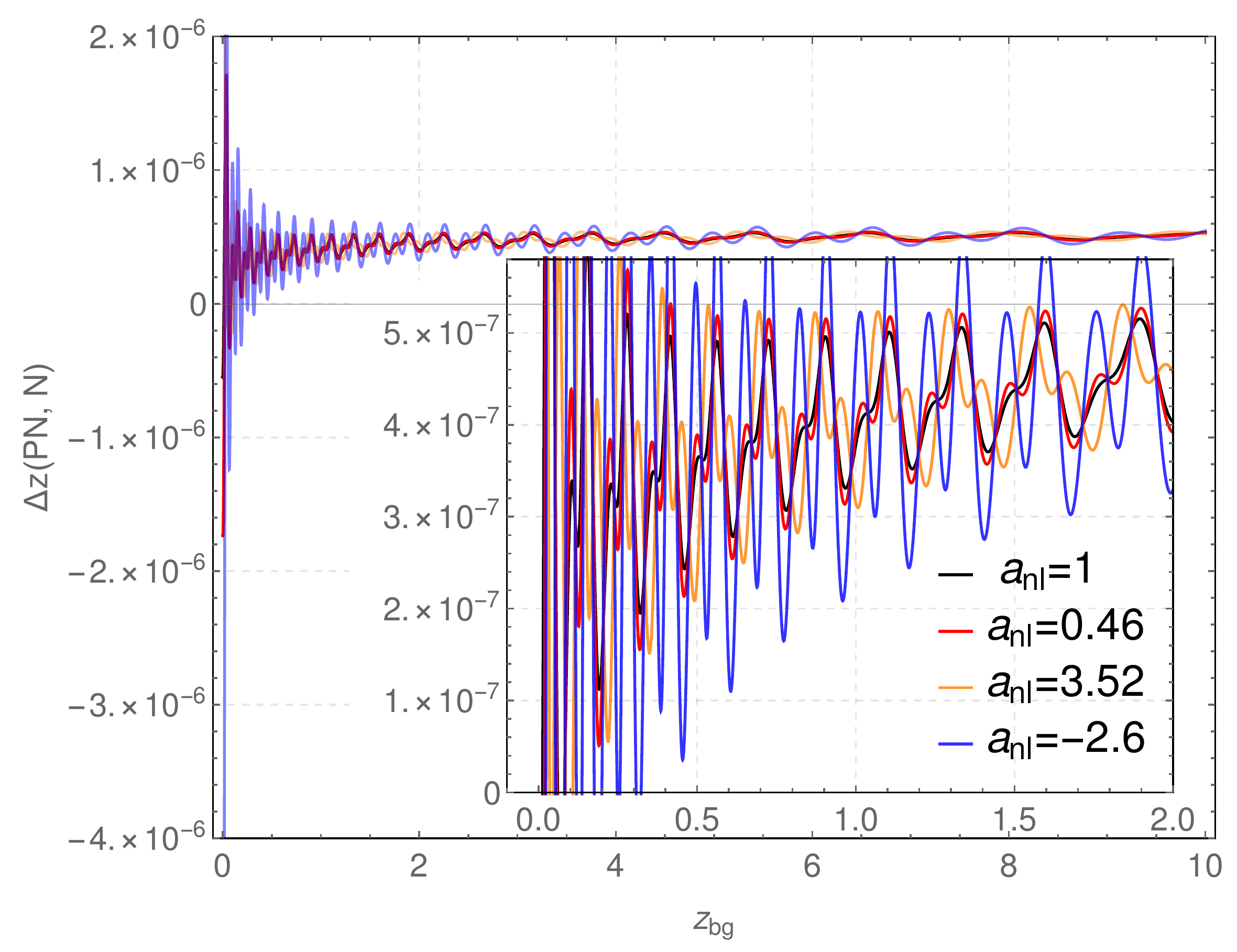}
       \caption{$\Delta z (\rm PN, \, N)$}
        \label{fig:dz_PNvsN_for_anl}
    \end{subfigure}
    \caption{The effect of varying primordial non-Gaussianity for the redshift: the variation in Eq.~\eqref{eq:DeltaO_anls},  \eqref{fig:zPN_a1_vs_a2}, and PN correction for different values of $a_{\rm nl}$, \eqref{fig:dz_PNvsN_for_anl}. We find that $\Delta z^{\rm PN} (a_{\rm nl_{\rm 1}}, a_{\rm nl_{\rm 2}}) \lesssim  \, \Delta z (PN, N)$. The variable on the horizontal axis is the $\Lambda CDM$ redshift.}\label{fig:dz_anl}
\end{figure}
%%%%%%%%%%%%%%%%%%%%%%%%%%%%%%%%%%%%%%%%%%%%%%%%%%%%%%%%%%%%%%%%%%%%%%%%%%%%%%%%%%%%%%%%%%%%%%%%%%%%%%%
\begin{figure}[!ht]
    \centering
    \begin{subfigure}[h]{0.49\textwidth}
        \includegraphics[width=\textwidth]{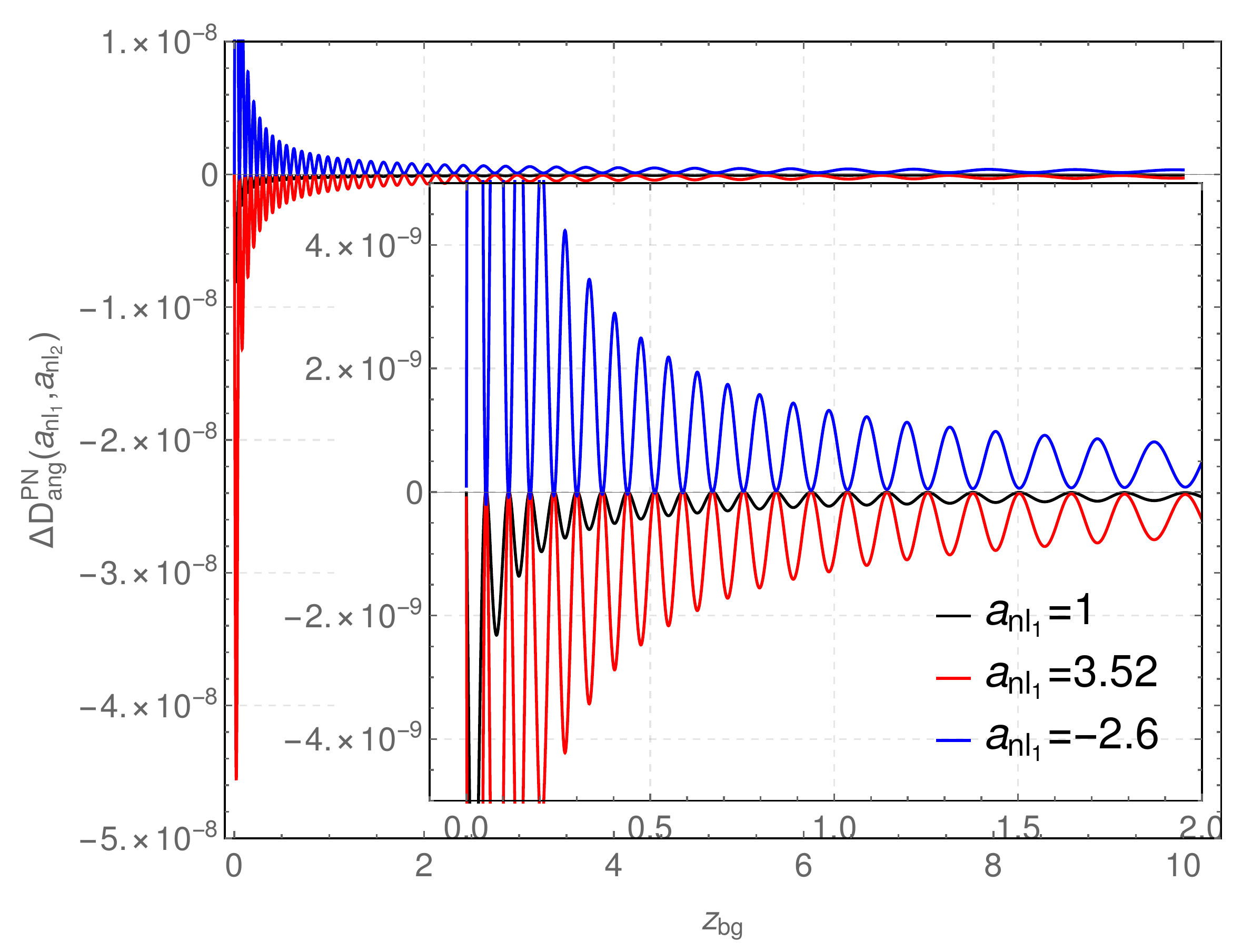}
        \caption{ $\Delta D_{\rm ang}^{\rm PN} (a_{\rm nl_{\rm 1}},\, a_{\rm nl_{\rm 2}})$}
        \label{fig:DPN_a1_vs_a2}
    \end{subfigure}
    \begin{subfigure}[h]{0.49\textwidth}
        \includegraphics[width=\textwidth]{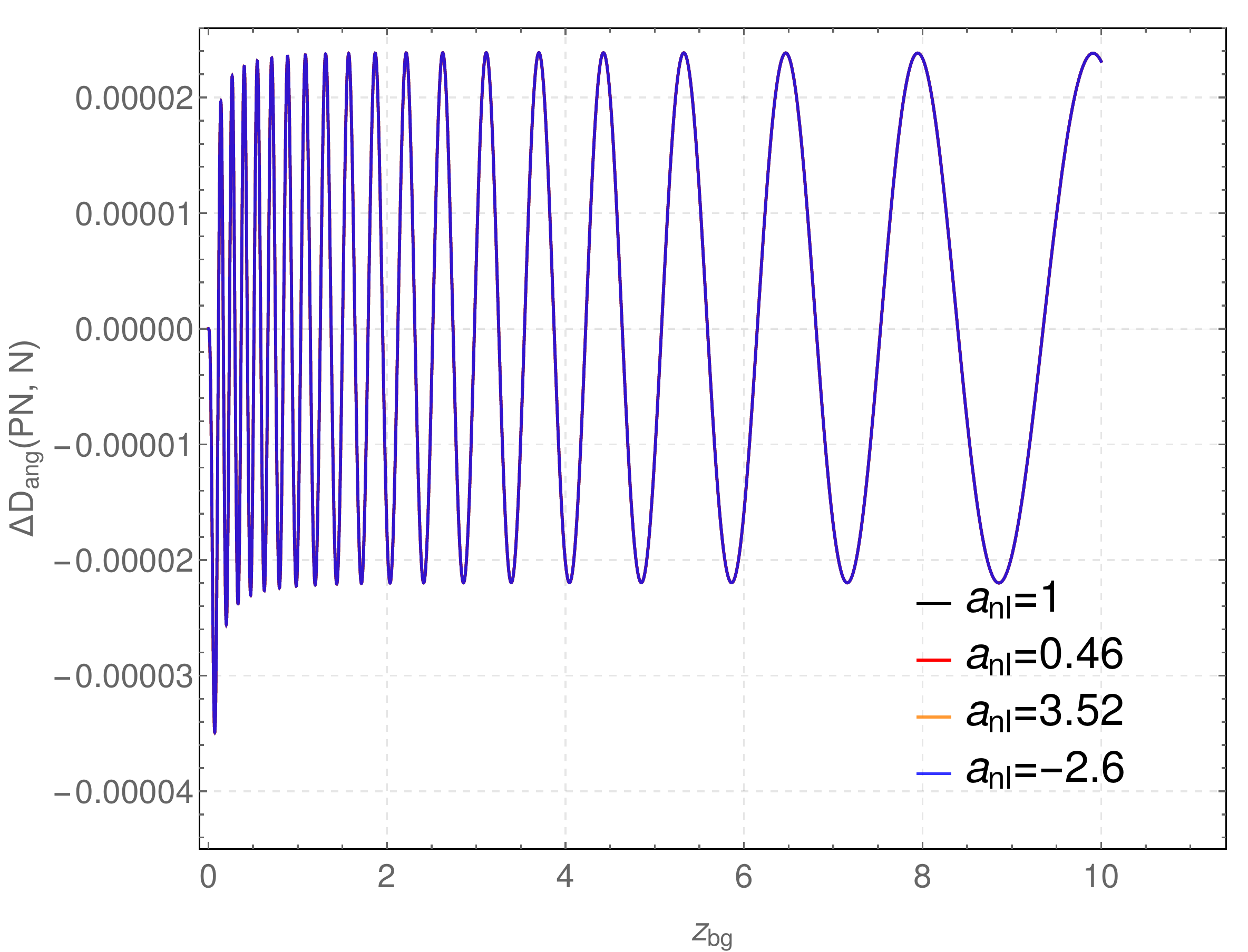}
       \caption{$\Delta D_{\rm ang}(\rm PN, \, N)$}
        \label{fig:dD_PNvsN_for_anl}
    \end{subfigure}
    \caption{The effect of varying primordial non-Gaussianity for the angular diameter distance: the variation in Eq.~\eqref{eq:DeltaO_anls},  \eqref{fig:DPN_a1_vs_a2},  and PN correction for different values of $a_{\rm nl}$, \eqref{fig:dD_PNvsN_for_anl}. We find that $\Delta D_{\rm ang}^{\rm PN} (a_{\rm nl_{\rm 1}}, a_{\rm nl_{\rm 2}}) \sim 10^{-4}  \, \Delta D_{\rm ang} (PN, N)$. The variable on the horizontal axis is the $\Lambda CDM$ redshift.}\label{fig:dD_anl}
\end{figure}
%%%%%%%%%%%%%%%%%%%%%%%%%%%%%%%%%%%%%%%%%%%%%%%%%%%%%%%%%%%%%%%%%%%%%%%%%%%%%%%%%%%%%%%%%%%%%%%%%%%%%%%
As we mentioned in Sec. \ref{sec:method}, different values of the primordial non-Gaussianity parameter $a_{\rm nl}$ tune some of the post-Newtonian terms in \eqref{eq:metricPN}, e.g. a perfect Gaussian initial perturbation ($a_{\rm nl}=1$) cancels out the third term in the PN part of the metric \eqref{eq:metricPN}. We then decided to quantify how the PN observables change when we vary the values of $a_{\rm nl}$ inside the condfidence interval measured by Planck, \cite{planck2019anl}. For this analysis, we choose $a_{\rm nl}= 1, 0.46, -2.6, 3.52$, corresponding to Gaussian perturbations ($a_{\rm nl}= 1$), Planck 2018 fiducial value ($a_{\rm nl}= 0.46$), and extremes of confidence interval ($a_{\rm nl}= -2.6, 3.52$). We start the discussion of our results by looking at Figs. \ref{fig:zPN_a1_vs_a2} and \ref{fig:DPN_a1_vs_a2}, in which we plot for the PN observables the quantity
\begin{equation}
\Delta O^{\rm PN} (a_{\rm nl_{\rm 1}}, a_{\rm nl_{\rm 2}})= \frac{O^{\rm{PN}}_{a_{\rm nl_{\rm 1}}}-O^{\rm{PN}}_{a_{\rm nl_{\rm 2}}}}{O^{\rm{PN}}_{a_{\rm nl_{\rm 2}}}}\, ,
\label{eq:DeltaO_anls}
\end{equation} 
where we fix $a_{\rm nl_{\rm 2}}$ to the Planck best-fit value and we vary $a_{\rm nl_{\rm 1}}$.
The effect is different for the redshift and for the angular diameter distance: the variation in $D_{\rm ang}^{\rm PN}$ is $\sim 10^{-9}$, two orders of magnitude smaller than the one in $z^{\rm PN}$. This very difference is evident when we plot the PN corrections for different values of $a_{\rm nl}$, see Figs. \ref{fig:dz_PNvsN_for_anl} and \ref{fig:dD_PNvsN_for_anl}. The effect of tuning primordial non-Gaussianity is roughly of the same order as the PN correction for the redshift and also changes its shape. On the contrary, $D_{\rm ang}$ is completely insensitive to the variation of the non-Gaussianity parameter, since $\Delta D_{\rm ang}^{\rm PN} (a_{\rm nl_{\rm 1}}, a_{\rm nl_{\rm 2}}) \sim 10^{-4}  \, \Delta D_{\rm ang} (PN, N)$.

To conclude our analysis, we isolate and quantify the contribution of the linear PN initial seed proportional to the gravitational potential, i.e. $\gamma_{i j}= - \frac{10}{3 c^2}\phi_{\rm 0} \delta_{i j}$ in the spacetime metric \eqref{eq:metricPN}. To do so, we have computed the angular diameter distance starting from the Newtonian metric plus the initial seed, i.e.
\begin{equation} \label{eq:metricNt} 
\begin{split}
\gamma^{\tilde{\rm{N}}}_{11} =& \left(1-\frac{2}{3}\frac{ \mathcal{D} \partial_{\rm q_{\rm 1}}^2 \phi_0}{ \stuff} \right)^2 - \frac{10}{3 c^2}\phi_{\rm 0} \\
\gamma^{\tilde{\rm{N}}}_{22}= & 1 - \frac{10}{3 c^2}\phi_{\rm 0}\\
\gamma^{\tilde{\rm{N}}}_{33}=& 1 - \frac{10}{3 c^2}\phi_{\rm 0}\, ,
 \end{split}
\end{equation}
and we have compared the result with the Newtonian $D_{\rm ang}^{\rm N}$ calculated from \eqref{eq:metricNWT}.
%%%%%%%%%%%%%%%%%%%%%%%%%%%%%%%%%%%%%%%%%%%%%%%%%%%%%%%%%%%%%%%%%%%%%%%%%%%%%%%%%%%%%%%%%%%%%%%%%%%%%%%
\begin{figure}[!ht]
    \centering
     \begin{subfigure}[h]{0.49\textwidth}
        \includegraphics[width=\textwidth]{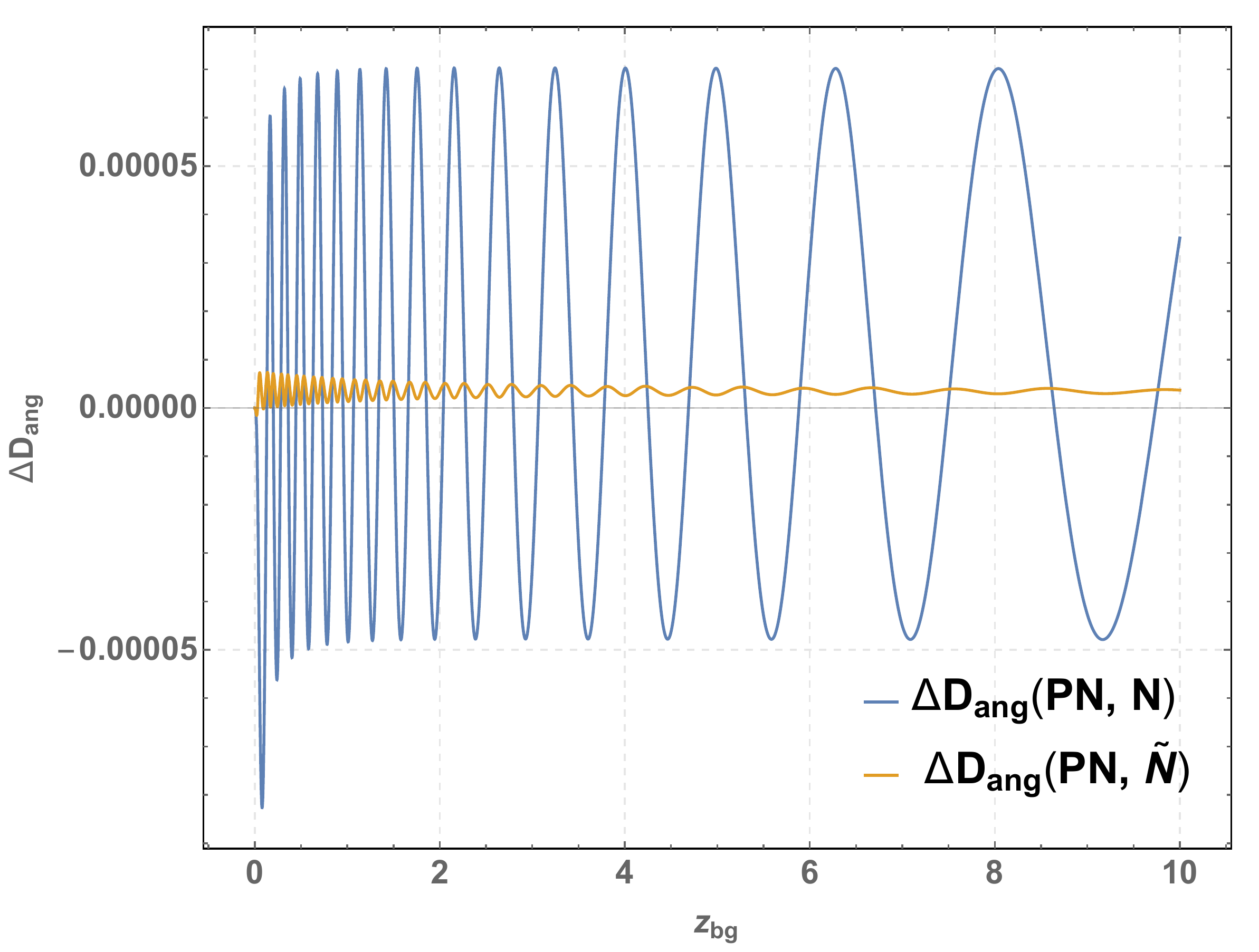}
        \caption{Comparison PN vs N (blue) and PN vs $\tilde{N}$ (orange) for the angular diameter distance, as defined in Eq.~\eqref{eq:variation_res}, on $ k= 300 \, \rm Mpc$. The variation $\Delta D_{ \rm ang} (PN, \tilde{N}) \sim \, 10^{-6}$ is two orders of magnitude smaller than $\Delta D_{\rm ang} (PN, N)$. The variable on the horizontal axis is the $\Lambda CDM$ redshift.}
        \label{fig:dD_PNvsRN_k=300}
    \end{subfigure}
    \\
    \begin{subfigure}[h]{0.49\textwidth}
        \includegraphics[width=\textwidth]{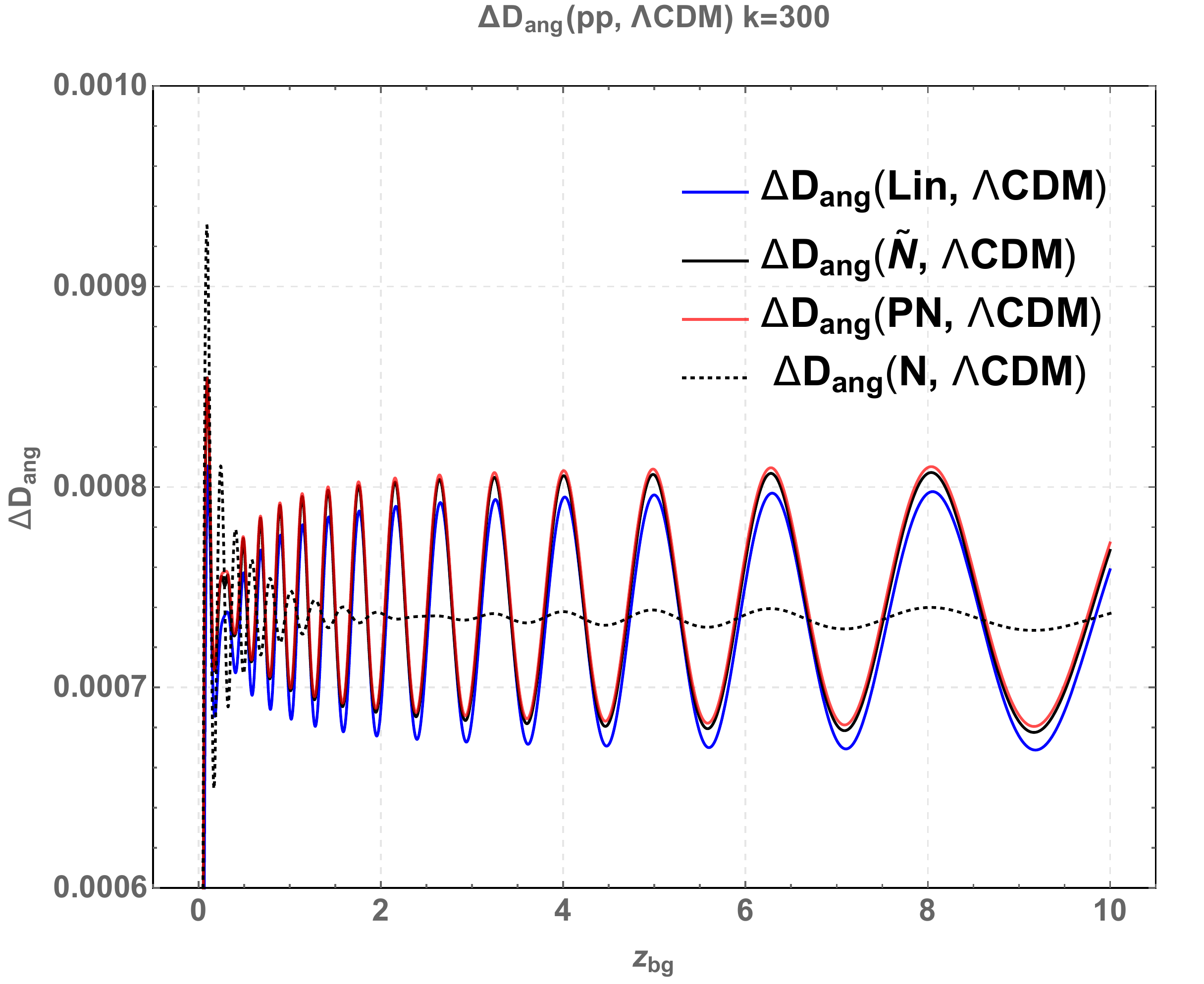}
       \caption{Comparison between the different approximations $\rm Lin$, N, $\tilde{N}$, PN and the $\Lambda CDM$ background for the angular diameter distance.}
        \label{fig:dD_PPvsLCDM_withRN_k=300}
    \end{subfigure}
    \caption{Results for the contribution of the initial seeds to the angular diameter distance. For the definition of $\tilde{N}$ see Eq. \eqref{eq:metricNt}.}\label{fig:point_3_LCDM}
\end{figure}
%%%%%%%%%%%%%%%%%%%%%%%%%%%%%%%%%%%%%%%%%%%%%%%%%%%%%%%%%%%%%%%%%%%%%%%%%%%%%%%%%%%%%%%%%%%%%%%%%%%%%%%
The inclusion of the initial seed in the modified Newtonian model is such that the PN variation is reduced by two orders of magnitude, see Fig. \ref{fig:dD_PNvsRN_k=300}. In other words the initial seed is the leading order of the post-Newtonian correction. The effect is even more evident when we consider the variations of each of the approximation $\rm Lin$, N, $\tilde{N}$, PN respect to the $\Lambda CDM$ background: we can clearly distinguish between the two approximations, observing that $\tilde{N}$ behaves as expected very close to the PN approximation.

%%%%%%%%%%%%%%%%%%%%%%%%%%%%%%%%%%%%%%%%%%%%%%%%%%%%%%%%%%%%%%%%%%%%%%%
%%%%%%%%%%%%%%%%%%%%%%%%%%%%%%%%%%%%%%%%%%%%%%%%%%%%%%%%%%%%%%%%%%%%%%%
\section{Conclusions} \label{sec:concl}
%%%%%%%%%%%%%%%%%%%%%%%%%%%%%%%%%%%%%%%%%%%%%%%%%%%%%%%%%%%%%%%%%%%%%%%
%%%%%%%%%%%%%%%%%%%%%%%%%%%%%%%%%%%%%%%%%%%%%%%%%%%%%%%%%%%%%%%%%%%%%%%
In this paper we use the new BGO framework for light propagation in General Relativity, presented in \cite{Grasso:2018mei} and applied to $\Lambda$CDM cosmology in \cite{Korzynski:2019oal}. We encoded the new framework in the {\tt Mathematica} package called {\tt BiGONLight} ({\color{blue}{\tt {https://github.com/MicGrasso/bigonlight1.0}}}) that is designed to compute optical observables numerically, once the spacetime metric components and the observer and source kinematics are provided as input. The code is adaptable to work in any gauge and with analytical as well as numerical inputs. A short description of the package is given in App. \ref{apx:bigonlight} here and we will give a more extensive discussion on {\tt BiGONLight} in \cite{Grasso:BGO}.
In the present work we focus on two observables in the cosmological context: redshift and angular diameter distance. We concentrate our analysis on a one-dimensional toy model in which the density perturbations around the $\Lambda$CDM background are distributed along parallel planes. In other words our perturbations depend on time and one spatial coordinate only. The purpose of our investigation is to isolate the contribution of non-linearities by considering the relative differences in the observables $\Delta O$, as defined in Eq. \eqref{eq:variation_res}, computed within three approximations: linear cosmological perturbation theory, Newtonian and post-Newtonian approximation. Although the plane-parallel universe is a simple model, let us remark that the spacetime metric in Eq.\eqref{eq:metricPN} is particularly well-suited for this kind of analysis, since the terms coming from all the three approximations are clearly identified and they can be directly used as input in our code for light propagation. 

We now present our findings relative to the different features that we examined: the dependence on the scale of perturbations, on primordial non-Gaussianity and the role of initial conditions.
We start by pointing out that the redshift and the angular diameter distance have different behaviour: for the redshift the Newtonian corrections are the leading order for the non-linearities with $\Delta z (Lin, N) \sim 10^2 \Delta z (PN, N)$, see Fig. \ref{fig:dz_point_1}, whereas for the angular diameter distance Newtonian and post-Newtonian contributions are of the same order $\Delta D_{ \rm ang} (Lin, N) \sim \Delta D_{\rm ang} (PN, N)$, as clear from Fig. \ref{fig:dD_point_1}. Our results confirm previous studies in the literature \citep{adamek2014distance}, in particular we have found that the non-linearities from the approximations we considered are well below $1 \%$ for both the observables. To be more precise, we have that the variation linear vs Newtonian approximations is of order $10^{-3}$ for $z$ and $10^{-5}$ for $D_{\rm ang}$, while the variation post-Newtonian vs Newtonian is of order $10^{-6}$ for $z$ and $10^{-5}$ for $D_{\rm ang}$, both on $k=100\, {\rm Mpc}$ and $\delta_0^{\rm max}=1$.

In addition, we analysed the dependence of the various contributions on the inhomogeneities scale, finding again a slightly different trend for $z$ and $D_{\rm ang}$, see Figs. \ref{fig:dz_point_2} and \ref{fig:dD_point_2}. For both observables, the change in the scale modifies the amplitude of the oscillations by more than one order of magnitude, but the amplitude of $\Delta D_{\rm ang}$ decreases monotonically with the scale $k$, while that of $\Delta z$ has a maximum around $k=100\, {\rm Mpc}$.
For the angular diameter distance the contributions from non-linearities span from $10^{-4}$ on $k=500\, {\rm Mpc}$ to $10^{-6}$ on $k=30\, {\rm Mpc}$.

To complete our investigation, we took advantage of having an analytical expression of our input, i.e. the spacetime metric.
As one can easily verify, some of the PN terms are triggered or cancelled out for specific values of the primordial non-Gaussianity parameter $a_{\rm nl}$ inside Planck confidence interval. Therefore, we decided to examine the response of the post-Newtonian observables to the variation of $a_{\rm nl}$. It turned out that the tuning of primordial non-Gaussianity has negligible effects on the PN observables. The only effect is the change in the shape of the PN part of the redshift but the net contribution is of the order of $10^{-6}$ see Figs. \ref{fig:dz_anl} and \ref{fig:dD_anl}.

Finally, we have estimated the relative variations of the three approximations with respect to the $\Lambda$CDM background in Fig. \ref{fig:dD_PPvsLCDM_withRN_k=300}. We find that the post-Newtonian contribution to $D_{\rm ang}$ comes almost exclusively from the linear post-Newtonian initial seed. The other, i.e. non-linear, post-Newtonian corrections are below $1\%$, in agreement with previous results in the literature.

\section*{Acknowledgements}

This work was supported by the National Science Centre, Poland (NCN) via the SONATA BIS programme, grant No~2016/22/E/ST9/00578 for the project
\emph{``Local relativistic perturbative framework in hydrodynamics and General Relativity and its application to cosmology''}.

%%%%%%%%%%%%%%%%%%%%%%%%%%%%%%%%%%%%%%%%%%%%%%%%%%%%%%%%%%%%%%%%%%%%%%%%%%%%%
%%%%%%%%%%%%%%%%%%%%%%%%%%%%%%%%%%%%%%%%%%%%%%%%%%%%%%%%%%%%%%%%%%%%%%%%%%%%%
\appendix

\section{\texttt{BiGONLight}: presenting and testing the code}
\label{apx:bigonlight}

In Sec.~\ref{sec:lightprop} we have introduced the BGO formalism, emphasizing that it provides a unified framework to compute all possible optical observables and how it extends the Sachs formalism: it includes the case of observations occurring for a prolonged period of time, when repeated observations are made, e.g. parallax and drifts. In this appendix, we present the  \texttt{BiGONLight.m} package used in this paper to calculate the Newtonian and post-Newtonian observables numerically within the BGO framework.
The main achievement of our package is to simulate light propagation in numerical relativity to extract observables.
\texttt{BiGONLight} works in any gauge and with any coordinate system and it requires the spacetime metric components and the source and observer kinematics (four-velocities and four-accelerations) as input\footnote{Note that the input is not in full tensorial form, but in the form of components. Let us remark that, of course, once we give the metric and the four-velocities components in practice we are making a gauge and a coordinate system choice. Nevertheless, the code can work with any choice.}. The flexibility of the code allows us to use two types of inputs: analytic expressions or the output of a (relativistic) numerical simulation.
In fact, in order to make our code compatible with the majority of the codes in numerical relativity, we recasted the BGO framework in $3+1$ formalism.
We decided to develop \texttt{BiGONLight} within \texttt{Mathematica} for several reasons. One of the advantages is that one can choose between a large variety of numerical methods to solve ODE without the need of modifying the code. Another useful feature in \texttt{Mathematica} are the quite detailed build-in precision control options, which allow the user to set precision and accuracy of the numerical calculations efficiently.

We devote a companion paper, \cite{Grasso:BGO}, to the comprehensive description of the modules of package, including also benchmark testing performed in the context of several exact and simulated cosmological models and addressing more obsevables than in the present work. In the rest of this appendix, we report the tests for the angular diameter distance only in the well-known $\Lambda$CDM and Szekeres models.

\subsection{$\Lambda$CDM model}
Let us start by considering the test in the flat $\Lambda$CDM model. It consists of a universe filled with a cosmological constant $\Lambda$ and an homogeneous and isotropic distribution of non-interacting matter (dust) and it represents the background model of all the approximations used in this work.

We test our code by computing the variation between the angular diameter distance calculated numerically using the \texttt{BiGONLight.m} package and its analytic expression. We plot our results in terms of
\begin{equation}
\Delta D_{\rm ang} (\rm BGO, an) = \frac{D_{\rm ang}^{\rm BGO} - D_{ \rm ang}^{\rm an}}{D_{ \rm ang}^{\rm an}}\,.
\end{equation}
The angular diameter distance is defined as
\begin{equation}
D_{ \rm ang}=\frac{D_{com}}{1+z}
\end{equation}
where the comoving distance $D_{com}$ in the flat $\Lambda$CDM model is
\begin{equation}
D_{com}(z)= \int^z_0 \frac{d z'}{\mathcal{H}_{\rm 0} \, \sqrt{\Omega_{\rm m0}(1+z')^3+\Omega_{\rm \Lambda}}}\, ,
\end{equation}
where $\mathcal{H}_{\rm 0}=\left. \frac{\dot{a}}{a} \right|_{\eta_{\rm 0}}$ and $\Omega_{\rm m0}+ \Omega_{\Lambda}=1$.
By solving the integral, \cite{gradshteyn2014table}, the analytic expression of the angular diameter distance is:
\begin{equation}
D^{\rm an}_{\rm ang}(z)= \frac{
{\rm F}\big[ \chi(z) |\mathit{r} \big] - {\rm F}\big[ \chi(0)|\mathit{r} \big]}{(1+z) \mathcal{H}_{\rm 0} (\Omega_{\rm m0})^{\frac{1}{3}}(\Omega_{\rm \Lambda})^{\frac{1}{6}} 3^{\frac{1}{4}}}
\label{eq:Dang(z)_LCDM}
\end{equation}
where ${\rm F}\big[ \chi(z) | \mathit{r} \big]$ is the elliptic integral of the first kind, with arguments $\mathit{r}=\sqrt{\frac{2+\sqrt{3}}{4}}$ and $\chi(z)=\arccos\left( \frac{2\sqrt{3}}{1+\sqrt{3}+(1+z)\sqrt[3]{\frac{\Omega_{\rm m0}}{\Omega_{\rm \Lambda}}}}-1\right)$.
\begin{figure}
    \centering
        \includegraphics[width=\columnwidth]{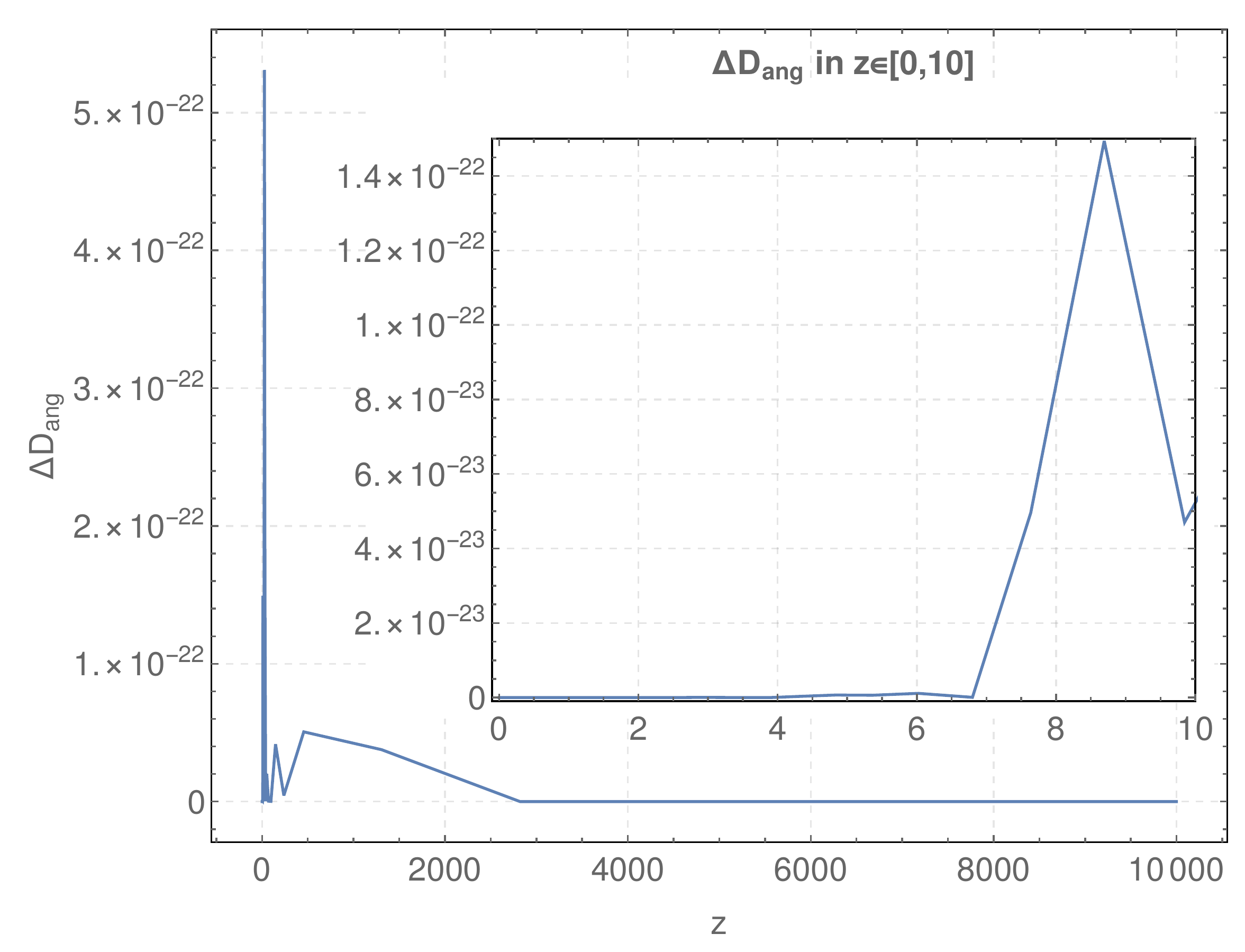}
        \caption{$ \Delta D_{\rm ang}(\mathrm{BGO}, \mathrm{an})$ in the $\Lambda$CDM model. The light geodesic is traced backwards in time up to $\eta_{\rm in}\approx \frac{1}{100}\eta_{\rm 0}= 5.1738\times 10^{8} {\rm yr}$. The values for the cosmological parameters $\Omega_{\rm m0}=0.3153$, $\Omega_{\rm \Lambda}=0.6847$ and $\mathcal{H}_{\rm 0}=67.36$ are taken from Planck~\cite{planck2018param}.}
        \label{fig:LCDM_dDa}
\end{figure}
The plot in Fig.~\ref{fig:LCDM_dDa} shows a deviation of the order of $10^{-22}$ between the numerical and the analytical calculation, highlighting the high precision reached by our code. Such a precision was possible thanks to the precision control options implemented in \texttt{Mathematica}.
We plot up to $z=10000$ to show that the deviation stays small over the whole simulation.

\subsection{Szekeres model}
As second test-bed for the code, we decided to use a more complicated spacetime. We chose the inhomogeneous dust Szekeres model plus a cosmological constant as presented in~\cite{Meures:2011ke}, discussed in Appendix~\ref{par:crfSzekeres} and briefly summarised here. The line element of the model is
\begin{equation}
ds^2=a(\eta)^2 \left[- d\eta^2 + X(\eta,q_{\rm 1},q_{\rm 2},q_{\rm 3})^2dq_{\rm 1}^2+dq_{\rm 2}^2 +dq_{\rm 3}^2\right]
\label{eq:Sz_line}
\end{equation}
In particular, here we consider the case with axial symmetry around the $q_{\rm 1}$ axis in which the function $X$ has the form
\begin{equation}
X(\eta, q_{\rm 1},q_{\rm 2},q_{\rm 3})=1+\beta_{\rm +}(q_{\rm 1}) \mathcal{D}(\eta)+\beta_{\rm +}(q_{\rm 1}) B \left(q_{\rm 2}^2+q_{\rm 3}^2\right)
\end{equation}
with the constant $B$ given by \eqref{eq:link_on_B}. 
The function $\beta_{\rm +}$ is the free function of the model and it is linked with the gravitational potential $\phi_{\rm 0}$ via
\begin{equation}
\beta_{\rm +}=-\frac{2}{3}\frac{\nabla^2\phi_{\rm 0}}{\stuff}
\label{eq:b+_app}
\end{equation}
as we show explicitly in Appendix~\ref{par:crfSzekeres}.
For our test, we set up $\beta_{+}$ using Eq. \eqref{eq:b+_app} with $\phi_{\rm 0}=\mathcal{I}\sin(\omega \, q_{\rm 1})$, where $\omega=\frac{2 \pi}{500\, {\rm Mpc}}$, the amplitude $\mathcal{I}$ is determined such that ${\rm Max} \big( \delta^{Sz}_{\rm 0} \big)=0.1$ and $\Omega_{\rm m0}=0.3153$ and $\mathcal{H}_{\rm 0}=67.36$ are taken from Planck~\cite{planck2018param}. Contrary to the $\Lambda$CDM case, where the spatial orientation of the geodesic is irrelevant due to the intrinsic homogeneity and isotropy of the model, in the inhomogeneous Szekeres model the light propagates differently in different directions. In order to facilitate the comparison with the literature, we decided to follow Ref.~\citep{Meures:2011gp} and consider geodesics traveling along the symmetry axis ${\rm q_{\rm 1}}$. The observer is placed at $q^{\mu}_{\rm \mathcal{O}}=(\eta_{\rm \mathcal{O}},0,0,0)$ such that $\delta|_{\rm \mathcal{O}}=0$.

The testing procedure for the \texttt{BiGONLight.m} package in this case is to compare the angular diameter distance calculated numerically using the BGO formalism implemented in \texttt{BiGONLight.m} and numerically, as well, but solving the Sachs focusing equation
\footnote{Here, the Ricci part of the optical tidal matrix is substitute using the Einstein equation $R_{\mu \nu}\ell^{\mu}\ell^{\nu}=\frac{3\mathcal{H}_{\rm 0}^2\Omega_{\rm m0}}{a}(\delta+1)(\ell^{0})^2$.}:
\begin{equation}
\begin{array}{l}
\ddot{D}_{\rm ang}+\frac{\dot{\tilde{\ell}}^0}{\tilde{\ell}^0} \dot{D}_{\rm ang}=-\frac{1}{\tilde{\ell}^{0^2}} \left(|\tilde{\sigma}|^2 + \frac{3}{2}\frac{\mathcal{H}_{\rm 0}^2\Omega_{\rm m0}}{a}(\delta+1) \right)D_{\rm ang}\\
\tilde{\sigma}=\frac{\tilde{\sigma}_{\mathcal{O}} \left.D^2_{\rm ang}\right|_{\mathcal{O}}}{D^2_{\rm ang}}\,,
\end{array}
\label{eq:Szekeres_Da}
\end{equation}
and the initial conditions are given considering that the light bundle has a vertex at the observation point and such that: 
\begin{align}
\tilde{\sigma}_{\rm \calO}=&0 \\
D_{\rm ang} \left|_{\rm \calO}\right.=&0 \\
\dot{D}_{\rm ang} \left|_{\rm \calO}\right.=&\frac{g_{\mu \nu} \ell^{\mu} u^{\nu}\left|_{\mathcal{O}}\right.}{\ell^0_{\rm \calO}}\,.
\end{align}
In the above the dot indicates derivative with respect to conformal time and $\tilde{\sigma}$ is the complex shear\footnote{In general, the equation for the shear contains an additive term $\Psi_{0}$, which is the Weyl focusing term. However, in the Szekeres model we have that $\Psi_{0}=0$, as shown in \cite{Meures:2011gp}.}. All we need to solve Eq.~\eqref{eq:Szekeres_Da} is the expression for $\delta$,
\begin{equation}
\delta = - \frac{\beta_{\rm +} \mathcal{D}}{X}\,,
\end{equation}
and the equation for $\ell^0$
\begin{equation}
\frac{\dot{\ell}^0}{\ell^0}= - \frac{\beta_{\rm +} \dot{\mathcal{D}}}{X} - 2 \mathcal{H}\,.
\end{equation}

We present the results of the comparison by plotting the variation
\begin{equation}
\Delta D_{\rm ang} (\rm BGO, Sachs) = \frac{\Delta D_{\rm ang}^{\rm BGO}-\Delta D_{\rm ang}^{\rm Sachs}}{\Delta D_{\rm ang}^{\rm Sachs}}\,.
\end{equation}
%%%%%%%%%%%%%%%%%%%%%%%%%%%%%%%%%%%%%%%%%%%%%%%%%%%%%%
\begin{figure}[h]
    \centering
        \includegraphics[width=\columnwidth]{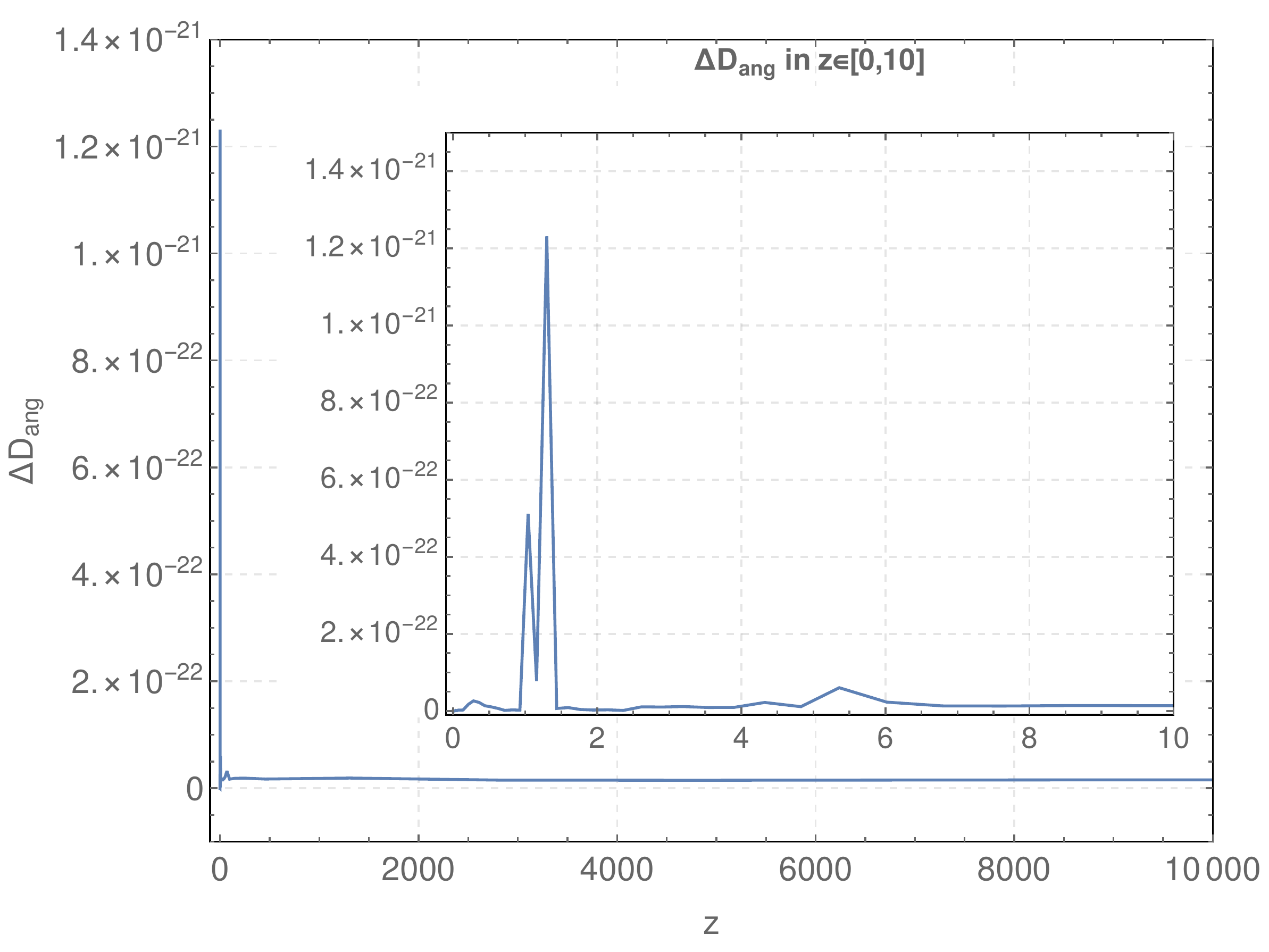}  
\caption{Deviation $\Delta D_{ \rm ang}(\mathrm{BGO},\mathrm{Sachs})$ for the angular diameter distance in the Szekeres spacetime. The observables are evaluated for a light bundle moving along the $q_{\rm 3}$-axis and received by the observer placed at $q^{\mu}_{\rm \mathcal{O}}=(\eta_{\rm \mathcal{O}},0,0,0)$ in a region where $\left.\delta\right|_{\mathcal{O}}=0$.}
 \label{fig:Szekeres_dDa}
\end{figure}
From Fig.~\ref{fig:Szekeres_dDa} we can conclude that also in the Szekeres model we have a good agreement. 
However, contrary to the $\Lambda$CDM case, the variation $\Delta D_{\rm ang}$ plotted represents the comparison between two numerical computations and it cannot be considered as numerical error over the observable. Nevertheless, the very good agreement between the \texttt{BiGONLight} code and the traditional method using the Sachs focusing equation is another piece of evidence that our code is a reliable tool for studying light propagation also in more complicated spacetimes.

%%%%%%%%%%%%%%%%%%%%%%%%%%%%%%%%%%%%%%%%%%%%%%%%%%%%%%%%%%%%%%%%%%%%%%%%%%%%%%%%%%%%%%%%%%%%%%%%%%%%%%%%%

\section{Linearising the BGO formalism: solutions for the plane-parallel universe}
\label{apx:linear_obs}

In this appendix we consider the flat FLRW background with linear perturbations in the synchronous-comoving gauge. We obtain the linearised evolution equations for the BGO, Eq.~\eqref{eq:1PT_BGO_syst}, the general expressions for their solutions, Eq.~\eqref{eq:1PT_BGO}, and the linear angular diameter distance $D_{\rm ang}^{\rm Lin}$ written in terms of the BGO, Eq.\eqref{eq:Dlin_lambda}.
We then specialise the general solutions to the $\Lambda$CDM background with perturbations at first order in standard cosmological perturbation theory and within our plane-parallel toy model and finally obtain the analytic expressions for $z^{\rm Lin}$ and $D_{\rm ang}^{\rm Lin}$ that we used in Section~\ref{sec:method}. 

The spacetime metric has the form
\begin{equation}
\tilde{g}_{\mu \nu}=a^2 g_{\mu \nu}
\end{equation}
and is expanded at first order as
\begin{equation}
\tilde{g}_{\mu \nu}= a^2 (\bar{g}_{\mu \nu}+\delta g_{\mu \nu})
\end{equation}
where $\bar{g}_{\mu \nu}$ is the conformal flat FLRW background, i.e. the Minkowski metric and $\delta g_{\mu \nu}$ represents the first-order scalar perturbations in the synchronous-comoving gauge, in general given by $\delta g_{\mu \nu}= {\rm Diag}(0, \delta g_{11}, \delta g_{22}, \delta g_{33})$.

The first observable that we study in this work is the redshift, defined as
\begin{equation}
1+z=\frac{\left.\tilde{g}_{\mu \nu}\tilde{\ell}^{\mu}\tilde{u}^{\nu}\right|_{\mathcal{S}}}{\left.\tilde{g}_{\mu \nu}\tilde{\ell}^{\mu}\tilde{u}^{\nu}\right|_{\mathcal{O}}}\,.
\label{eq:redshift_def}
\end{equation}
In the above expression ${u}^{\mu} $ is the four-velocity of the observer (source) and $\ell^{\mu}$ is the tangent vector to the photon geodesics. For our coordinates choice, all observers and sources are comoving with the cosmic flow with four-velocity given by 
\begin{equation}
\tilde{u}^{\mu}\equiv \frac{1}{a}\bar{u}^{\mu} =\frac{1}{a}\left(1 ,0, 0, 0\right)
\end{equation}
at all orders and the null tangent vector is expanded as
\begin{equation}
\tilde{\ell}^{\mu}=\frac{a^2_{\rm \mathcal{O}}}{a^2}\left(\bar{\ell}^{\mu}+\delta \ell^{\mu}\right)
\end{equation}
with $\tilde{\bar{\ell}}^{\mu}=\frac{a^2_{\rm \mathcal{O}}}{a^2}\left(\bar{\ell}^{0}, \bar{\ell}^{i}\right)$\footnote{The conformal tangent vector in the FLRW background $\bar{\ell}^{\mu}$ is constant. Note that usually the normalisation $\bar{\ell}^{0}\pm 1$ is used for the temporal component. Here, however, we leave it un-normalised.}. The linear redshift is then given by
 \begin{equation}
 1+z=\frac{a_{\mathcal{O}}}{a_{\mathcal{S}}}\left[ 1 +\frac{1}{\bar{\ell}^0} \left( \left.  \delta \ell^{0}\right|_{\mathcal{S}} - \left.  \delta \ell^{0}\right|_{\mathcal{O}} \right) \right]\, ,
 \label{eq:linear_z}
 \end{equation}
 where $\delta \ell^{0}$ is founded from the first-order geodesic equation
 \begin{equation}
 \frac{d \delta \ell^{\mu}}{d \lambda}= \frac{1}{2}\bar{g}^{\mu \sigma} \partial_{\sigma} \delta g_{\alpha \beta} \bar{\ell}^{\alpha}\bar{\ell}^{\beta}-\bar{g}^{\mu \sigma}\bar{\ell}^{\alpha} \partial_{\alpha} \delta g_{\sigma \beta}\bar{\ell}^{\beta}\,.
 \label{eq:lin_geod}
 \end{equation}
 
The second observable is the angular diameter distance $\tilde{D}_{\rm ang}$. However, it is more convenient to expand the conformal angular distance $D_{\rm ang}$ which we write here in terms of the BGO as (for a derivation see \citep{Grasso:2018mei} and \citep{Korzynski:2019oal})
\begin{equation}
D_{\rm ang}=\left. \ell^{\mu}u_{\mu}\right|_{\mathcal{O}}|\mathrm{det}(\WXL{}\UD{\boldsymbol{A}}{\boldsymbol{B}})|^{\frac{1}{2}}
\label{eq:dang_def_app_lin}
\end{equation}
and then obtain $\tilde{D}_{\rm ang}$ from the very well-known conformal transformation $\tilde{D}_{\rm ang}= \frac{a}{a_{\mathcal{O}}} D_{\rm ang}$, that we verified for Eq. \eqref{eq:dang_def_app_lin}.
The BGO $\WXL{}\UD{\boldsymbol{A}}{\boldsymbol{B}}$ are expanded as
\begin{equation}
\WXL{}\UD{\boldsymbol{A}}{\boldsymbol{B}}=\overline{\WXL}{}\UD{\boldsymbol{A}}{\boldsymbol{B}}+ \delta \WXL{}\UD{\boldsymbol{A}}{\boldsymbol{B}}\,,
\end{equation}
where $\overline{\WXL}$ and $\delta \WXL$ are found by solving the linearised GDE \eqref{eq:GDE_for_BGO} in conformal space
\begin{equation}
\frac{d}{d \lambda} \mathcal{W}=\begin{pmatrix}
0 & \mathbb{1}_{4 \times 4}\\
R_{\ell \ell} & 0
\end{pmatrix} \mathcal{W}
\end{equation} 
Notice that the optical tidal matrix in the frame is purely a first-order quantity - the conformal Riemann tensor vanishes in the background - and it is given by:
\begin{equation}
R\UDDD{\boldsymbol{\mu}}{\ell}{\ell}{\boldsymbol{\nu}}=\bar{\phi}^{\rho \boldsymbol{\mu}} \delta R_{\rho \alpha \beta \sigma}\bar{\ell}^{\alpha} \bar{\ell}^{\beta}
 \bar{\phi}\UD{\sigma}{\boldsymbol{\nu}}
\end{equation}
where $\delta R_{\rho \alpha \beta \sigma}$ is the first-order Riemann tensor and $\bar{\phi}\UD{\mu}{\boldsymbol{\alpha}}=(u^{\mu}, \bar{\phi}\UD{\mu}{\boldsymbol{A}}, \bar{\ell}^{\mu})$ is the background parallel transported frame along the background geodesic.

Let us start by solving the background GDE, which reads:
\begin{equation}
\left\{\begin{array}{l}
\frac{d \, \overline{\WXX}{}\UD{\boldsymbol{\mu}}{\boldsymbol{\nu}}}{d \lambda}=\overline{\WLX}{}\UD{\boldsymbol{\mu}}{\boldsymbol{\nu}}\\
\frac{d \, \overline{\WLX}{}\UD{\boldsymbol{\mu}}{\boldsymbol{\nu}}}{d \lambda}=0\\
\frac{d \, \overline{\WXL}{}\UD{\boldsymbol{\mu}}{\boldsymbol{\nu}}}{d \lambda}= \overline{\WLL}{}\UD{\boldsymbol{\mu}}{\boldsymbol{\nu}}\\
\frac{d \, \overline{\WLL}{}\UD{\boldsymbol{\mu}}{\boldsymbol{\nu}}}{d \lambda}=0
\end{array}\right.
\label{eq:background_BGO_syst}
\end{equation}
with initial conditions $\overline{\mathcal{W}}=\mathbb{1}_{8 \times 8}$. The solution is
\begin{equation}
\overline{\mathcal{W}}=\begin{pmatrix}
\delta\UD{\boldsymbol{\mu}}{\boldsymbol{\nu}} & (\lambda - \lambda_{\mathcal{O}})\delta\UD{\boldsymbol{\mu}}{\boldsymbol{\nu}}\\
0 & \delta\UD{\boldsymbol{\mu}}{\boldsymbol{\nu}}
\end{pmatrix}
\label{eq:BGO_background}
\end{equation}
Next we find the first-order BGO from:
\begin{equation}
\left\{\begin{array}{l}
\frac{d \, \delta \WXX{}\UD{\boldsymbol{\mu}}{\boldsymbol{\nu}}}{d \lambda}=\delta \WLX{}\UD{\boldsymbol{\mu}}{\boldsymbol{\nu}}\\
\frac{d \, \delta \WLX{}\UD{\boldsymbol{\mu}}{\boldsymbol{\nu}}}{d \lambda}=R\UDDD{\boldsymbol{\mu}}{\ell}{ \ell}{ \boldsymbol{\nu}} \\
\frac{d \, \delta \WXL{}\UD{\boldsymbol{\mu}}{\boldsymbol{\nu}}}{d \lambda}= \delta \WLL{}\UD{\boldsymbol{\mu}}{\boldsymbol{\nu}}\\
\frac{d \,  \delta \WLL {}\UD{\boldsymbol{\mu}}{\boldsymbol{\nu}}}{d \lambda}= (\lambda - \lambda_{\mathcal{O}})R\UDDD{\boldsymbol{\mu}}{\ell}{ \ell}{ \boldsymbol{\nu}} 
\end{array}\right.
\label{eq:1PT_BGO_syst}
\end{equation}
with initial conditions $\delta \mathcal{W}=\boldsymbol{0}_{8 \times 8}$, where we have replaced the background solutions \eqref{eq:BGO_background}.
We obtain
\begin{equation}
\left\{\begin{array}{l}
\delta \WXX{}\UD{\boldsymbol{\mu}}{ \boldsymbol{\nu}}=\int^{\lambda_{\mathcal{O}}}_{\lambda}\int^{\lambda_{\mathcal{O}}}_{\lambda'}R\UDDD{\boldsymbol{\mu}}{\ell}{ \ell}{ \boldsymbol{\nu}} d\lambda'd\lambda''\\
\delta \WXL{}\UD{\boldsymbol{\mu}}{ \boldsymbol{\nu}}= \int^{\lambda_{\mathcal{O}}}_{\lambda}(\lambda_{\mathcal{O}}-\lambda')(\lambda- \lambda')R\UDDD{\boldsymbol{\mu}}{\ell}{ \ell}{ \boldsymbol{\nu}} d\lambda'\\
\delta \WLX{}\UD{\boldsymbol{\mu}}{ \boldsymbol{\nu}}= -\int^{\lambda_{\mathcal{O}}}_{\lambda}R\UDDD{\boldsymbol{\mu}}{\ell}{ \ell}{ \boldsymbol{\nu}} d\lambda'  \\
\delta \WLL{}\UD{\boldsymbol{\mu}}{ \boldsymbol{\nu}}= \int^{\lambda_{\mathcal{O}}}_{\lambda}(\lambda_{\mathcal{O}}-\lambda')R\UDDD{\boldsymbol{\mu}}{\ell}{ \ell}{ \boldsymbol{\nu}} d\lambda'
\end{array}\right.\,.
\label{eq:1PT_BGO}
\end{equation}
In order to find $D_{\rm ang}^{\rm Lin}$ from the expansion of Eq.~\eqref{eq:dang_def_app_lin} we need the second of the above solutions and we recall that the expansion of the square root of the determinant is given by
\begin{equation}
\sqrt{\mathrm{det} W_{XL}}  = \sqrt{\mathrm{det} \left(\overline{W_{XL}}\right)}\left[ 1+\frac{1}{2}\mathrm{tr}\left(\overline{W_{XL}}^{-1} \delta W_{XL} \right) \right]
\end{equation}
Now, looking at Eqs.~\eqref{eq:BGO_background} and~\eqref{eq:1PT_BGO} we have that
\begin{equation}
\left\{\begin{array}{l}
\sqrt{\mathrm{det} \left(\overline{W_{XL}}\right)}=(\lambda-\lambda_{\cal O})\\
\mathrm{tr}\left(\overline{W_{XL}}^{-1} \delta W_{XL} \right)=\frac{\int^{\lambda_{\mathcal{O}}}_{\lambda}(\lambda_{\mathcal{O}}-\lambda')(\lambda- \lambda')R\UDDD{\boldsymbol{A}}{\ell}{ \ell}{ \boldsymbol{A}} d\lambda'}{(\lambda-\lambda_{\cal O})}
\end{array}\right.
\end{equation}
The final result for $D^{\rm Lin}_{\rm ang}$ is
\begin{equation}
\begin{array}{c}
D_{\rm ang}^{\rm Lin}=(\ell^{0}_{\mathcal{O}}+\delta \ell^0_{\mathcal{O}})(\lambda_{\mathcal{O}}-\lambda)\\
-\frac{\ell^{0}_{\mathcal{O}}}{2}\int^{\lambda_{\mathcal{O}}}_{\lambda}(\lambda_{\mathcal{O}}-\lambda')(\lambda- \lambda')\mathrm{tr}\left(R\UDDD{\boldsymbol{A}}{\ell}{ \ell}{ \boldsymbol{B}}\right) d\lambda'\,.
\end{array}
\label{eq:Dlin_lambda}
\end{equation}
It is important to stress that all the quantities are evaluated along the background geodesic, i.e. $\eta \equiv \bar{\eta}$ and $\bar{q}_{\rm 1}(\eta)\equiv \frac{\bar{\ell}^1}{\bar{\ell}^0}(\bar{\eta}_{\rm \mathcal{O}}-\bar{\eta})+\bar{q}_{\rm 1}(\bar{\eta}_{\rm \mathcal{O}})$

We checked that our result coincides with the standard result in the literature, e.g. \cite{di2016curvature}, by simply noting that the quantity $-\frac{1}{2}\mathrm{tr}\left(R\UDDD{\boldsymbol{A}}{\ell}{ \ell}{ \boldsymbol{B}}\right)$ is nothing more than the Ricci part of the optical tidal matrix $\mathcal{R}$, usually defined as
\begin{equation}
\mathcal{R}=\frac{1}{2}R_{\alpha \beta} \ell^{\alpha} \ell^{\beta}=-\frac{1}{2}R\UDDD{\mu}{\alpha}{\beta}{\mu} \ell^{\alpha} \ell^{\beta}\,,
\end{equation}
and evaluated at first order. 

We finally specialise the above results for our plane-parallel model. The linear perturbation of the spacetime metric around the flat $\Lambda$CDM background is
\begin{equation}
\delta g_{\mu \nu}=\begin{pmatrix}
0 & 0 & 0 & 0\\
0 & -\mathcal{F}-\frac{10}{3 c^2}\phi_{\rm 0} & 0 & 0\\
0 & 0 & -\frac{10}{3 c^2}\phi_{\rm 0} & 0\\
0& 0 & 0 & -\frac{10}{3 c^2}\phi_{\rm 0}
\end{pmatrix}\, ,
\label{eq:sync_gauge_quantities}
\end{equation}
where we define 
\begin{equation}
\mathcal{F}(\eta, q_{\rm 1})=\frac{4}{3}\frac{\partial^2_{\rm q_{\rm 1}}\phi_{\rm 0}(q_{\rm 1})}{\stuff}\mathcal{D}(\eta)
\end{equation}
A straightforward substitution gives for the redshift
 \begin{equation}
 \begin{array}{l}
 1+z^{\rm Lin}=\\
 \vspace*{0.5 cm} \frac{a_{\mathcal{O}}}{a_{\mathcal{S}}}\left[ 1 -\left(\frac{\bar{\ell}^1}{\bar{\ell}^0}\right)^2 \int^{\bar{\eta}_{\mathcal{O}}}_{\bar{\eta}_{\mathcal{S}}} \frac{2}{3}\frac{\partial_{\rm q_{\rm 1}} \phi_{\rm 0}(q_{\rm 1}(\bar{\eta}'))}{\stuff}\partial_{0}{\mathcal{D}}(\bar{\eta}')   d \bar{\eta}' \right]
 \end{array}
 \label{eq:zlin_plane-parallel_2}
 \end{equation}
and for the angular diameter distance
\bwt
\begin{equation}
\tilde{D}^{\rm Lin}_{\rm ang}(\eta)=\frac{a}{a_{\mathcal{O}}}\left[(\bar{\eta}_{\rm \mathcal{O}}-\bar{\eta})+ \frac{\bar{\ell}^{1^2}}{2 \bar{\ell}^{0^2}}\int^{\eta_{\rm \mathcal{O}}}_{\eta} \int^{\eta_{\rm \mathcal{O}}}_{\eta'}  \dot{\mathcal{F}} d\eta'd\eta''+\int^{\eta_{\rm \mathcal{O}}}_{\eta}(\eta_{\rm \mathcal{O}}-\eta')(\eta-\eta') \frac{\mathcal{R}(\eta')}{\bar{\ell}^{0^2}}  d\eta'\right]
\label{eq:Dang_time_conf}
\end{equation}
\ewt
The two last relations are those we use in section~\ref{sec:method} for our comparison.

%%%%%%%%%%%%%%%%%%%%%%%%%%%%%%%%%%%%%%%%%%%%%%%%%%%%%%%%%%%%%%%%%%%%%%%%%%%%%%%%%%%%%%%%%%%%%%%%%%%%%%%%%
%%%%%%%%%%%%%%%%%%%%%%%%%%%%%%%%%%%%%%%%%%%%%5
\section{Comparison with the Szekeres metric}
\label{par:crfSzekeres}

In this section we compare the Szekeres spacetime with the plane-parallel case considered in this work. In his original paper \cite{Szekeres:1974ct}, Szekeres studied all the solutions to the Einstein field equation with irrational dust for line elements having the form
\begin{equation}
ds_{\rm Sz}^2=-d t^2 + e^{2 \alpha(t,q_{\rm 1},q_{\rm 2},q_{\rm 3})} dq_{\rm 1}^2 + e^{2 \beta(t,q_{\rm 1},q_{\rm 2},q_{\rm 3})} (dq_{\rm 2}^2+ dq_{\rm 3}^2)
\label{eq:Old_Szekeres_metric}
\end{equation}
Two different classes of solutions can be distinguished: class-\textsc{I} solutions are a generalization of the  Lema\^{i}tre-Bondi-Tolman model while class-\textsc{II} solutions are a generalization of Kantowski-Sachs and FLRW model. 
Subsequently, Barrow and Stein-Schabes \cite{Barrow:1984zz} generalized the Szekeres solutions by adding a cosmological constant $\Lambda$ to the dust. 
More recently, Bruni and Meures \cite{Meures:2011ke} presented a new formulation of the class-\textsc{II} Szekeres solutions in which the separation between inhomogeneities and the FLRW background is explicitly provided and thus the spacetime metric is presented in a more convenient form for cosmological applications. We compare our plane-parallel metric with their formulation. We begin by summarizing the results presented in \cite{Meures:2011ke}.

The authors focused their analysis on the Szekeres solutions which admit a flat FLRW background and such that the line element\footnote{We choose here to use  our notation instead that of that of  \cite{Meures:2011ke}. The line element (\ref{eq:Szekeres_metric}) is different from the one presented in \cite{Meures:2011ke} since we use conformal time and we have chosen a different axis of symmetry. Of course this does not affect any results, since it is easy to show that the two metrics are equivalent under a coordinate transformation.} can be written as
\begin{equation}
ds_{\rm Sz}^2=a^2 \left( -d \eta^2 + \gamma^{\rm{Sz}}_{\rm 11} dq_{\rm 1}^2 + \gamma^{\rm{Sz}}_{\rm 22} dq_{\rm 2}^2+ \gamma^{\rm{Sz}}_{\rm 33}dq_{\rm 3}^2 \right)
\end{equation}
where
\begin{equation} \label{eq:Szekeres_metric}
\begin{split}
\gamma^{\rm{Sz}}_{\rm 11} =& X^2(\eta,q_{\rm 1},q_{\rm 2},q_{\rm 3}) \\
\gamma^{\rm{Sz}}_{\rm 22}= & 1\\
\gamma^{\rm{Sz}}_{\rm 33}=& 1.
 \end{split}
\end{equation}
As it is shown in \cite{Meures:2011ke}, thanks to the symmetry of the problem, the function $X(\eta,q_{\rm 1},q_{\rm 2},q_{\rm 3})$ can be decomposed as 
\begin{equation}\label{eq:sz_X}
X(\eta,q_{\rm 1},q_{\rm 2},q_{\rm 3})=F(\eta,q_{\rm 1})+A(q_{\rm 1},q_{\rm 2},q_{\rm 3})\,,
\end{equation}
where the function $F(\eta, q_{\rm 1})$ satisfies the Newtonian evolution equation for the first-order density contrast\footnote{This was shown implicitly in section $5$ of the Szekeres' original paper \cite{Szekeres:1974ct} and subsequently by many other authors as those of \cite{bonnor:1977pp}. However, it was Goode and Wainwright who recognized explicitly that the relativistic equations for the density fluctuations in Szekeres model are the same as in Newtonian gravity, \cite{Goode:1982pg}. They also provide a new formulation of the Szekeres solutions, much more useful in cosmology, in which the relationship with the FLRW solution is clarified.}
\begin{equation} \label{eqforF}
\ddot{F} +\mathcal{H}\dot{F}-\frac{3}{2}\mathcal{H}_0^2\Omega_{m_0}\frac{F}{a}=0\,,
\end{equation}
which admits two linearly independent solutions, the growing and decaying modes, as is well known. Then $F(\eta, q_{\rm 1})$ coincides with the linear density contrast and more precisely we have $F(\eta, q_{\rm 1})= -\delta_{\rm Lin}(\eta, q_{\rm 1})$ \footnote{The minus sign between $F$ and $\delta$ follows from the fact that in eq. ($A8$) of \cite{Meures:2011ke} the authors set, in full generality,  $\delta_{in}=-\frac{F_{in}}{F_{in}+A}$.}. Neglecting the decaying modes it is possible, without loss of generality, to factorize $F(\eta, q_{\rm 1})$ as\footnote{The time-dependent-only growing mode is denoted by $f_+$ in \cite{Meures:2011ke} and it is given in a dimensionless time variable $\tau$ in Eq.~(11b). To match $\cal D$ in \eqref{eq:growth_fact} and  $f_+$ one needs to: first transform $f_+(\tau)$ to conformal time $f_{+}(\eta)$ and then normalise such that $f_{+}(\eta_{\rm 0})=1$. The final result is $F(\eta, q_{\rm 1})$ as in \eqref{eq:F_Sz}.}
\begin{equation}
\label{eq:F_Sz}
F(\eta, q_{\rm 1})= {\cal D}(\eta) \beta_+(q_{\rm 1})\, ,
\end{equation}
where $\mathcal{D}$ is the growing mode solution for the density contrast given by (see e.g. Eq.~($5.13$) in \cite{Villa:2015ppa}, where we have already normalized in order to have $\mathcal{D}_0=1$)
\begin{equation}
\mathcal{D} (\eta) = \frac{a}{\frac{5}{2}\Omega_{\rm{m 0}} } \sqrt{1+\frac{\Omega_{\rm{\Lambda 0}}}{\Omega_{\rm{m 0}}}a^3}\,  {}_2 F_{1} \left( \frac{3}{2}, \frac{5}{6}, \frac{11}{6}, -\frac{\Omega_{\rm{\Lambda 0}}}{\Omega_{\rm{m 0}}}a^3 \right)\,,
\label{eq:growth_fact}
\end{equation}
with ${}_2 F_{1} \left(a,b,c, x\right)$ being the Gaussian (or ordinary) hypergeometric function. 

On the other hand, our Newtonian plane-parallel metric is given by:
\begin{equation} \label{eq:app_metricNWT}
\begin{split}
\gamma^{\rm{N}}_{11} =& \left(1-\frac{2}{3}\frac{ \mathcal{D} \partial_{\rm q_{\rm 1}}^2 \phi_0}{ \stuff} \right)^2 \\
\gamma^{\rm{N}}_{22}= & 1\\
\gamma^{\rm{N}}_{33}=& 1.
 \end{split}
\end{equation}

We now investigate the link between \eqref{eq:app_metricNWT} and \eqref{eq:Szekeres_metric} by comparing the two forms of $\gamma_{\rm 11}$ and referring to Eqs.~\eqref{eq:sz_X} and \eqref{eq:F_Sz}. Let us start from~\eqref{eq:F_Sz}: to fix the time-independent function $\beta_{\rm +}(q_{\rm 1})$, one can take advantage from the fact that $\delta_{\rm Lin}(\eta, q_{\rm 1})=-{\cal D}(\eta) \beta_+(q_{\rm 1})$ and use the cosmological Poisson equation \eqref{eq:Poisson_eq} to find
\begin{equation}
\beta_+(q_{\rm 1}) = - \frac{2}{3}\frac{\partial_{\rm q_{\rm 1}}^2 \phi_{\rm 0}(q_{\rm 1})}{\stuff}\, .
\label{eq:beta+}
\end{equation}
At this point we have completely fixed $F(\eta, q_{\rm 1})$. Now, by looking at Eq.~\eqref{eq:sz_X} it is straightforward to conclude that the two metrics \eqref{eq:app_metricNWT} and \eqref{eq:Szekeres_metric} are fully equivalent if $A(q_{\rm 1},q_{\rm 2},q_{\rm 3})= 1$\footnote{To be more precise it would be enough to neglect the dependence on $(q_{\rm 2}, q_{\rm 3})$  in~\eqref{eq:sz_X} by imposing that $A(q_{\rm 1},q_{\rm 2},q_{\rm 3})\equiv A(q_{\rm 1})$, i.e. $$X(\eta, q_{\rm 1})= A(q_{\rm 1})+F(\eta, q_{\rm 1})\, .$$ 
However, if this were the case, it is easy to show that performing the coordinate transformation $\tilde{q_{\rm 1}}=\int A(q_{\rm 1}) d q_{\rm 1}$ and rescaling $\tilde{F}(\eta, \tilde{q}_{\rm 1})=\frac{F(\eta, q_{\rm 1})}{A(q_{\rm 1})}$, we obtain again $X(\eta, \tilde{q}_{\rm 1})= 1 + \tilde{F}(\eta, \tilde{q}_{\rm 1})$. }.
However, this cannot be the case, as we will now show. Let us start by noticing that planar symmetry implies that the metric components can depend only on the coordinates $(\eta, q_{\rm 1})$, while in the Szekeres symmetry the metric can depend in general on all spatial coordinates (and time). In~\cite{Meures:2011ke} this dependence is encoded in $A(q_{\rm 1},q_{\rm 2},q_{\rm 3})$ in Eq.~\eqref{eq:sz_X} and has the form
\begin{widetext}
\begin{equation}
A= 1+ B \beta_{\rm +}(q_{\rm 1})\left[\left(q_{\rm2}+\gamma(q_{\rm 1})\right)^2+\left(q_{\rm 3}+\omega(q_{\rm 1})\right)^2 \right] \, ,
\label{eq:Sz_A}
\end{equation}
\end{widetext}
thus the only possibility to have $A=1$ is that $B=0$.
After simple manipulations of the Einstein equations together with Eq.~\eqref{eq:Sz_A} we find that 
\begin{equation}
2 B=\frac{3}{2}\frac{\stuff}{a} \mathcal{D} + \mathcal{H} \dot{\mathcal{D}} \, .
\end{equation}
In \cite{Villa:2015ppa} it is shown that the R.H.S. is constant and it is always different from zero. Indeed, from Eq.($5.52$) in \cite{Villa:2015ppa} we find
\begin{equation}
B= \frac{5}{4} \stuff \frac{\cal D_{\rm in}}{a_{\rm in}}\, ,
\label{eq:link_on_B}
\end{equation}
where $\mathcal{D}_{\rm in}= a_{\rm in}$ for Einstein-de Sitter initial conditions. We finally conclude that the two metrics \eqref{eq:Szekeres_metric} and \eqref{eq:app_metricNWT} cannot coincide.

% To modify and compile the bibliography:
%- go to https://inspirehep.net/ or similar
%- find the paper
%- click on BibTex 
%- copy and paste on the .bib file (please put new references on top). Save
%- to compile: pdfLatex + bibTex + pdfLatex a couple of times for the .tex file

\bibliography{ms}

\end{document}